\documentclass[12pt,longbibliography]{article}

\usepackage[utf8]{inputenc}

\usepackage[T1]{fontenc}

\usepackage[a4paper, margin=2.7cm]{geometry}

\usepackage{amsmath, amsthm, amsfonts, amssymb}
\usepackage{float}
\usepackage{mathrsfs}           
\usepackage{graphicx}
\usepackage[usenames]{color}
\usepackage{soul}
\usepackage[colorlinks=true,linkcolor=blue,citecolor=blue,urlcolor=blue,breaklinks]{hyperref}

\usepackage{bbm} 

\usepackage{todonotes}
\usepackage{bbold}

\usepackage{caption,subcaption}
\usepackage{geometry}

\usepackage{url}

\def\N{\mathbb{N}}
\def\R{\mathbb{R}}

\newcommand {\f}   {\frac}

\newtheorem{thm}{Theorem}[section]

\theoremstyle{definition}
\newtheorem{dfn}[thm]{Definition}
\theoremstyle{remark}
\newtheorem{rem}[thm]{Remark}

\title{
  An understanding of the physical solutions
	and the blow-up phenomenon for
	Nonlinear Noisy Leaky Integrate and Fire neuronal models
}
\author{María J Cáceres\thanks{Departamento de Matem\'atica Aplicada,
Universidad de Granada, 18071 Granada, Spain.
\texttt{caceresg@ugr.es}} \and Alejandro Ramos-Lora\thanks{Departamento de Matem\'atica Aplicada,
Universidad de Granada, 18071 Granada, Spain.
\texttt{ramoslora@ugr.es}}}

\begin{document}

	\maketitle
	
	\begin{abstract}

          The Nonlinear Noisy Leaky Integrate and Fire neuronal models
          are  mathematical models
          that describe the activity of neural networks.   
          These models have been studied 
           at a microscopic level, using Stochastic Differential
          Equations, and at a mesoscopic/macroscopic level, through
          the mean field limits using Fokker-Planck type equations.
	  The aim of this paper is to improve their understanding,
	        using a numerical study of their particle systems. We analyse in depth the behaviour
	        of the classical and physical solutions of the Stochastic Differential
	        Equations and, we compare it with
	        what is already known about the Fokker-Planck equation.
	        This allows us to better understand what happens in the neural
	        network when an explosion occurs in finite time.
                After firing all neurons at the same time,  if the system is
                weakly connected,
                the neural network converges
                towards its unique steady state.  Otherwise,  its behaviour is
                more complex, because it can tend towards
                a stationary state or a ``plateau'' distribution.

	\end{abstract}

	\section{Introduction}
	Nonlinear Noisy Leaky Integrate and Fire neuronal (NNLIF) models are one of the simplest
	models used to describe the behaviour of a neuronal network 
	\cite{lapicque,sirovich,omurtag,brunel2000dynamics,BrGe,brunel1999fast}.
	In recent years,  NNLIF models
        have been studied from a mathematical point of view;
	 at the microscopic level, using  Stochastic Differential Equations (SDE)
	\cite{delarue2015global,delarue2015particle,liu2020rigorous}, and at
        a mesoscopic/macroscopic level, through
          the mean field limits using Fokker-Planck type equations (FPE)
        \cite{caceres2011analysis,CGGS,caceres2014beyond,caceres2017blow,caceres2018analysis,caceres2019global,hu2019structure,roux2020towards}.
	The considerable amount of publications and unanswered questions
        on these models reveal
	their high mathematical complexity, despite their simplicity.

\
	
	The aim of this paper is to advance the understanding of the
        NNLIF models.	
        We analyse in depth the behaviour
	of the classical and physical solutions of the Stochastic Differential
	Equations and we compare it with
	what is already known about the Fokker-Planck equation,
        using a numerical study of their particle systems.
	This allows us to better understand what happens in the neural
	network when an explosion occurs in finite time, which
         is one of the most
	important open problems about this kind of models.

	\subsection{Stochastic Differential Equation and Fokker-Planck Equation}
	\label{sec:model}
	
	Let us consider a large set of $\mathcal{N}$ 
	identical neurons which are connected to each other in a network
	and described by the  Nonlinear Noisy Leaky Integrate and Fire
        (NNLIF) model.
	This model represents  the network activity
        in relation to        
        the
	{\em membrane potential}, which is
	the potential difference on both sides of the neuronal membrane. 
	The membrane potential $V_i(t)$ of a single neuron
	$i$ 
        is given by  \cite{brunel1999fast, brunel2000dynamics,  compte2000synaptic, renart2004mean}:
	\begin{equation}
	C_m\dot{V}_i(t)=-g_L(V_i(t)-V_L)+I_i(t),
	\quad  i=1, \ldots , \mathcal{N},
	\label{eq: *}
	\end{equation}
	where $C_m$ is the capacitance of the membrane, $g_L$ is the leak conductance,
	$V_L$ is the resting potential and $I_i(t)$ are the synaptic currents.
	These currents are produced by the local and external synapses, i.e. they are the sum of spikes received from $C$ neurons
	(inside and outside the neuron network):
	\begin{equation}
	I_i(t)=\sum_{j}\sum_{k}J_{ij}\delta(t-t_{ik}^j-\delta).
	\end{equation}
        The Dirac delta
	$\delta(t-t_{ik}^j-\delta)$ models the input contribution of each spike;
        $t_{ik}^j$ is the time when the $k$-th
	spike of the $j$-th neuron took place,
        $J_{ij}$ is the synaptic strength (positive value for
        excitatory neurons and negative value
	for inhibitory ones), and 
        $\delta$ in the argument is the synaptic delay.
        
	\	
	
	A  neuron spikes when its membrane voltage reaches the firing
	threshold value $V_F$. Then, the neuron discharges
	itself by sending a spike perturbation over the network, and its
        membrane potential is set to the reset value $V_R$.
	The relation between the three values  $V_L$, $V_F$
	and $V_R$ is  the following:  $V_L<V_R<V_F$.
	
\
	
	We consider networks in which pair correlations
	can be neglected, as a consequence of the network sparse random
        connectivity,
	i.e.,  
        $C/\mathcal{N} \ll 1$.
	We also consider each input as a small contribution
	compared to the firing threshold
	($|J_{ij}| \ll V_F $).
	Under these conditions (see \cite{brunel1999fast} for more details),
        if we assume
	every neuron in the network to spikes according to Poisson processes
	with a common instantaneous probability of emitting a spike per unit
	time $\nu $ (firing rate),  
	 we can describe the synaptic current of a neuron as
	\begin{equation}
	I_i(t)\simeq \mu_c(t)+\sigma_c\eta_i(t)
	\end{equation}
	where $\mu_c=b\nu(t-\delta)$ 
	represents the average value of the current
	(taking into account the excitatory and inhibitory neurons in the
	network), $ \sigma^2_c=\alpha\nu(t-\delta) $ its variance, 
	and $\eta_i(t)$ are independent Gaussian white noises.
	The constant $\alpha$ depends on the number and the strength of
        excitatory and inhibitory neurons.
	The parameter $b$  will be called the {\em connectivity parameter} and
	its value shows how excited or inhibited the network is.
	If $b>0 $ the network is average-excitatory;
	a high value of $b$ means that neurons are highly connected, on the other hand $b<0$ describes a network which is average-inhibitory. The limiting case $b=0$ means that neurons are not connected with each other and the system becomes linear.
	
\		
	
	As we mentioned above, the equation \eqref{eq: *}
	is completed with the following boundary condition: 
	$
	V_i(t_s^-)=V_F$ and $V_i(t_s^+)=V_R$,
        with
	$t_s$  the firing time of neuron $i$.

\
        
	Without loss of generality, we will take $C_m=g_L=1$. We rewrite
	$\nu(t)=\nu_{ext}+N(t)$, splitting according to external,
        $\nu_{ext}$, and
	internal, $N(t)$, activity; where $\nu_{ext}$ is a constant and
	$N(t)$ is the firing rate of the network, and
	$ \sigma_c = \sqrt{2a(N)} $, with $a(N)=a_0+a_1N$ ($0\le a_0,a_1$).
        With these considerations and the translation of $V$ by a factor
	$V_L+b\nu_{ext}$, the equation \eqref{eq: *} becomes 
	\begin{equation}\label{eq:model-main-delay}
	\dot{V_i}(t)=-V_i(t)+bN(t-\delta)+\sqrt{2(a_0+a_1N(t-\delta))}\eta_i(t),
	\end{equation}
	or, in the case without transmission delay  ($\delta = 0$)
	\begin{equation}\label{eq:model-main}
	\dot{V_i}(t)=-V_i(t)+bN(t)+\sqrt{2(a_0+a_1N(t))}\eta_i(t).
	\end{equation}
	Following the mean-field limit  when the number of neurons tends
        to infinity,
	$\mathcal{N}\to \infty$, given in 
	\cite{delarue2015particle,delarue2015global,liu2020rigorous}
	we can define
	$e(t)$
	as the theoretical expected number of times that the firing threshold $V_F$
	has been reached, by any neuron of the network before time $t$.
	And assuming that the neurons become asymptotically independent,
	$e(t)$ can be found as the following limit
	\begin{equation}
	  e(t) = \lim_{\mathcal{N}\to \infty} \frac{1}{\mathcal{N}}
          \sum_{i}\sum_{j}\sum_{k}\mathbb{1}_{\{t_{ik}^j\leq t\}}=
	\lim_{\mathcal{N}\to \infty} \frac{1}{\mathcal{N}}\sum_{j}^N\sum_{k}\mathbb{1}_{\{\tau_{k}^j\leq t\}},
	\label{eq: expectation}
	\end{equation}
	where $\left(\tau_{k}^j\right)_{k\ge 1}$ denotes the sequence
	of spike times of the $j$ neuron  of the network.
	In \cite{delarue2015particle,delarue2015global} was proved that firing rate $N(t)$ is the time derivative of the expectation:
	$ N(t) = e'(t) $.
	In this way, if we  choose a constant drift parameter $ a=a_0 $, 
        Eq.(\ref{eq:model-main-delay}) becomes the nonlinear stochastic
        mean-field equation:
	\begin{equation}\label{eq:model-simplify-delay}
	\dot{V}(t)=-V(t)+be'(t-\delta)+\sqrt{2a_0}\eta(t),
	\end{equation}
	and Eq.(\ref{eq:model-main}), without delay, becomes 
	\begin{equation}\label{eq:model-simplify}
	\dot{V}(t)=-V(t)+be'(t)+\sqrt{2a_0}\eta(t).
	\end{equation}
	This microscopic stochastic description has a related
	nonlinear Fokker-Planck equation
	\cite{brunel2000dynamics,delarue2015particle,delarue2015global,liu2020rigorous}:
	\begin{equation}\label{eq:edp-delay}
	\frac{\partial p}{\partial t}(v, t)+\frac{\partial}{\partial v}\left[h\left(v,N(t-\delta)\right)p(v,t)\right]-a\left(N(t-\delta)\right)\frac{\partial^2p}{\partial v^2}(v,t)=\delta (v-V_R)N(t),
	\end{equation}
	whose expression without delay is
	\begin{equation}\label{eq:edp}
	\frac{\partial p}{\partial t}(v, t)+\frac{\partial}{\partial v}\left[h\left(v,N(t)\right)p(v,t)\right]-a\left(N(t)\right)\frac{\partial^2p}{\partial v^2}(v,t)=\delta (v-V_R)N(t).
	\end{equation}
    This Partial Differential Equation (PDE)      
	provides  the evolution
	in time $t\geq 0$ of the probability density
	$p(v,t)>0$ of finding neurons at voltage $v\in\left(-\infty,V_F\right]$.
	The drift and diffusion coefficients are
	$h(v, N)=-v+bN$ and
	$a(N)=\sigma_c^2/2$, respectively.
	The right hand side represents the fact that neurons return to
        the reset potential $V_R$,
	just after they reach the threshold potential $V_F$ .
	This means that $p(V_F,t)=0$. Moreover, 
	$p(-\infty, t)=0$ is assumed.

\

	We point out the system is nonlinear, for $b\neq 0$, because
	the firing rate is computed as follows:
	$
	N(t):=-a\frac{\partial p}{\partial v}(V_F,t)\ge 0,
	$
        which guarantees that
	the solution of the related Cauchy problem satisfies the conservation
        law:
    \vspace{0.25cm}     
	$$
	\int_{-\infty}^{V_F}p(v,t)dv=\int_{-\infty}^{V_F}p^0(v)dv,
	\quad \mbox{for} \ t\in [0, T),
	$$
	for a given initial datum  $p(v,0)=p^0(v)\geq0$ and
	$T>0$ the maximal time of existence
	\cite{CGGS,caceres2019global,roux2020towards}.
	Without loss of generality we will take $\int_{-\infty}^{V_F}p^0(v)dv=1$.

	\

	Parallel to the studies on the NNLIF models, others have been developed for different PDE families, and their related SDE, which also describe the activity of a neural network at the level of the membrane potential. For instance:
	Population density models of Integrate and Fire neurons with jumps 
	\cite{omurtag, henry13, henry12, dumont2017mean},
	Fokker-Planck equations for uncoupled neurons \cite{NKKRC10,NKKZRC10};
	Fokker-Planck equations including conductance variables,
	\cite{Caikinetic,CCTa, perthame2013voltage, perthame2019derivation},
	time elapsed models \cite{PPD,PPD2,pakdaman2014adaptation} which
	have been recently derived as mean-field
	limits of Hawkes processes
	\cite{chevallier2015microscopic,chevallier2015mean}, 
	McKean-Vlasov equations \cite{acebron2004noisy, mischler2016kinetic},
	which are the mean-field equations related to 
	the behaviour of Fitzhugh-Nagumo neurons \cite{fitzhugh1961impulses}, 
	etc.
	
	\vspace{0.3cm}
	\subsection{What is  known so far about NNLIF models and aims of the article}
	\label{subsec: known}
	\vspace{0.3cm}
	In recent years, there have been many advances in the studies of
	NNLIF models, both numerically and analytically.
	The results obtained can be divided into those related to the global in
	time existence of the solution versus  the blow-up phenomenon,
	and those about steady states and long time behaviour of the solutions.
	All of these questions have been addressed for the Fokker-Planck
	equation \eqref{eq:edp}, however 
	the stability and asymptotic behaviour of the  SDE \eqref{eq:model-simplify}
	solutions remain open.
	Our work tries to shed light in that direction.
	
	\
	
	The existence theory for the Fokker-Planck
	equation \eqref{eq:edp}, and its extension for
	the complete model, which takes into account two separate neurons
        populations
	(excitatory and inhibitory), with and without refractory states,
	was developed in
	\cite{CGGS,carrillo2014qualitative,caceres2017blow,caceres2018analysis, caceres2019global,roux2020towards}.
        For the simplest model,
        in  the average-inhibitory case  ($b<0$) there is
	global existence. However, for the average-excitatory case ($b>0$),
	the time of existence is determined by the time in which the firing
	rate $N(t)$ does not diverge. The blow-up phenomenon appears
        \cite{caceres2011analysis} if
	there is no transmission delay, while it is avoided
	if some transmission delay or 
        stochastic discharge potential
	are considered \cite{caceres2014beyond}.        
        The analogous criteria for existence and
	blow-up phenomena  were  studied
	   in
	\cite{delarue2015global}, for the associated microscopic 
	system \eqref{eq:model-simplify}.
	Later, 
        \cite{delarue2015particle} proved that physical solutions
	to SDE do not present explosion in finite time.
	Understanding what physical solution means for the Fokker-Planck
	equation is an open problem. 
	
\	
	
        Results about steady states and long time
        behaviour were
	developed in \cite{caceres2011analysis,carrillo2014qualitative,caceres2014beyond,caceres2017blow,caceres2018analysis,roux2020towards},
        for the Fokker-Planck models.
        There is a unique steady state
        in the average-inhibitory  networks, whereas
        for an average-excitatory  network, there are different
	situations, depending on the value of the connectivity parameter $b$:
	for small values, there is a unique steady state,
	for large values, there are not steady states, and for intermediate values there are at least two.
        Long-term behavioural studies
        show that the unique steady state of  systems with low connectivity
        is stable and the exponential convergence of the solution to it is
        known.
        Moreover,
         \cite{caceres2019global,roux2020towards}
        proved that there are not periodic solutions, if
        $b$ is big enough,  $V_F\leq0$ and a transmission delay is considered.
        However, for the complete model and the stochastic discharge
        potential model,
        the results in \cite{caceres2018analysis,caceres2014beyond}
        numerically show periodic solutions.
	
\
		
        Parallel to the analytical study, numerical solvers are also developed
        to better understand the complex dynamics of the Fokker-Planck models
	\cite{caceres2011analysis,caceres2017blow,caceres2018analysis,hu2019structure}.
	This article aims to complete the numerical analysis.
	We provide numerical simulations for the microscopic description,
        with a triple intention: To compare the two families
        of models (Fokker-Planck equations and  Stochastic Differential
        equations),  better understand the
	notion of physical solution proposed
        in \cite{delarue2015particle} and the blow-up phenomenon, and 
        understand how it translates to the Fokker-Planck equation.

\
	
	The rest of the paper is structured as follows:  Sect.\ref{sec: solutions}
	summarizes the notions of 
	solution to the Stochastic Differential Equation
	\eqref{eq:model-simplify}, which were
	analysed in \cite{delarue2015global, delarue2015particle}; in Sect.\ref{sec: Numerical scheme},
	we explain the numerical schemes developed to obtain our
	numerical results, which are described in Sect.\ref{sec: Numerical results};
	and finally in Sect.\ref{sec: Conclusions},
        we collect the conclusions and perspectives about  future works.
	
	\vspace{0.3cm}
	\section{Classical and physical solutions to the SDE}
	\label{sec: solutions}
	\vspace{0.4cm}
	In this section we summarize the notions of solution to the
	Stochastic Differential Equation  \eqref{eq:model-simplify}, which were
	analysed in \cite{delarue2015global,delarue2015particle}.
	The main difference between {\em classical}
	and {\em physical} solutions
        lies in the re\-gu\-larity of the expectation $e(t)$
        (see \eqref{eq: expectation}). For classical solutions
	 $e(t)$ is continuous, while for physical ones, $e(t)$ can present
	certain positive jump discontinuities.	
	The blow-up phenomenon appears for classical solutions without
	transmission delay \cite{delarue2015global}. However,
        physical solutions
	 do not present explosion in finite time \cite{delarue2015particle}.
	 From a neurophysiological point of view,
        the classical notion implies that
          neurons only  fire 
	  when they reach the threshold potential. With
           the physical definition, they
	 can fire if their membrane potentials are close to the threshold
         value.

\
	
       Here we present both notions of  solution, which are required
       to understand our study. (See
       \cite{delarue2015global, delarue2015particle}  	
        for more precise definitions and an extensive explanation).
\newpage       
        	
	\begin{dfn}[\emph{Classical solution}]
		\label{def:1}
		A classical solution $V(t)$ to equation (\ref{eq:model-simplify}),
		with  $0<t<T$,  is a strong solution in this interval, which satisfies: 	
		\begin{enumerate}
			\item The sequence of spiking times is given by
			\begin{equation*}
                          \tau_0 = 0, \
                          \tau_k =  \inf{\{t>\tau_{k-1}:V(t^-)\geq V_F\}},
				\quad k\ge 1.
			\end{equation*}
			And the membrane potential is reset to $V_R$ immediately after: $V(t)=V_R$.
			
			\item  The firing rate $N(t)=e'(t)$ is a
			continuous function in $ 0<t<T$.
		\end{enumerate}
	\end{dfn}
	
	\begin{dfn} [\emph{Physical solution}]
		\label{def:2}
		A physical solution  $V(t)$ to equation (\ref{eq:model-simplify}),
		with $0<t<T$, is a weak solution which satisfies:
		
		\begin{enumerate}
			\item  The sequence of spiking times  is given as follows
			\begin{equation*}
			  \tau_0  =0, \ \tau_k = \inf{\{t>\tau_{k-1}:V(t^-)+
					b\Delta e(t)\geq V_F\}}, \quad k\ge 1.
			\end{equation*}
                        It is strictly increasing and cannot
			accumulate in finite time.
                        The membrane potential is reset to:
			$V(t)=V(t^-)-(V_F-V_R)+b\Delta e(t)$.
			\item The discontinuity points of the function $e$ satisfy:
			\begin{equation*}
			\Delta e(t) =
			\inf\left\{\alpha\geq0:\mathbb{P}\left(V(t^-)+b\alpha\geq V_F\right)<
			\alpha\right\},
			\end{equation*}
			where $\mathbb{P}$ is the probability function
                        of the associated probability
			space and
                        $\Delta e(t)$ is the variation of the expectation,
                        $\Delta e(t) := e(t) - e(t^-)$.
		\end{enumerate}
	\end{dfn}	
	In view of these two definitions, we can realize that
	a classical solution is also a physical solution without jumps,
	i.e. $\Delta e(t)=0$, since $e\in C^1([0, T)]$.
	Moreover, the membrane potential is reset to
	$
	V(t)=V(t^-)-(V_F-V_R)+b\Delta e(t)$, which is 
	$V_R$ since $\Delta e(t)=0$ and $V(t^-)=V_F$.
	However, physical solutions may have some jumps
	with size  $\Delta e(t)$, and
	 neurons may spike before they strictly reach the firing voltage
	$V_F$. After that, they are reset to
	$V(t^-)-(V_F-V_R)+b\Delta e(t)$, which is bigger than
	$V_R$.

	\begin{rem}
		\label{rem:b}
		The notion of physical solution requires that neurons cannot fire
		more than one spike at the same time. This fact holds if the connectivity
		parameter $b$ is less than $V_F-V_R$, as we will show in Sect. \ref{sec:cascade}.
		We recall that in this case there is a unique steady state
		\cite{caceres2011analysis}.
	\end{rem}
	
	We will  see numerically  how classical solutions
	are not able to avoid blow-up si\-tua\-tions,	
	which was proved in \cite{caceres2011analysis}
	for the Fokker-Planck equation and for the SDE  in \cite{delarue2015global}.
	However, for $b<V_F-V_R$ solutions with jumps, i.e. physical solutions,  
	will be able to avoid blow-up phenomena, as it was shown
	in \cite{delarue2015particle}.
	In the case of the delayed equation \eqref{eq:model-simplify-delay},
	physical solutions are actually classical.
	The delay itself guarantees global existence of a classical solution (therefore
	physical)
	without  jumps, i.e. $\Delta e(t-\delta)$=0, because
	$e(t-\delta)$ is always continuous in this case
	(see \cite{delarue2015particle,caceres2019global}).

\	
	
	The proof of the existence of physical solutions was given in
	\cite{delarue2015particle} following two different strategies:
	considering the limit of a particle system when the size of the network tends to
	infinite and, approximating \eqref{eq:model-simplify} by delayed equation
	\eqref{eq:model-simplify-delay} with
	transmission delay tending to zero, $\delta\to 0$.

	\subsection{Spikes cascade mechanism for an $\mathcal{N}$ particle system}
	\label{sec:cascade}
	
	As we already explained, the SDE \eqref{eq:model-simplify}  arises as
	mean field limit of networks composed by $\mathcal{N}$ neurons,
        with $\mathcal{N}\to \infty$.
        Therefore, in our simulations, we consider a particle system with $\mathcal{N}$
        neurons, that approximates the SDE \eqref{eq:model-simplify}.
        Before explaining our numerical schemes we  describe
        the cascade mechanism introduced in \cite{delarue2015particle},
	which will be implemented in our algorithm. 
	The cascade mechanism establishes how to determine the number
        of neurons in the network that fire at  time $t$,
	so that the uniqueness of the physical solution is guaranteed (see
		\cite{delarue2015particle} for details).
	
	\

	Let us define the set
	$\Gamma_0:=\{i\in\{1,...,\mathcal{N}\}:V_i(t^-)=V_F\}$.
	If that set  is not empty,  $ t $ is  a
        spike time, and neurons in $ \Gamma_0 $ spike at
        same time $ t $. Then, we consider a second time dimension, the
        "cascade time axis". We order the spikes that the rest of neurons
        in the network emits, as a consequence of receiving a
        kick of size $ b\left|\Gamma_0\right|/\mathcal{N} $:
    \vspace{0.25cm}    
	\begin{equation*}
	\Gamma_1:=\left\{i\in\{1,...,\mathcal{N}\}\setminus \Gamma_0:V_i(t^-)+b\frac{\left|\Gamma_0\right|}{\mathcal{N}}\geq V_F\right\}.
	\end{equation*}
        Iteratively,
        for $ k\in\mathbb{N}_0 $
        \vspace{0.25cm}	
	\begin{equation*}
	\Gamma_{k+1}:=\left\{i\in\{1,...,\mathcal{N}\}\setminus \Gamma_0\cup...\cup\Gamma_k:V_i(t^-)+b\frac{\left|\Gamma_0\cup...\cup\Gamma_k\right|}{\mathcal{N}}\geq V_F\right\}.
	\end{equation*}
	The cascade continues until $ \Gamma_l=\emptyset $ for some $ l\in\{1,...,\mathcal{N}\} $.
	Along the cascade time axis, a 'virtual' time axis located at $ t $,
        neurons in $ \Gamma_{k+1}$ spike after neurons in $ \Gamma_0\cup...\cup\Gamma_k $, with $k+1<l$. We can then define
    \vspace{0.25cm}    
	$$
	\Gamma:=\bigcup_{0\leq k\leq \mathcal{N}-1}\Gamma_k,
	$$
	which is the set of all neurons that spike at time $ t $,
        according to the natural order of the cascade. Finally, we update the membrane potential of each
        neuron in the network by setting:
        \vspace{0.25cm}
	\begin{equation*}
	V_i(t) = V_i(t^-) + \frac{b\left|\Gamma\right|}{\mathcal{N}} \ \text{if}\; i\not\in\Gamma,\quad V_i(t)=V_i(t^-)+\frac{b\left|\Gamma\right|}{\mathcal{N}}-(V_F-V_R) \ \text{if}\;i\in\Gamma.
	\end{equation*}
	At the end of the 
	cascade process $V_i(t) \le  V_F$ for all $i\in\{1,...,\mathcal{N}\}$.
        If $i\not\in\Gamma$, it is clear that $V_i(t) < V_F$.
        And if $i\in\Gamma$, then
                  $V_i(t)=V_i(t^-)+\frac{b\left|\Gamma\right|}{\mathcal{N}}-(V_F-V_R)$.
                  Therefore, since $\left|\Gamma\right|<\mathcal{N}$ and
                  $b<V_F-V_R$ (see
                  Remark \ref{rem:b}),
                  we obtain
                  \vspace{0.25cm}
		$$
		V_i(t)= V_i(t^-)+\frac{b\left|\Gamma\right|}{\mathcal{N}}-(V_F-V_R)
		\le V_i(t^-)+b-(V_F-V_R) < V_i(t^-)\le V_F.
		$$
	We note that all the neurons feel the spikes from the cascade,
        a kick  of size $ \frac{b\left|\Gamma\right|}{\mathcal{N}}$.
	
\newpage	
	
	\section{Numerical scheme}
	\label{sec: Numerical scheme}
	In this section, we explain the algorithms
        developed to obtain
	our  numerical results.
	We consider the general equation \eqref{eq:model-simplify-delay}
        with a transmission delay $\delta$, which includes
        Eq.\eqref{eq:model-simplify} assuming $\delta=0$.

	Let us consider a system of $\mathcal{N}$ identical neurons, and
	a uniform mesh in  the time interval 
        $\left[0,T\right]$, 
	with $n+1$ nodes and  time step $\Delta t=\frac{T}{n}$:
	\vspace{0.15cm}
	$$
	t_0=0, \, \ldots, \, t_j=t_0+j\Delta t, \, \ldots \, t_{n}=T.
	$$
        If $\delta\neq 0$, 
	the value of $ \Delta t $ is chosen as a multiple of it.

\	
	
	The membrane potential of  each neuron $i \in\left\{1,...,\mathcal{N}
        \right\}$
	evolves over time following Eq.\eqref{eq:model-simplify-delay}.
	Thus,  we integrate this equation  over
	each interval $[t_{j},t_{j+1}]$ to obtain:
	\vspace{0.15cm}
	\begin{equation}
	V_i(t_{j+1}) = V_i(t_j) - \int_{t_j}^{t_{j+1}}V_i(s) \ ds  +
	\int_{t_j}^{t_{j+1}} be'(s-\delta) \ ds+
	\int_{t_j}^{t_{j+1}} \sqrt{2a_0}\eta(s) \ ds.
	\label{eq: integral}
	\end{equation}
	After that, using the Euler-Maruyama method for stochastic
        differential equations,
	we approximate Eq.\eqref{eq: integral} by
	\vspace{0.15cm}
	\begin{equation}\label{eq:v-time-t+dt}
	V_i(t_{j+1}) = V_i(t_j) - V_i(t_j)\Delta t + be'(t_j-\delta)\Delta t
	+ \sqrt{2a_0\Delta t}B_i\quad \text{for}\;i\in\left\{1,...,\mathcal{N}
        \right\},
	\end{equation}
	where $ V_i(t_j) $ is the approximate value of the membrane potential of the
	$i$th neuron  at time $t_j$, 
	and $ B_i $ is a Brownian motion with main value and variance given by $ \nu_0 = 0 $ and $ \sigma^2 = \Delta t $.
	Moreover, we  approximate the time derivative of the expectation in
	terms of the variation of the expectation:
	\vspace{0.15cm} 
	$
	e'(t_j-\delta)
	\approx \frac{\Delta e(t_j-\delta)}{\Delta t}.
	$
	Taking into account that the system is nonlinear,
	the variation of the expectation is computed in two different ways,
        according
	to classical and physical solutions. In the first case, we count
        the number of
	particles having spiked in the interval
	$\left[t_{j-1}-\delta,t_j-\delta\right] $. For the physical solution, we consider the number of neurons which reach the firing threshold,
	following the  spikes cascade mechanism
	described in Sect.\ref{sec:cascade}
	at time $t_j$, since in this case $\delta=0$.
        In both cases the calculated quantity is divided
        by the total number of particles $\mathcal{N}$.

\

        Considering the initial condition, as well as the firing and reset conditions,
        we write  our algorithms as follows: 	
	\begin{equation}\label{eq:numerical-scheme}
	\left\{\begin{array}{ll}
	V_i^{j+1}=V_i^j-V_i^j\Delta t+b\frac{n_\delta^j}{\mathcal{N}}+\sqrt{2a_0\Delta t}B_i,\quad \text{for}\;i\in\{1,...,\mathcal{N}\},\; j\in\{0,...,n\}\\
	V_i^0 \;\text{initial condition}
	\\
	\mbox{If} \ V_i^{j+1}\approx\tilde{V}_F \Rightarrow
	V_i^{j+1}\leftarrow\tilde{V}_R,
	\end{array}\right.
	\end{equation}
	where $n_\delta^j/\mathcal{N}$ represents the variation of the expectation, and
	$V_i^j$ the approximate value of the membrane potential of the
	$i$th neuron at time $t_j$. The firing and
        reset conditions will be different for each type of solution,
        as explained in Sect.\ref{sec: solutions}.
        There is, however, a common element;
	we say that a neuron spikes when its voltage reaches a firing threshold
	$\tilde{V}_F$, and immediately afterwards it returns to the reset potential  $ \tilde{V}_R $.
	
        \
        
	\emph{Classical solution}.
	In this case, 
	$\tilde{V}_F=V_F$ and $\tilde{V}_R=V_R$,  and the numerical
	scheme \eqref{eq:numerical-scheme} is rewritten as follows:
	\begin{equation}\label{eq:numerical-scheme-classical}
	\left\{\begin{array}{ll}
	V_i^{j+1}=V_i^j-V_i^j\Delta t+b\frac{n_\delta^j}{\mathcal{N}}+\sqrt{2a_0\Delta t}B_i,\quad \text{for}\;i\in\{1,...,\mathcal{N}\},\; j\in\{0,...,n\}\\
	V_i^0 \;\text{initial condition}
	\\
	\mbox{If} \ V_i^{j+1}\approx V_F \Rightarrow
	V_i^{j+1}\leftarrow V_R.
	\end{array}\right. 
	\end{equation}
	The number of neurons which reached the firing threshold in the interval
	$\left[t_{j-1}-\delta,t_j-\delta\right] $ is recorded in $n_\delta^j$, and their membrane voltages are set to $ V_R $.
	If $\delta\neq 0$, then, 
        $ \Delta e(t_j) =\Delta e(0) $, for  $ t_j\le\delta $.\\
	
	\emph{Physical solution}.
	We do not consider delayed systems for this case.
        The variation of the expectation, and the firing and 
        reset conditions follow the cascade
        mechanism explained in Sect.\ref{sec:cascade}. Thus, the numerical scheme \eqref{eq:numerical-scheme}
	becomes:
	\begin{equation}\label{eq:numerical-scheme-cascade}
	\left\{\begin{array}{ll}
	V_i^{j+1}=V_i^j-V_i^j\Delta t+b\frac{n^j}{\mathcal{N}}+\sqrt{2a_0\Delta t}B_i,\quad \text{for}\;i\in\{1,...,\mathcal{N}\},\; j\in\{0,...,n\}\\
	V_i^0 \;\text{initial condition}
	\\
	\mbox{If} \ V_i^{j+1}\ge V_F-b\frac{n^j}{\mathcal{N}} \Rightarrow
	V_i^{j+1}\rightarrow V_i^{j+1}+b\frac{n^j}{\mathcal{N}}-(V_F-V_R),
	\end{array}\right. 
	\end{equation}
	where $n^j$ is the record value that counts
	the number of neurons which reach the firing threshold at time $t_j$, according to the cascade mechanism.

\
	
	The condition $V_i^{j+1}\ge V_F-b\frac{n^j}{\mathcal{N}}$ means
	that neurons fire before reaching the threshold value $V_F$. If
	$\frac{n^j}{\mathcal{N}}$ is a considerable  amount, this creates a crucial difference with the process considered in the classic description.
	
\

        To compare the results obtained by the particle system with
        those known for the Fokker-Planck equation, we need a sufficiently
        large number of particles and a time  step small enough.
	After some tests, we find those values which guarantee that:
        $ \mathcal{N}=80000 $ for the number of neurons and $ \Delta t=10^{-6} $ or
	$  \Delta t=10^{-8} $ for the time step, depending on the nature of the problem.
        The comparison between the approximate values of the
         solution of
        the Fokker-Planck equation \eqref{eq:edp-delay}
        and our numerical results is made using  histograms
        of the voltages  from our simulations.

\section{Numerical results}
\label{sec: Numerical results}
\begin{figure}[H]
	\begin{center}
		\begin{minipage}[c]{0.38\linewidth}
			\begin{center}
				\includegraphics[width=\textwidth]{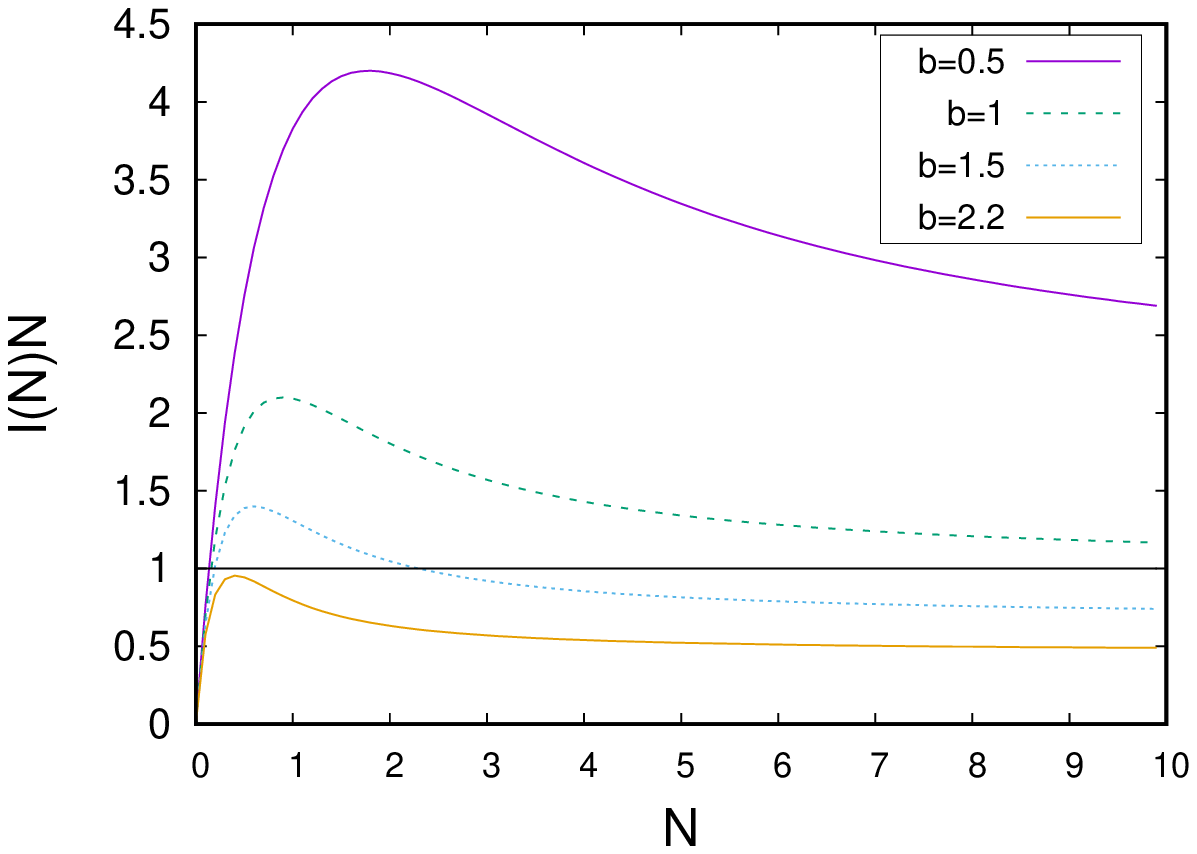}
			\end{center}
		\end{minipage}
		\begin{minipage}[c]{0.38\linewidth}
			\begin{center}
				\includegraphics[width=\textwidth]{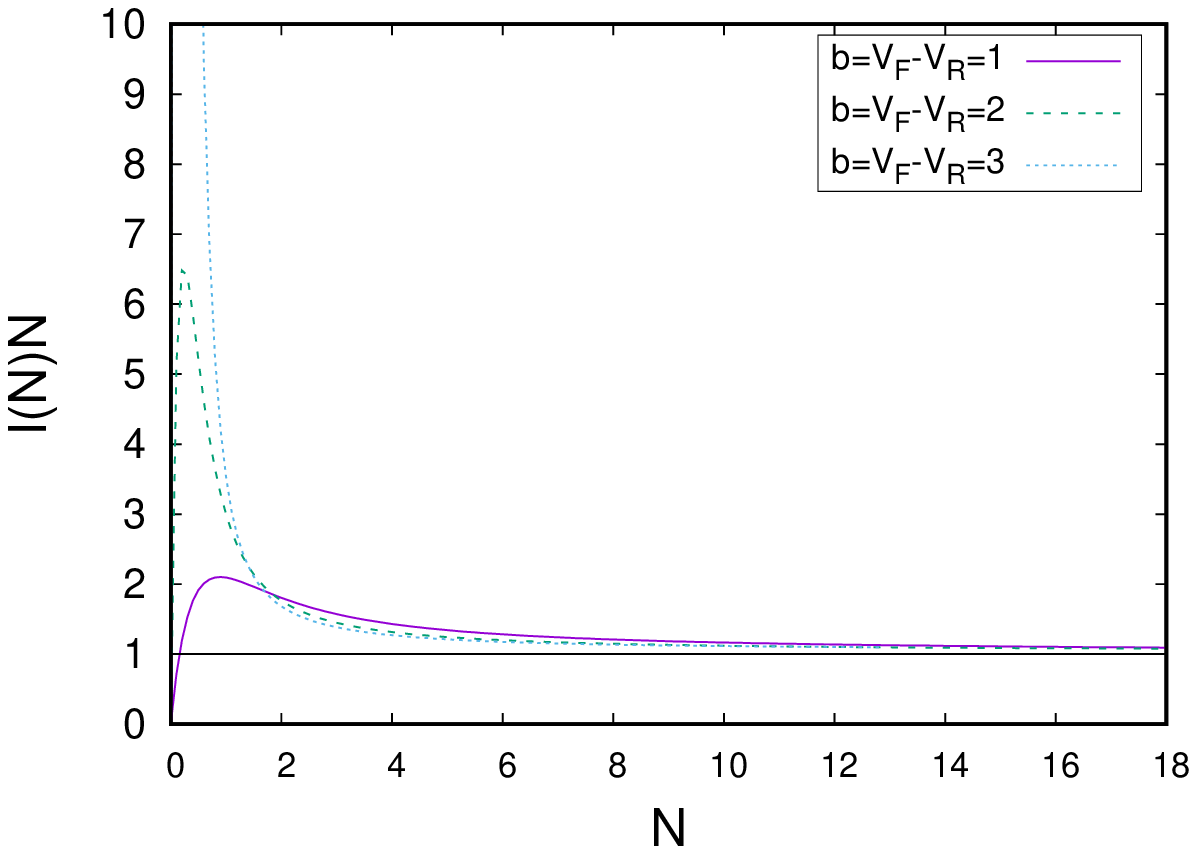}
			\end{center}
		\end{minipage}
	\end{center}
        \caption{
          {\bf Number of steady states  depending on the connectivity parameter
            $\mathbf{b}$}. Intersections between function $I(N)N$ and the
          straight line one provide the stationary firing rates of the
          Fokker-Planck equation \eqref{eq:model-simplify}.
  {\em Left}: Function $I(N)N$ for fixed values: $V_F=2$, $V_R=1$ and $a=1$ and different values of the connectivity parameter $b$.
          {\em Right}: Function $I(N)N$ for values of connectivity parameter exactly
          the distance between $V_F$ and $V_R$.
          $V_R=1$ and $ V_F $ takes the values $ \{2,3,4\} $.
                  }
	\label{fig:steady-states}
\end{figure}

As stated in Sect.\ref{subsec: known},
the behaviour of  solutions of the NNLIF models depends strongly
on the value of the connectivity parameter $b$.
Depending on that value, there is a different number of steady states. That
number is found solving Eq.(3.6) in \cite{caceres2011analysis}, which is
an implicit equation for the firing rate $N$: $NI(N)=1$, with $I(N)$ a function depending of the system parameters (see \cite{caceres2011analysis} for details and
 Sect.\ref{sec:result_justification-plateau}).
Fig.\ref{fig:steady-states} 
shows the intersections between $I(N)N$ and the straight line 1, which
provide the stationary firing rates of the Fokker-Planck equation \eqref{eq:model-simplify}.

\

We structure our results taking into account the restriction in the connectivity parameter so that the notion of physical solution makes sense.
First, we analyse the case $b<V_F-V_R$, where there is a unique stationary state
(see Fig.\ref{fig:steady-states}) and
the notion of physical solution makes sense. 
In the case $b\ge V_F-V_R$, where physical solutions do not make sense,
we address the study in two different subsections:
First, we  describe the case $b>V_F-V_R$, 
and secondly, we  analyse 
the 
limiting case ($b=V_F-V_R$),
right at the value where the notion of physical solution ceases to make sense.
Through the particle system,
we replicate the results obtained in \cite{caceres2011analysis} for the
Fokker-Planck 
and we study  the behaviour
of the system beyond blow-up phenomenon.
For connectivity values $b\ge V_F-V_R$, we observe the formation of certain distributions, which we call
"plateau".
To our knowledge, these types of distributions 
had not been shown before for the NNLIF models. We end
this section with a justification
for their formation, Sect.\ref{sec:result_justification-plateau}.

For every value of $b$, simulations have been performed with and without delay,
as well as using the cascade mechanism when the physical solution makes sense.
In the simulations, unless otherwise indicated,
we consider the following fixed values:
80000 neurons, i.e., $\mathcal{N}=80000$, 
$a=1$,  $V_R=1$ and $V_F=2$.

\subsection{Physical solutions: $b<V_F-V_R$}\label{sec:results-fisical-solution}

In this section we describe the behaviour of the physical solutions of the model. We also compare the results with those obtained for the delayed system and for the Fokker-Planck description.
We recall that  in this case  there is a unique
stationary solution (see Fig.\ref{fig:steady-states})
and the solutions can explode in a finite time,
if the initial data is highly concentrated around the threshold potential
\cite{caceres2011analysis}.
We numerically study these two properties of the model 
for the particle system.

\

\begin{figure}[H]
	\begin{center}
		\begin{minipage}[c]{0.38\linewidth}
			\begin{center}
				\includegraphics[width=\textwidth]{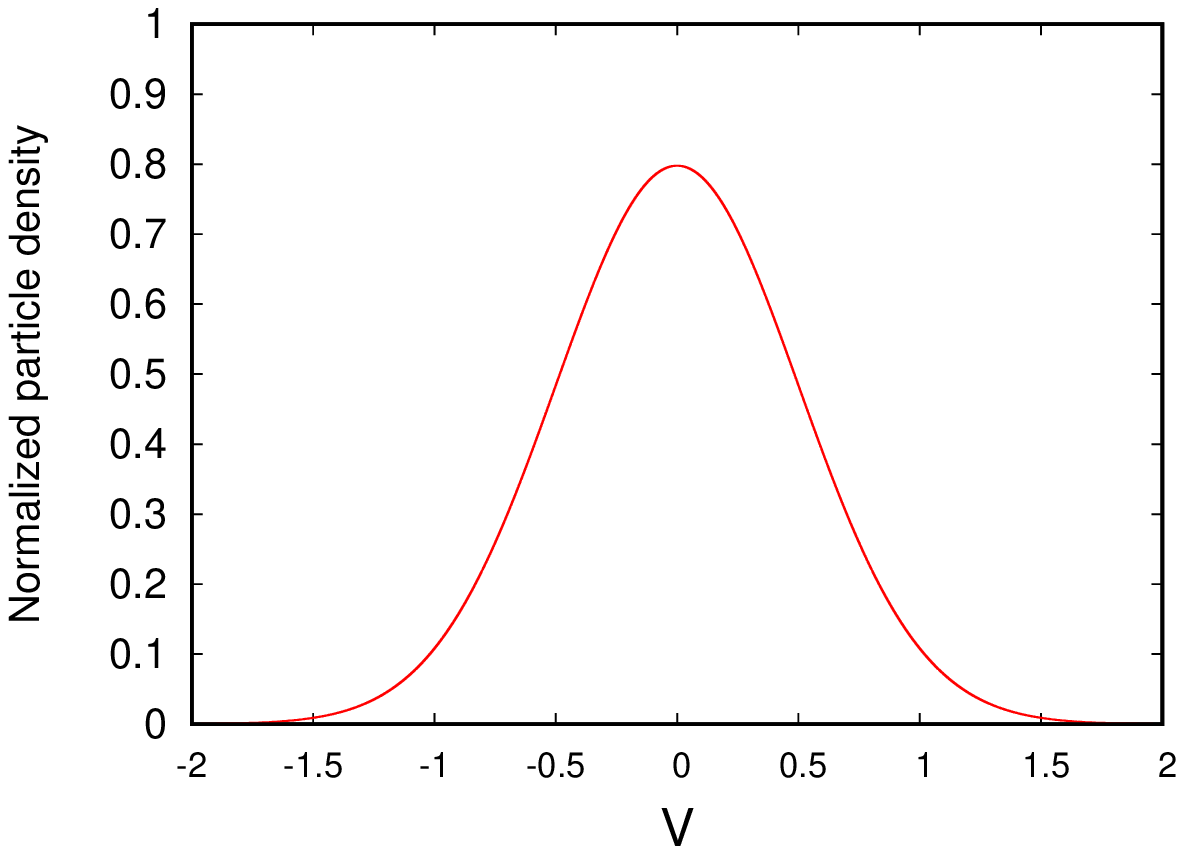}
			\end{center}
		\end{minipage}
		\begin{minipage}[c]{0.38\linewidth}
			\begin{center}
				\includegraphics[width=\textwidth]{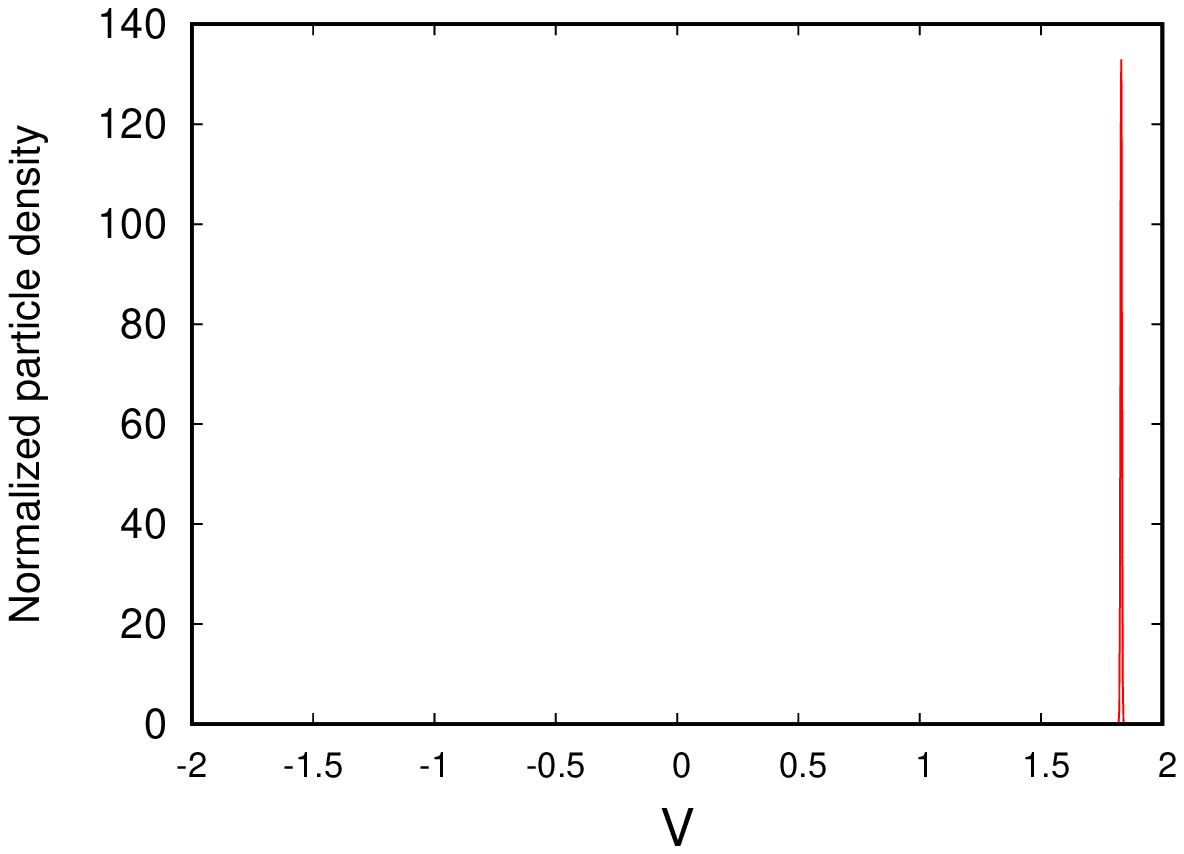}
			\end{center}
		\end{minipage}
	\end{center}
	\caption{{\bf Two initial distributions of the particle system}.
		{\em Left}: normal distribution with $ \nu_0 = 0 $ and $ \sigma^2 = 0.25$.
		{\em Right}: normal distribution with $ \nu_0 = 1.83 $ and
		$ \sigma = 0.003 $. 
	}
	\label{fig:initial-conditions}
\end{figure}

We consider $b=0.5$ in our simulations (the behaviour for different
values of $b$ is qualitatively equivalent), and  two different kind of initial data, whose
histograms are described in Fig.\ref{fig:initial-conditions}.
These histograms show the initial distribution through which we order the $80000$ neurons that compose the particle system,
	depending on the value of
	their membrane potential.
In the left plot we consider a normal distribution with $\nu_0 = 0$ and
	$\sigma^2 = 0.25$. We consider this initial condition for simulations
	which describe  global existence situations. 
	While in the right plot,
	we consider  a normal distribution with $ \nu_0 = 1.83 $ and $ \sigma = 0.003$, placed very close to the threshold potential
	($V_F=2$).
This initial condition is used to study 
 the blow-up phenomenon. 

\subsubsection{Convergence to the stationary solution}

\begin{figure}[H]
	\centering 
	\begin{subfigure}{0.38\textwidth}
		\includegraphics[width=\linewidth]{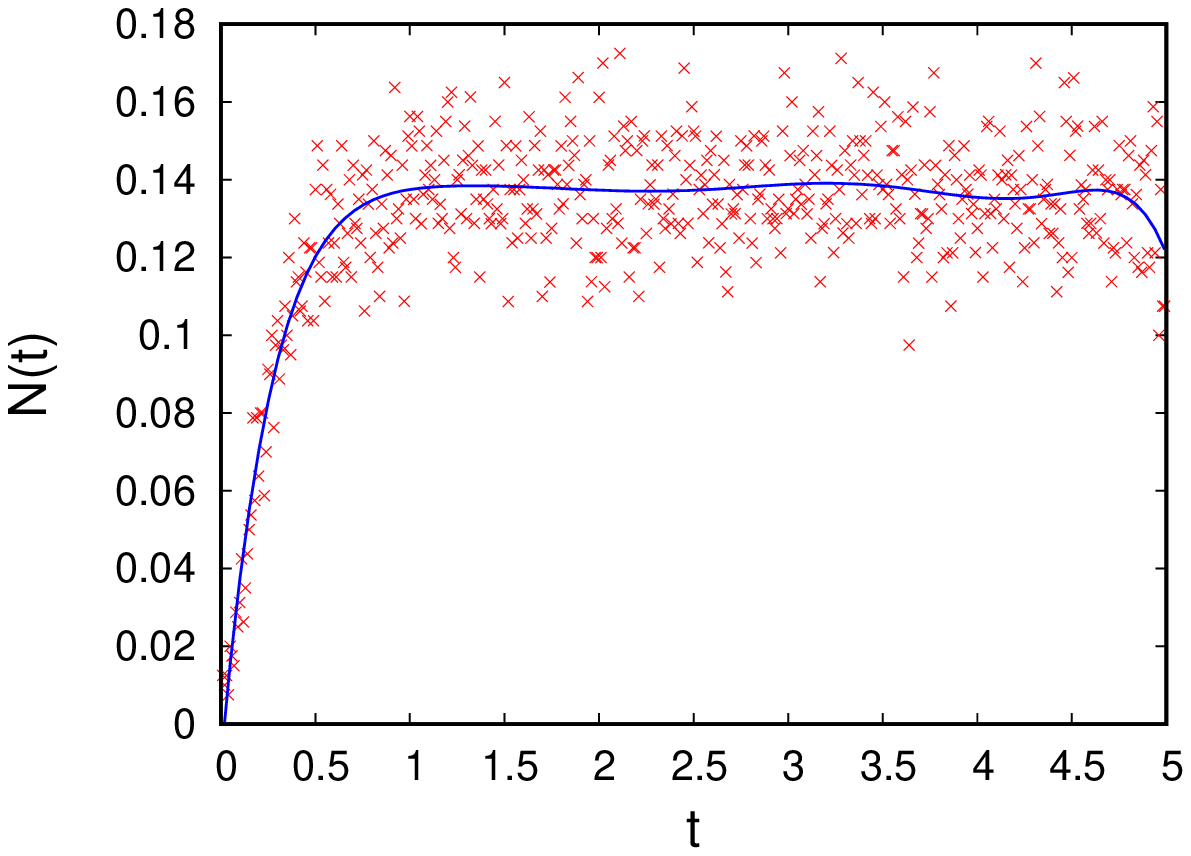}
		\caption{}
		\label{fig:firing-rate-b05-stationary}
	\end{subfigure}\hfil 
	\begin{subfigure}{0.38\textwidth}
		\includegraphics[width=\linewidth]{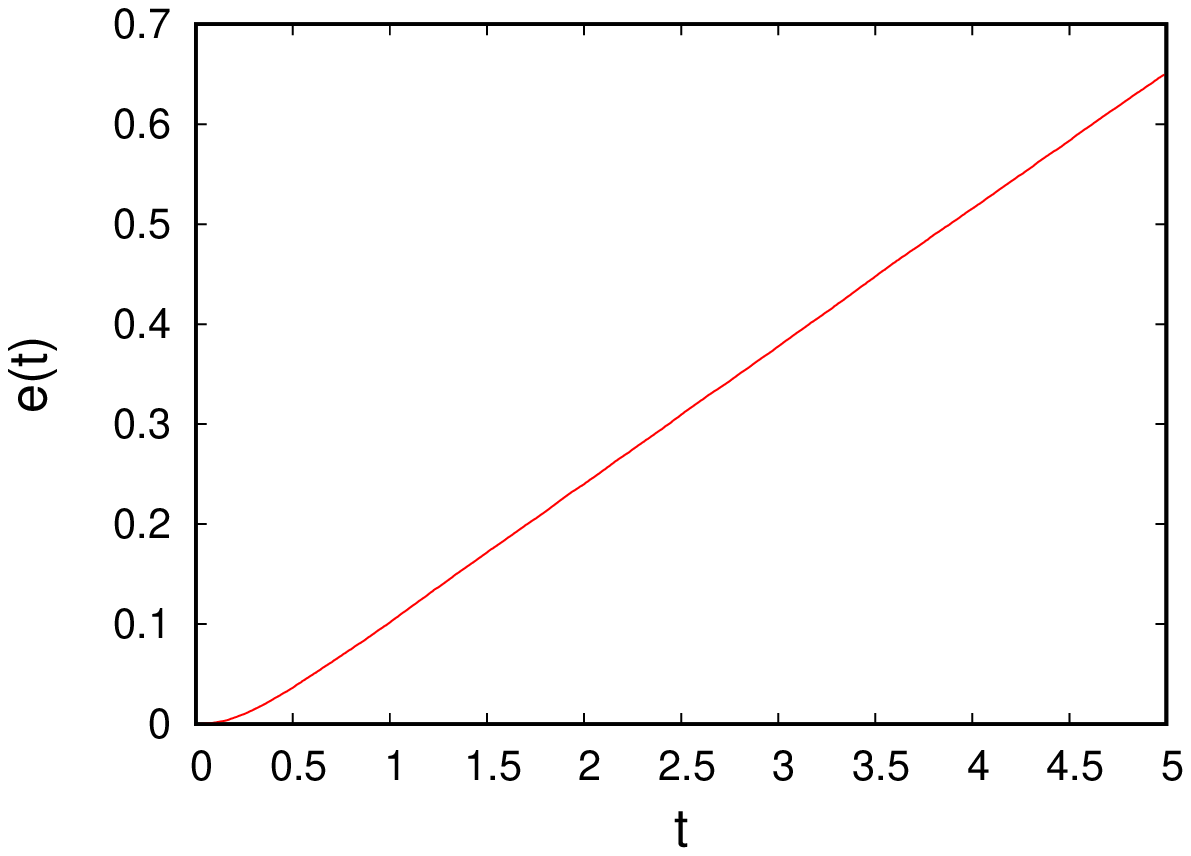}
		\caption{}
		\label{fig:expectation_b05_stationary}
	\end{subfigure}\hfil 
	
	\medskip
        \begin{subfigure}{0.38\textwidth}
	\includegraphics[width=\linewidth]{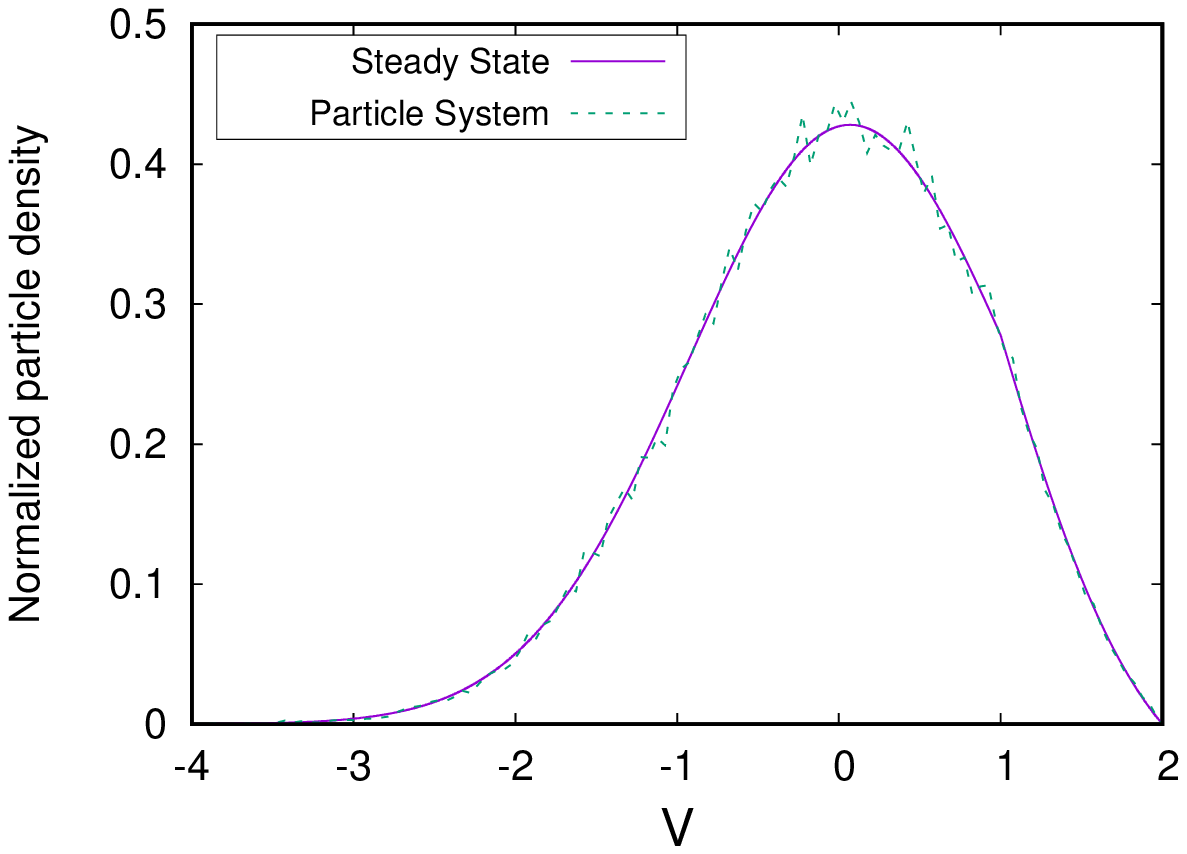}
 \caption{}
        \label{fig:comparison_edp_b05_stationary}
        \end{subfigure}
        \caption{          {\bf Convergence to the steady state}.
          $\mathbf{b=0.5}$.
          The initial condition is a normal distribution with $ \nu_0 = 0 $ and $ \sigma^2 = 0.25 $. 
          {\em (a)} Time evolution of the firing rate.
          {\em (b)} Time evolution of the expectation.
          {\em (c)} Particle voltage distribution at final time $5$ in
          comparison with  the stationary solution of  Fokker-Planck equation.}
	\label{fig:steadystate-b0.5}
\end{figure}

We consider the
initial datum which is far from  $V_F$,
Fig.\ref{fig:initial-conditions} Left, and  the algorithm for
classical solutions described  in Sect.\ref{sec: Numerical scheme}.
 Fig.\ref{fig:steadystate-b0.5} shows the evolution in time of the solution
until it reaches the steady state.

\newpage

We observe a very sparse firing rate value, Fig.\ref{fig:firing-rate-b05-stationary},
whose arithmetic mean is around the value $0.14$, in accordance with the value
found for the Fokker-Planck equation \cite{caceres2011analysis}.
Fig.\ref{fig:expectation_b05_stationary} shows the evolution of the expectation along the simulation. At the beginning the expectation remains almost zero,
because the initial condition guarantees that there are not neurons spiking.
After that the expectation increases due to the effect of the diffusion.
The distribution of the voltage values at the end of the simulation
is shown in Fig.\ref{fig:comparison_edp_b05_stationary} compared to
the the stationary solution of  Fokker-Planck equation given
in  \cite{caceres2011analysis}.

\subsubsection{Blow-up phenomenon}
\label{sec:result_b05_blow-up_longtime}
\vspace{0.2cm}

\begin{figure}[H]
	\centering 
	\begin{subfigure}{0.38\textwidth}
		\includegraphics[width=\linewidth]{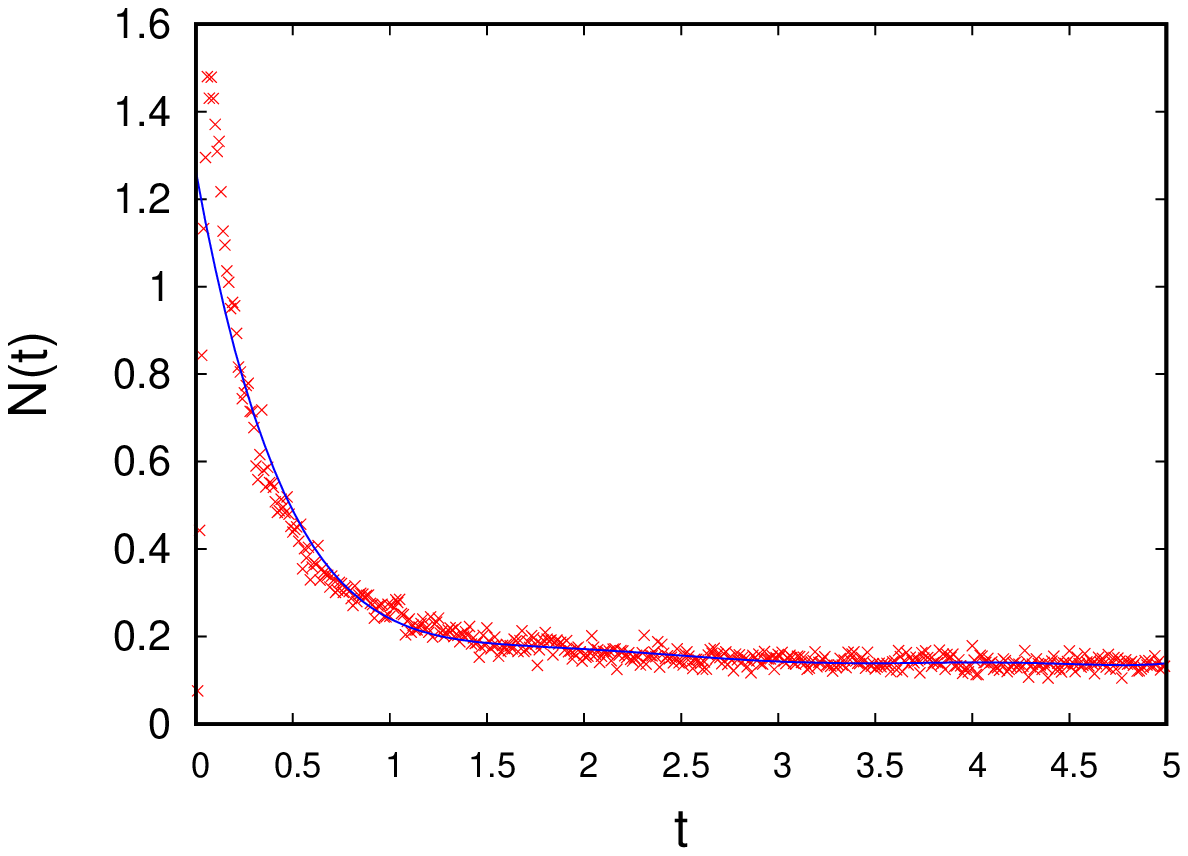}
		\caption{}
		\label{fig:firing_rate_b05_blow-up_longtime}
	\end{subfigure}\hfil 
	\begin{subfigure}{0.38\textwidth}
		\includegraphics[width=\linewidth]{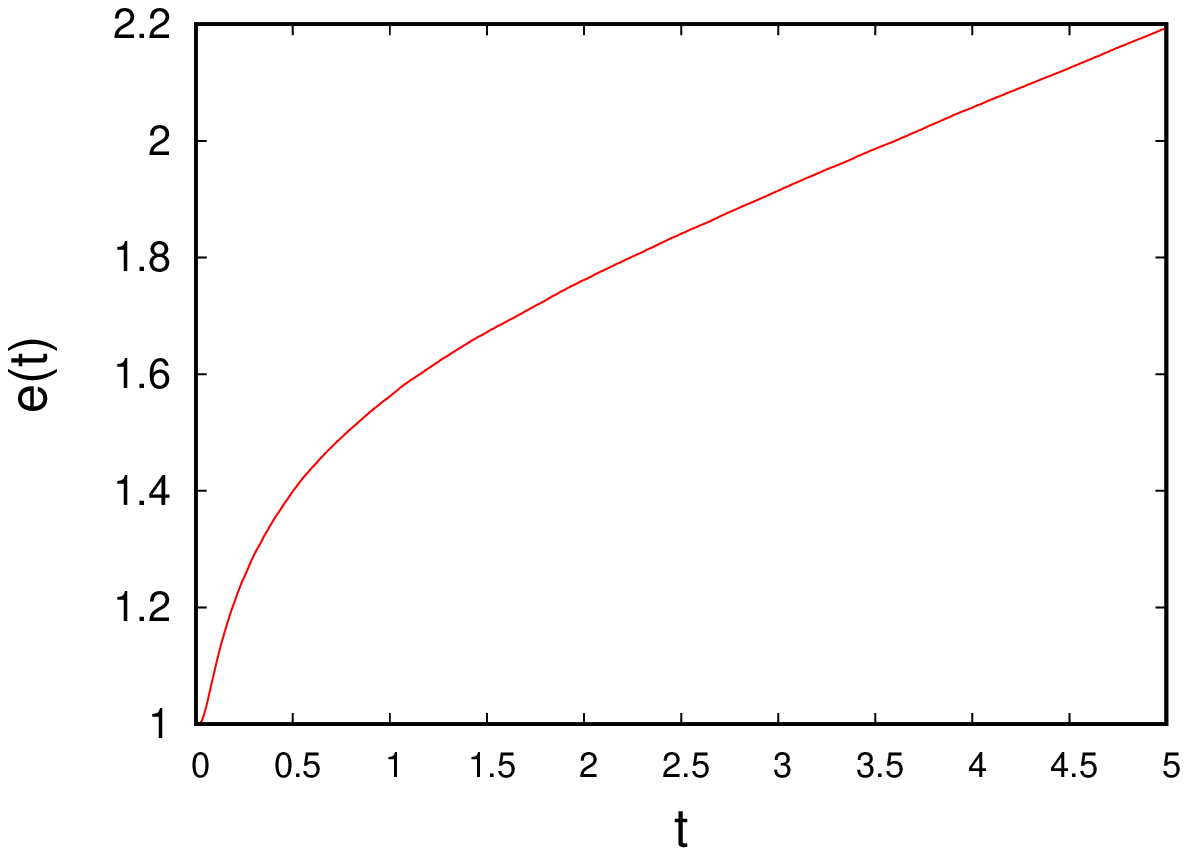}
		\caption{}
		\label{fig:expectation_b05_blow-up_longtime}
	\end{subfigure}\hfil 
	
	\medskip
	\begin{subfigure}{0.38\textwidth}
		\includegraphics[width=\linewidth]{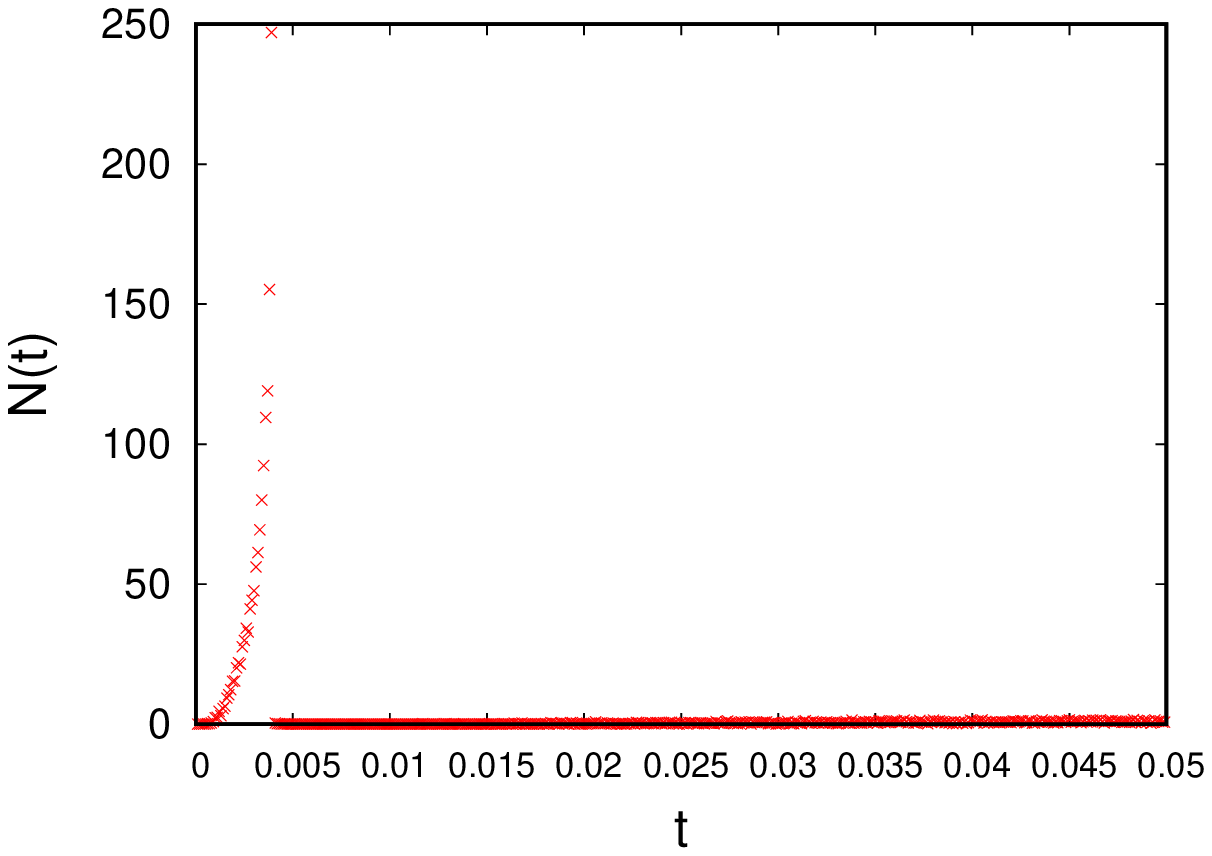}
		\caption{}
		\label{fig:firing_rate_b05_blow-up_zoom}
	\end{subfigure}\hfil 
	\begin{subfigure}{0.38\textwidth}
		\includegraphics[width=\linewidth]{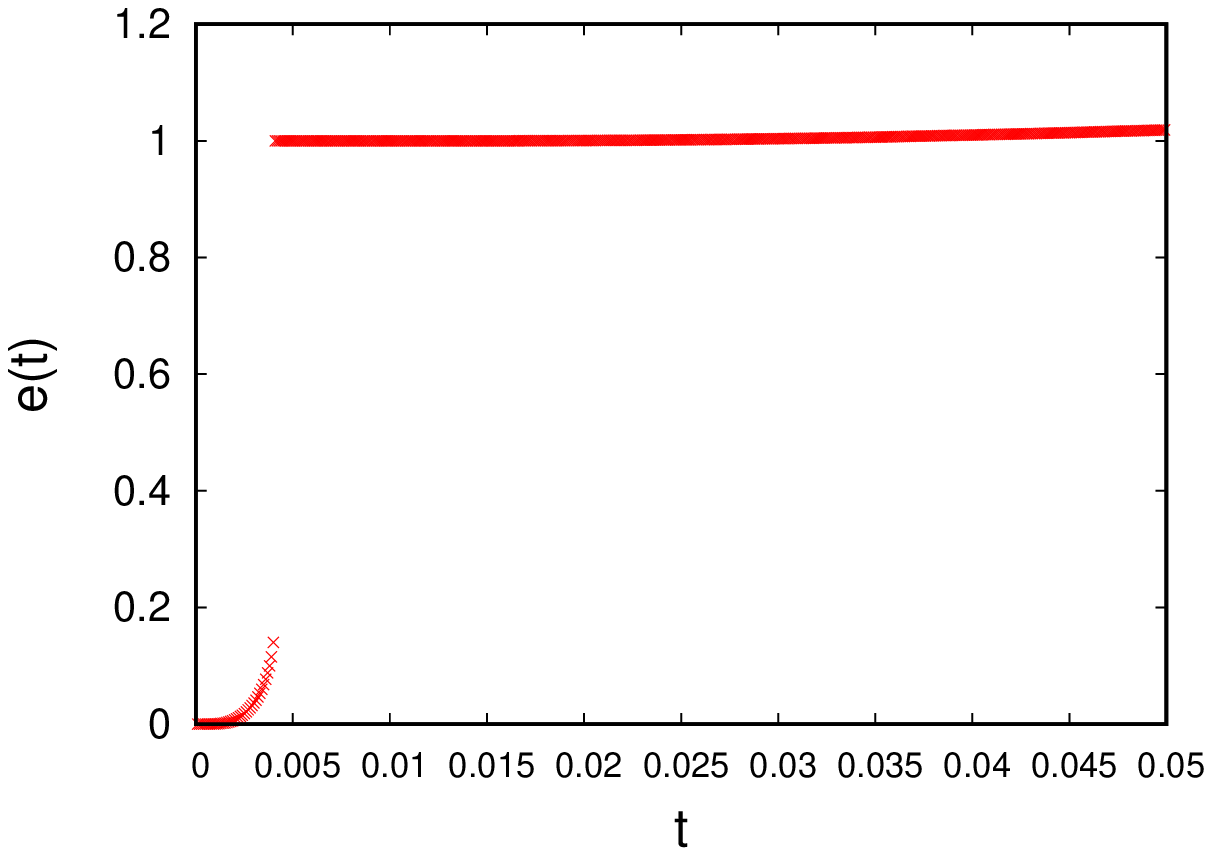}
		\caption{}
		\label{fig:expectation_b05_blow-up_zoom}
	\end{subfigure}\hfil 

	\medskip
	\begin{subfigure}{0.38\textwidth}
		\includegraphics[width=\linewidth]{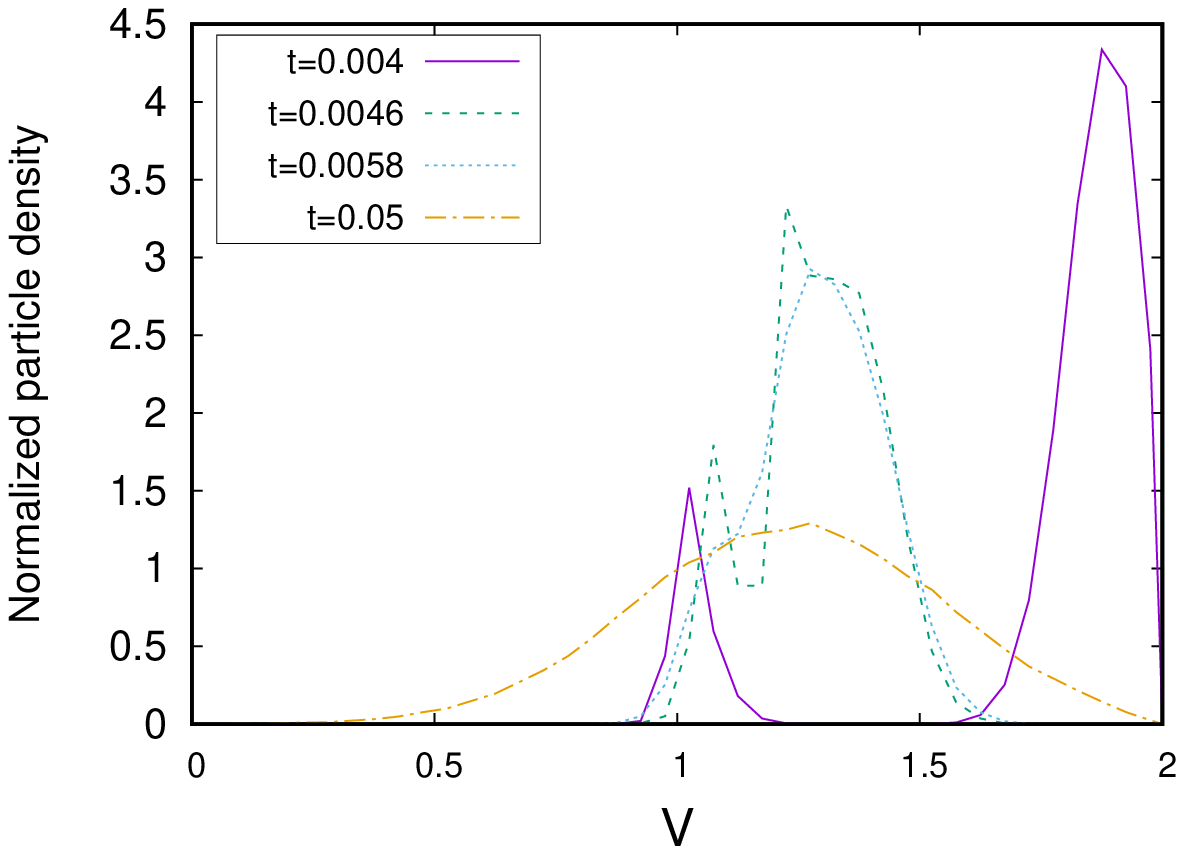}
		\caption{}
		\label{fig:b05_histograms}
	\end{subfigure}\hfil 
	\begin{subfigure}{0.38\textwidth}
		\includegraphics[width=\linewidth]{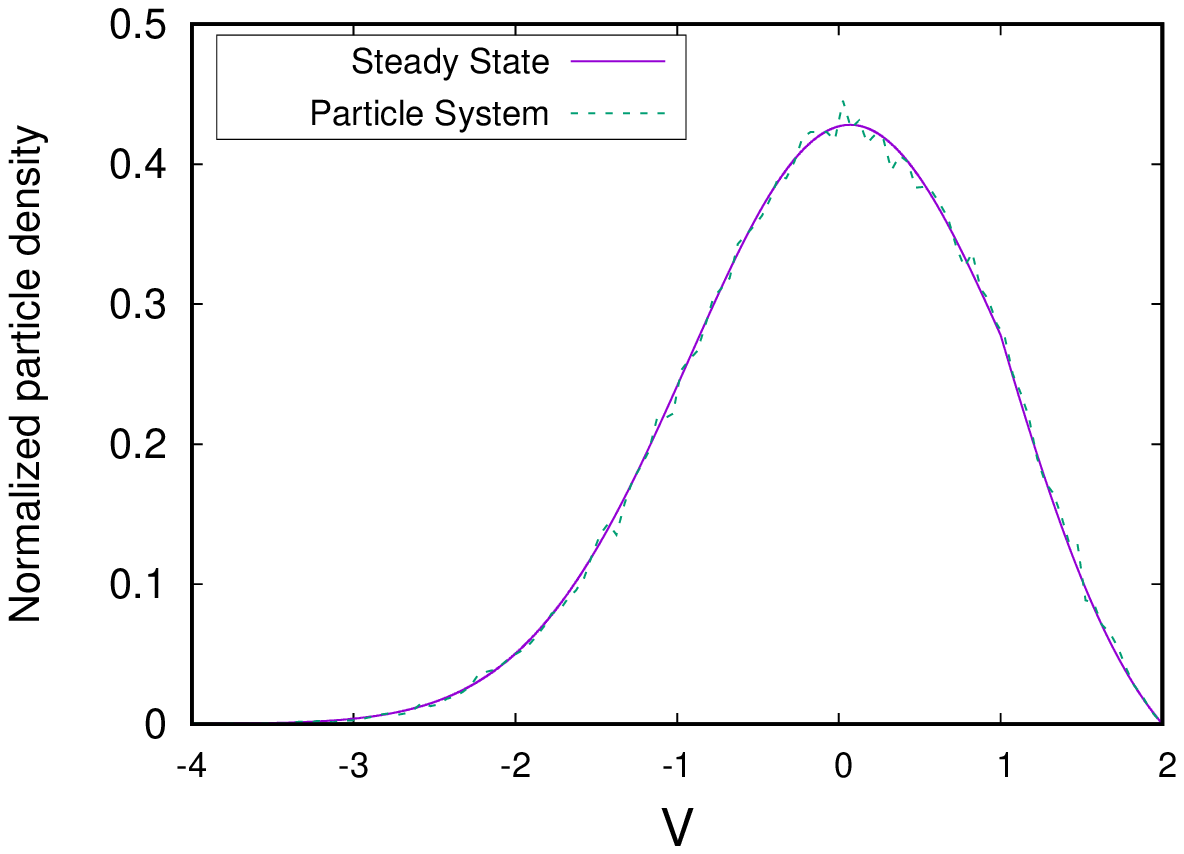}
		\caption{}
		\label{fig:b05_blowup_steady}
	\end{subfigure}\hfil 
	\caption{
          {\bf Blow-up phenomenon with classical approach}.
          $\mathbf{b=0.5}$.
The initial condition is a normal distribution with
$ \nu_0 = 1.83 $ and $ \sigma = 0.003 $.
          {\em (a)} Time evolution of the firing rate. 
          {\em (b)} Time evolution of the expectation.
          {\em (c)} Time evolution of the firing rate before and after
          the blow-up.
          {\em (d)} Time evolution  of the expectation before and after
          the blow-up.
          {\em (e)} Time evolution  of the particle voltage distribution
          before and after the blow-up.
          {\em (f)} Particle voltage distribution at time 5 in
                comparison with the stationary solution of Fokker-Planck
                equation.
}
	\label{fig:b05-blowup}
\end{figure}

To describe the blow-up phenomenon we consider the initial
datum given in Fig.\ref{fig:initial-conditions} Right, and,
as first step,  the algorithm for
classical solutions (see Sect.\ref{sec: Numerical scheme}).
The evolution in time of the solution is described in
Fig.\ref{fig:b05-blowup}.
Figs.\ref{fig:firing_rate_b05_blow-up_longtime} and \ref{fig:expectation_b05_blow-up_longtime} show
that blow-up occurs at the beginning of the simulation, 
the firing rate increases
very fast.
At blow-up time
the expectation is set to $ 1 $, which means that all neurons have already spiked, sense in which we understand the blow-up phenomenon.
After that the increasing rate of the expectation decreases. 

\

In Fig.\ref{fig:firing_rate_b05_blow-up_longtime} we see how the firing rate increases very fast
to its maximum value, and then decreases  to the stationary value.
Figs.\ref{fig:firing_rate_b05_blow-up_zoom} and
\ref{fig:expectation_b05_blow-up_zoom} show the singularity that occurs
during the blow-up time, when the derivative of the expectation, i.e.,
the firing rate, becomes infinity for the mean field limit. After that,
the expectation shows that every neuron has spiked and then it grows slowly. 
In the Fig.\ref{fig:b05_histograms}, we observe the value of the
voltages distribution before and after the blow-up phenomenon.
The distribution at $ t=0.004 $  is obtained before almost all the neurons spike at once. After that, the distribution shows a very singular shape ($ t=0.0046 $), to subsequently becomes more regular and evolves to the steady state.

\

Therefore, Fig.\ref{fig:b05-blowup} shows that the particle system
continues after the blow-up and converges to the unique stationary
distribution (see Fig.\ref{fig:b05_blowup_steady}). In this process, the notion of a classical solution is lost,
since the expectation has a jump and therefore, the solution really becomes
physical. After the synchronization of the system (explosion time),
the expectation becomes continuous again and the system evolves towards the
steady state, in a classical way.

\subsubsection{Cascade mechanism}
\begin{figure}[H]
	\centering 
	\begin{subfigure}{0.38\textwidth}
		\includegraphics[width=\linewidth]{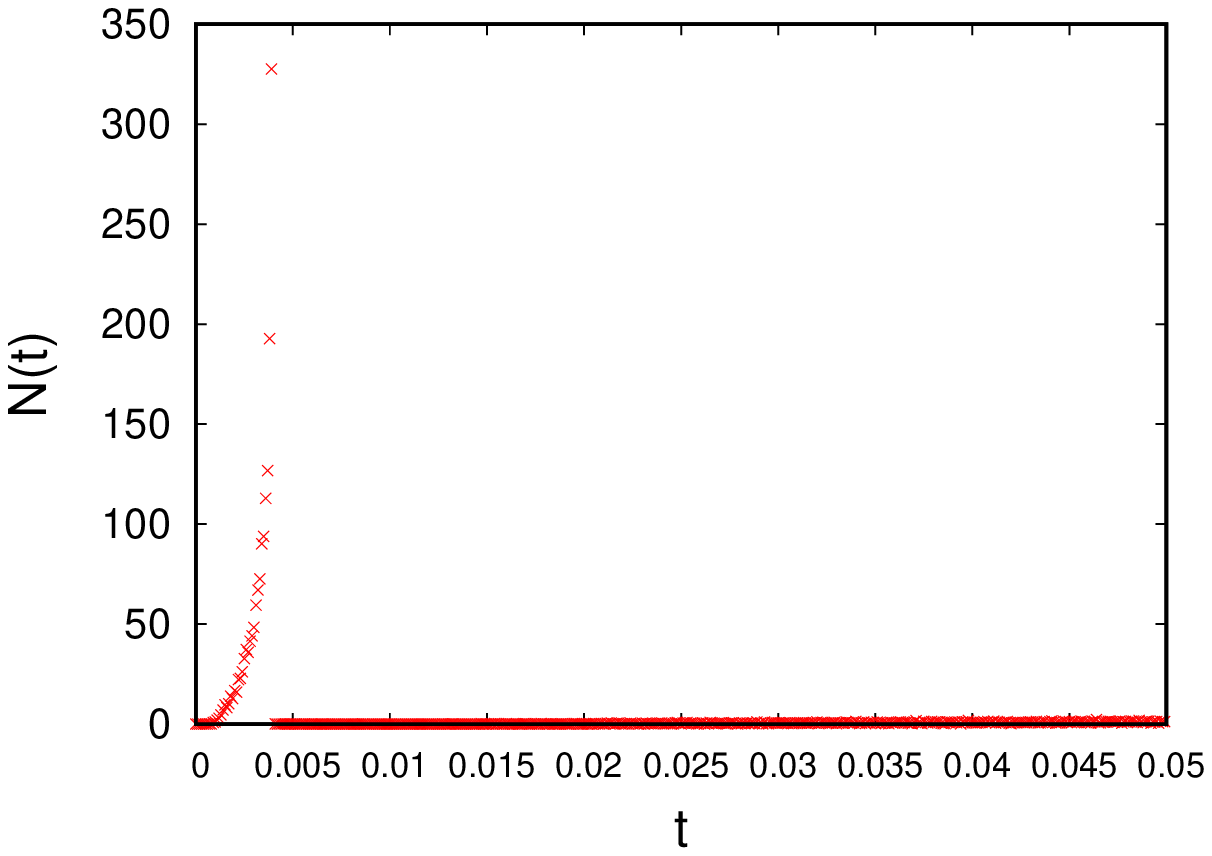}
		\caption{}
		\label{fig:b05_cascade_firing_rate}
	\end{subfigure}\hfil 
	\begin{subfigure}{0.38\textwidth}
		\includegraphics[width=\linewidth]{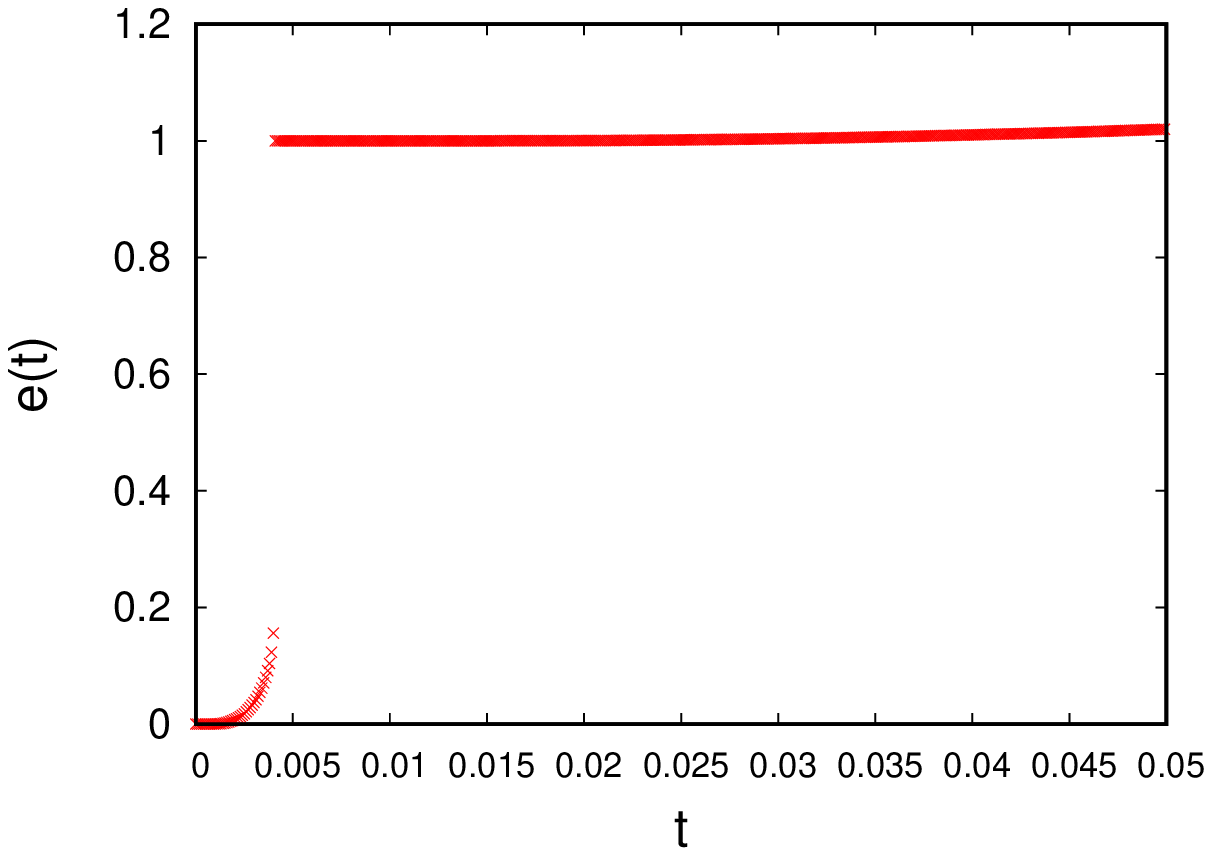}
		\caption{}
		\label{fig:b05_cascade_expectation}
	\end{subfigure}\hfil 
	
	\medskip
	\begin{subfigure}{0.38\textwidth}
		\includegraphics[width=\linewidth]{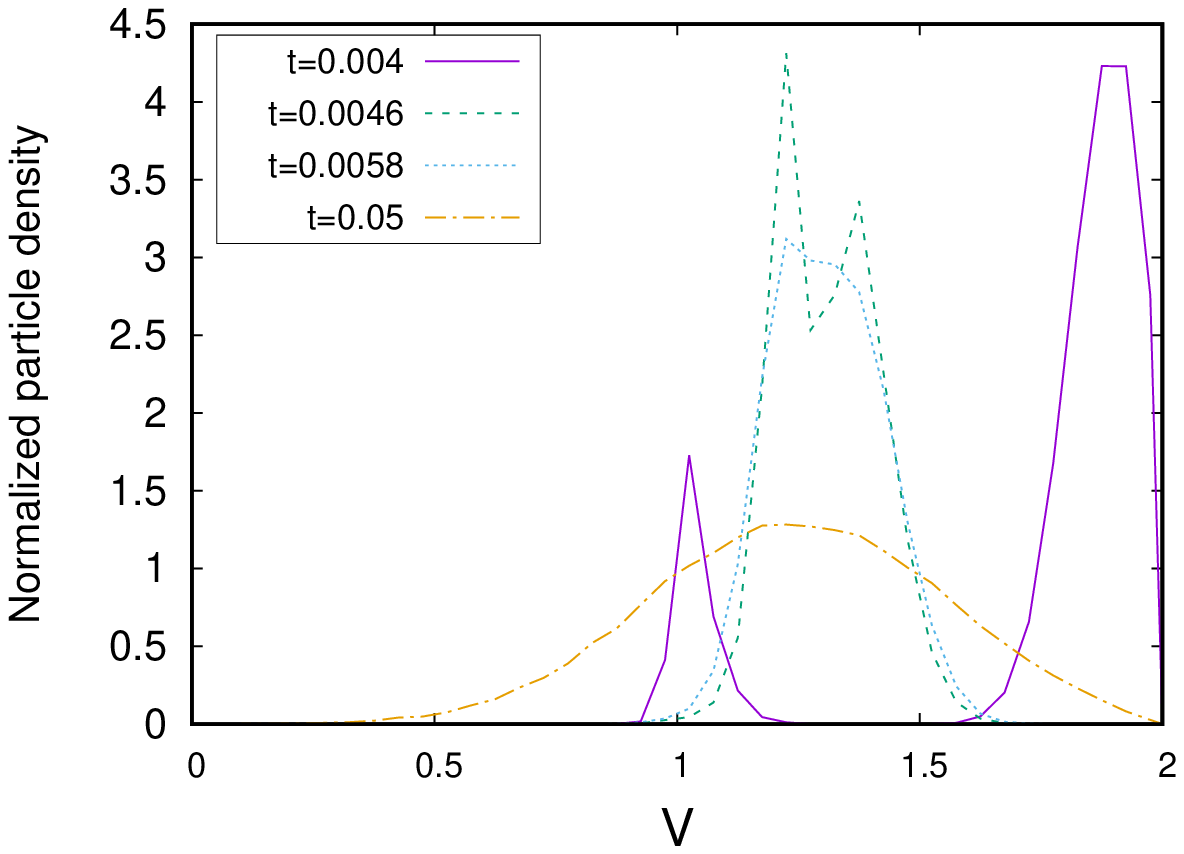}
		\caption{}
		\label{fig:b05_cascade_blowup}
	\end{subfigure}\hfil 
	\begin{subfigure}{0.38\textwidth}
		\includegraphics[width=\linewidth]{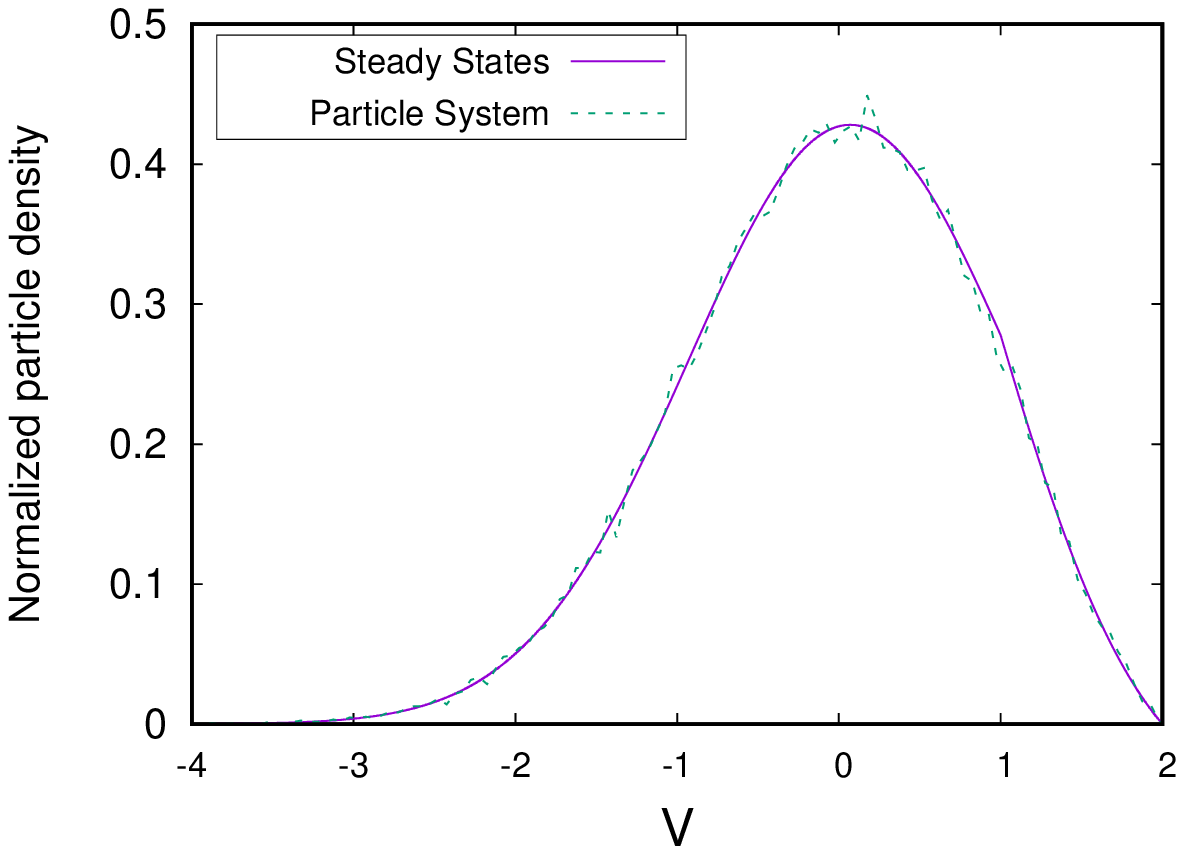}
		\caption{}
		\label{fig:b05_cascade_steady}
	\end{subfigure}\hfil 

	\medskip
	\begin{subfigure}{0.38\textwidth}
		\includegraphics[width=\linewidth]{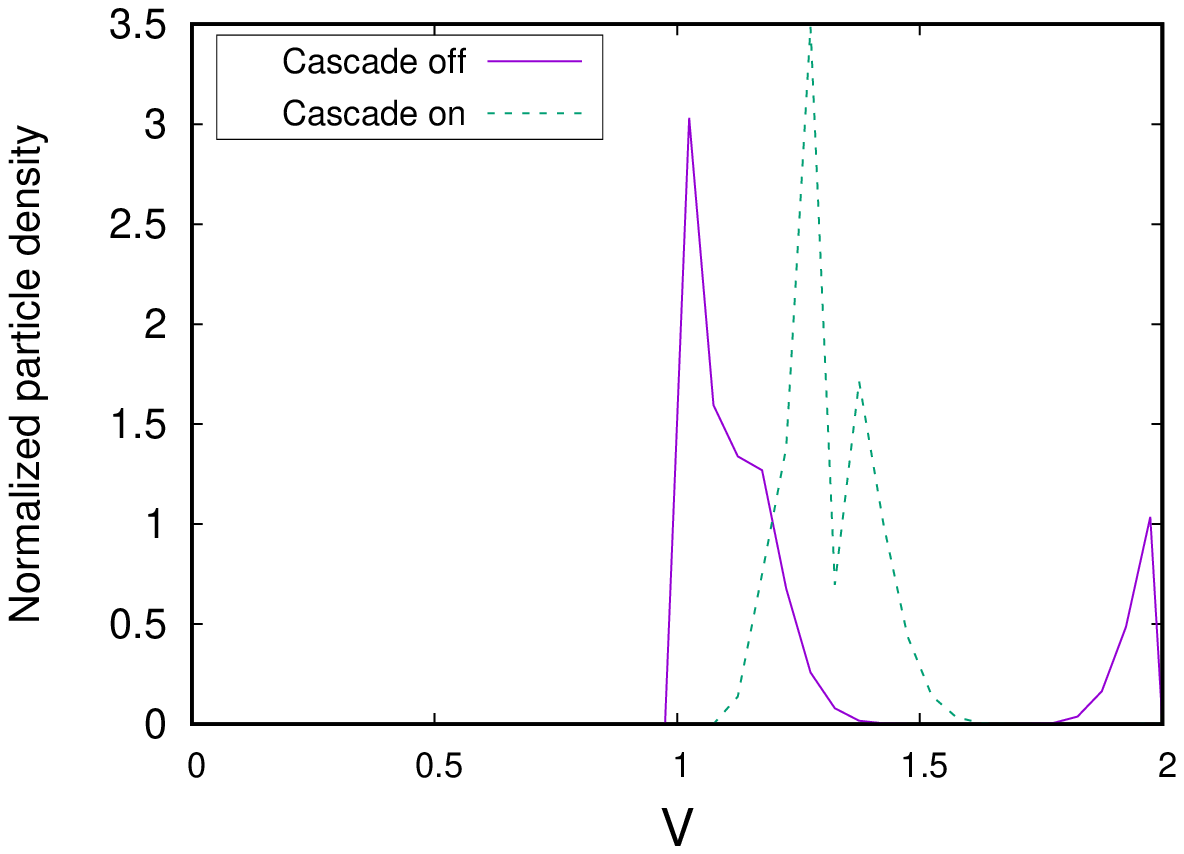}
		\caption{}
		\label{fig:b05_cascade_blowup_comparison}
	\end{subfigure}\hfil 
	
	\caption{
		{\bf Blow-up phenomenon with cascade mechanism}.
		$\mathbf{b=0.5}$.
		The initial condition is a normal distribution with
		$ \nu_0 = 1.83 $ and $ \sigma = 0.003 $.
		{\em (a)} Time evolution  of the firing rate before and
                after the blow-up.
		{\em (b)} Time evolution of the expectation before
                and after the blow-up.
		{\em (c)} Time evolution of the particle voltage
                distribution before and after the blow-up.
		{\em (d)} Particle voltage distribution at time 5 in
                comparison with the stationary solution of Fokker-Planck
                equation.
		{\em (e)} Comparison of the voltage distributions 
                in the blow-up moment with both approaches: with and without
                cascade mechanism.
	}
	\label{fig:b05_cascade}
\end{figure}

\vspace{0.2cm}
To better understand the notion of physical solution, we repeat
the previous experiment considering 
the cascade mechanism (see Sect.\ref{sec: Numerical scheme}).
We observe that the behaviour of the particle system is very similar with both
approaches. In Figs.\ref{fig:b05_cascade_firing_rate} and \ref{fig:b05_cascade_expectation}
the firing rate and the expectation perform a similar
behaviour to the blow-up one described with the algorithm
without cascade mechanism. The voltages distribution before and after the blow-up, Fig.\ref{fig:b05_cascade_blowup}, are also similar
to the one showed before. 
After that,
the system stabilizes to the
unique steady state (see Fig.\ref{fig:b05_cascade_steady}).

The main differences between both approaches are observed in
Fig.\ref{fig:b05_cascade_blowup_comparison}, which describes
the values of the voltages distributions in the  blow-up time.
The distribution with cascade mechanism is shifted to the right,
since the reset value is
$\tilde{V}_R=V\left(t^-\right)-\left(V_F-V_R\right)+b\Delta e(t)$,
which is different from $V_R$ because $\Delta e(t)\neq 0$. Moreover, 
we  appreciate a group of neurons that has
not yet fired, in the "cascade off" picture, while
the "cascade on" distribution has completely overtaken
the firing threshold. This happens because the cascade mechanism
allows neurons to fire at  $ \tilde{V}_F=V_F-b\Delta e(t) $. Also the expectation variation is different to the computed in the classical case, since it counts spikes which occur beyond  $ \Gamma_0 $ (see Sect.\ref{sec:cascade}). In particular, the variation of the expectation computed without cascade mechanism was $ \Delta e(t) = 0.303075 $, lower than that calculated in the cascade model, whose value is $ \Delta e(t)=0.574662 $.

\subsubsection{Transmission delay}
Finally, 
we analyse how  the transmission delay avoids the blow-up phenomenon.
With this aim we consider: a normal distribution with mean
$\nu_0 = 1.83$  and standard deviation $\sigma = 0.003$, as initial
distribution;
and three different transmission delays:
$ \delta=0.1 $, $ \delta=0.01$ and $ \delta=0.001$.

\begin{figure}[H]
	\centering 
	\begin{subfigure}{0.38\textwidth}
		\includegraphics[width=\linewidth]{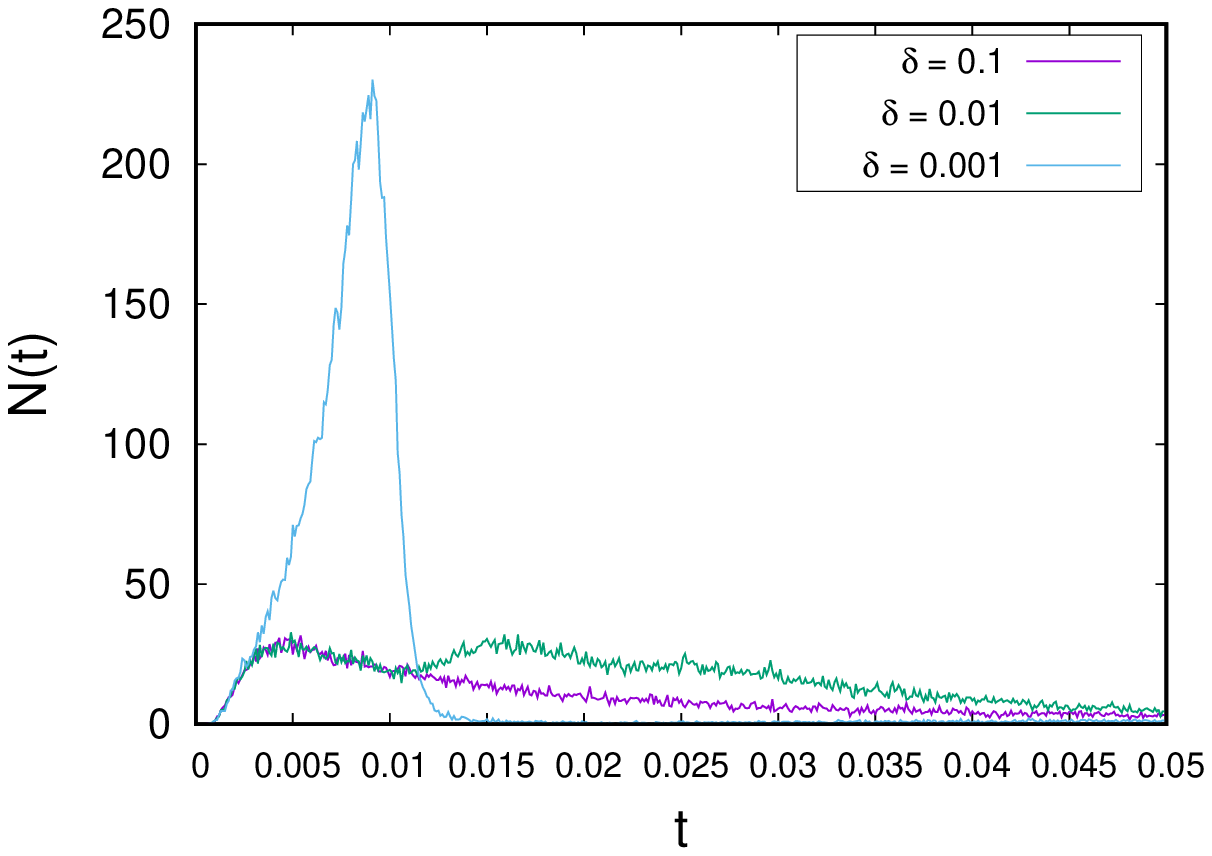}
		\caption{}
		\label{fig:firing_rate_b05_blow-up_zoom-delay}
	\end{subfigure}\hfil 
	\begin{subfigure}{0.38\textwidth}
		\includegraphics[width=\linewidth]{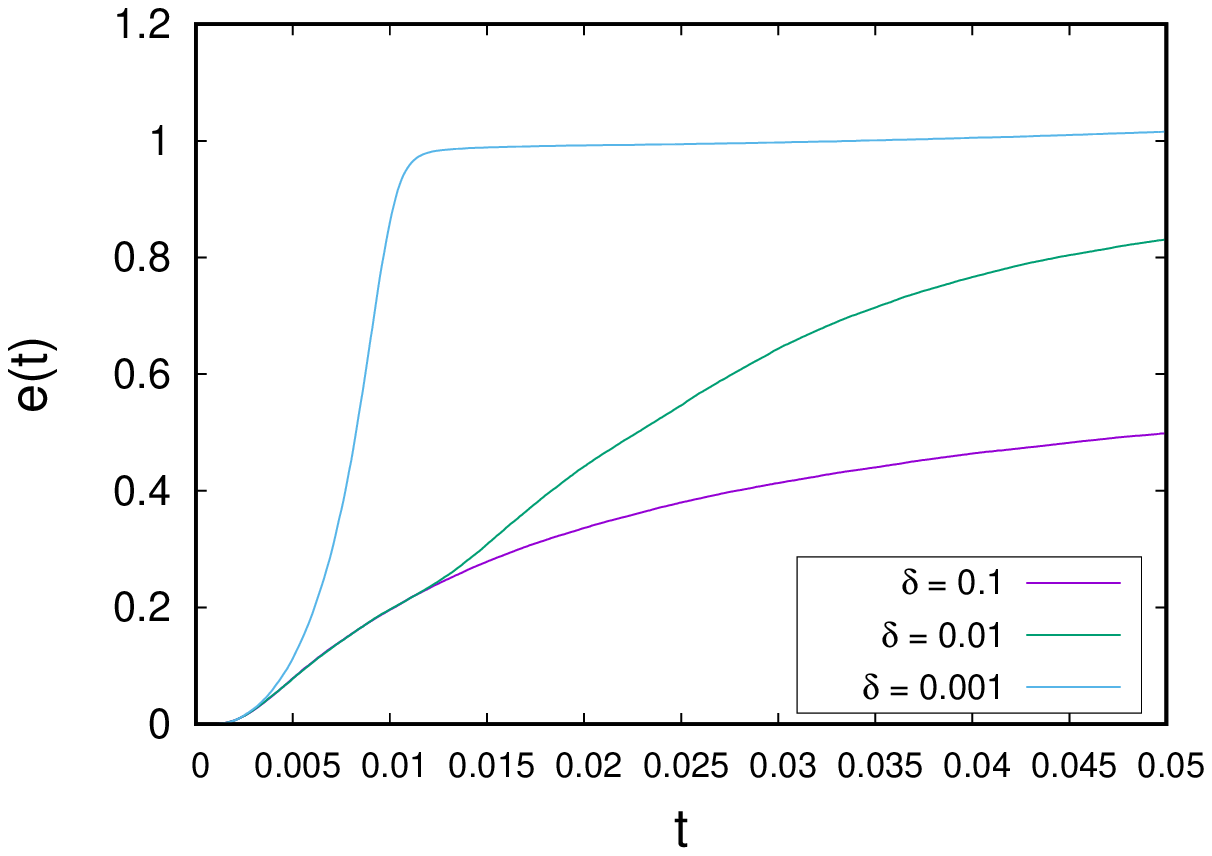}
		\caption{}
		\label{fig:expectation_b05_blow-up_zoom-delay}
	\end{subfigure}\hfil 
        \medskip
	\begin{subfigure}{0.38\textwidth}
		\includegraphics[width=\linewidth]{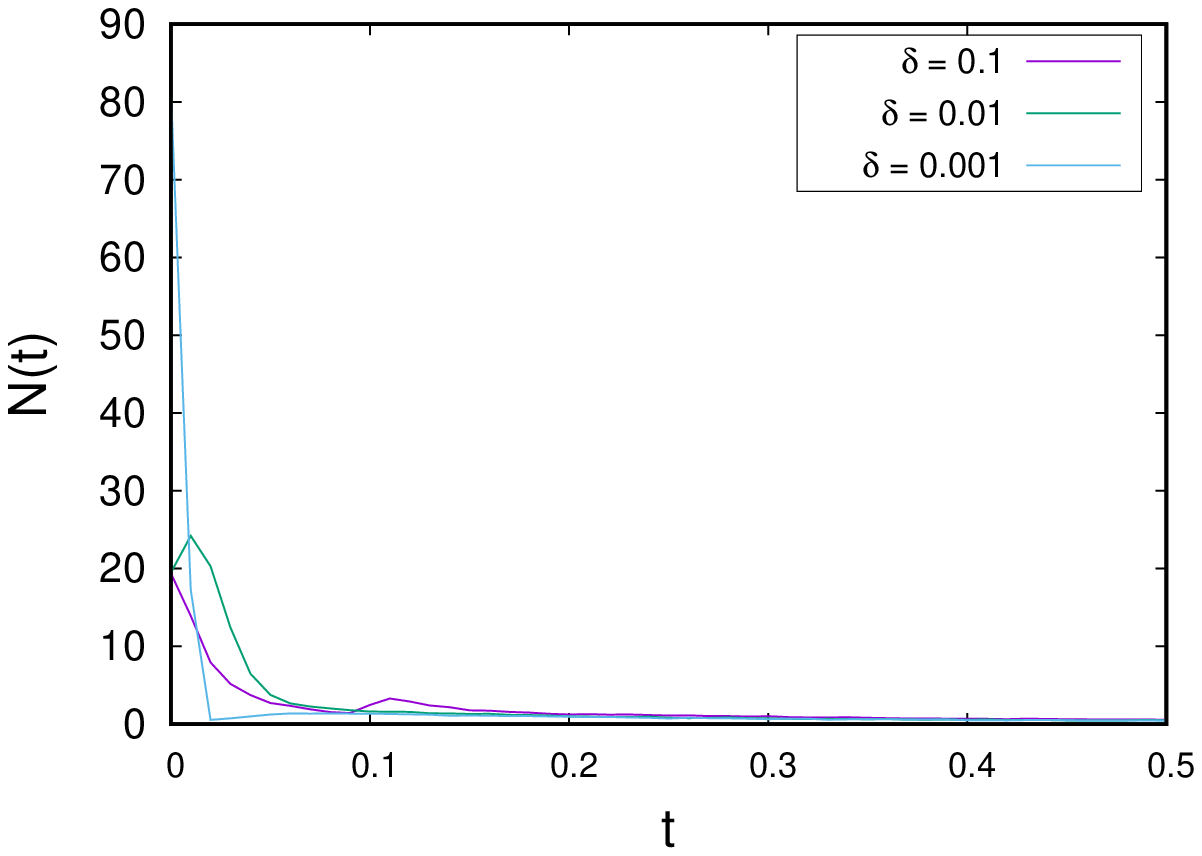}
		\caption{}
		\label{fig:firing_rate_b05_blow-up_longtime-delay}
	\end{subfigure}\hfil 
	\begin{subfigure}{0.38\textwidth}
		\includegraphics[width=\linewidth]{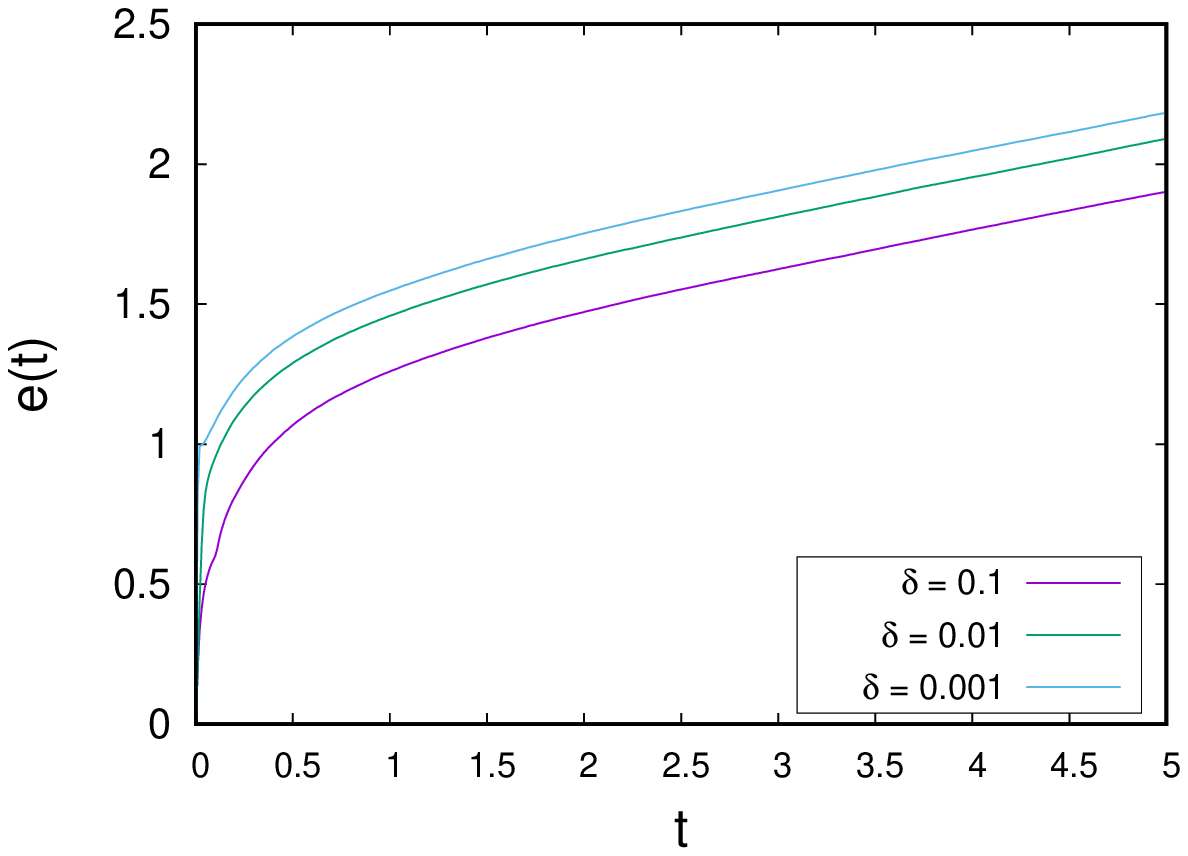}
		\caption{}
		\label{fig:expectation_b05_blow-up_longtime-delay}
	\end{subfigure}\hfil 
	\caption{
		{\bf Blow-up situation with delay}.
		$\mathbf{b=0.5}$.
		The initial condition is a normal distribution with
		$ \nu_0 = 1.83 $ and $ \sigma = 0.003 $. We considered different values for the delay.		
		{\em (a)} Time evolution of the firing rate until 0.05.
		{\em (b)} Time evolution  of the expectation until 0.05.
                {\em (c)} Time evolution  of the firing rate until time 0.5.
		{\em (d)} Time evolution  of the expectation until time 5.
	}
	\label{fig:b05-blowup-delay}
\end{figure}
Fig.\ref{fig:b05-blowup-delay}
shows 
that neurons achieve managing to dodge the blow-up
phenomenon, because
the firing rate and the expectation prevent the singularities
with the transmission  delay.
In Fig.\ref{fig:firing_rate_b05_blow-up_zoom-delay}
we observe how the firing rate increases until it reaches its maximum value
and then decreases.  Therefore, the
expectation does not present any singularity,
see Fig.\ref{fig:expectation_b05_blow-up_zoom-delay},
and the solution is classical.
However, if $\delta\to 0$ the firing rate diverges in finite time,
because its maximum value is higher when the value of
the transmission delay is reduced,
and the solution will be physical, instead of classical.
In the long time behaviour,
the firing rates converge to the stationary value, see
Fig.\ref{fig:firing_rate_b05_blow-up_longtime-delay}, since
the growth rates of the expectation  are the same, Fig.\ref{fig:expectation_b05_blow-up_longtime-delay}.
As a consequence,  the system converges to the unique steady state.

\vspace{0.1cm}
\subsection{Non-physical solutions $b>V_F-V_R$}\label{sec:results-non-physical-solutions}
\vspace{0.2cm}

In this section we focus on the case $b>V_F-V_R$, where the notion
of physical solutions does not make sense, since
the neuronal network is strongly excitatory and neurons could fire
more than one spike at the same time (see Remark \ref{rem:b}).
We remember that for this range of values of $b$,
the system has two steady states or none
(see Fig.\ref{fig:steady-states}).
We describe both situations with $b = 1.5$ (two steady states)
and $ b=2.2 $ (no steady states).

\subsubsection{Case with two steady states. Connectivity parameter
  $\mathbf{b=1.5}$}
\label{sec:results-b1.5}
For $b=1.5$, the long time behaviour for Fokker Planck Equation solutions
was numerically studied in \cite{caceres2011analysis}: the steady state with lowest firing
rate is stable, while the higher one is unstable, and the network
explodes at finite time for certain initial conditions.
Here we reproduce these phenomena at the level of the particle system. We
 discover that the particle system, in cases where blow-up occurs, tends to a "plateau distribution" if we consider a transmission delay.

\

\begin{figure}[H]
	\centering 
	\begin{subfigure}{0.38\textwidth}
		\includegraphics[width=\linewidth]{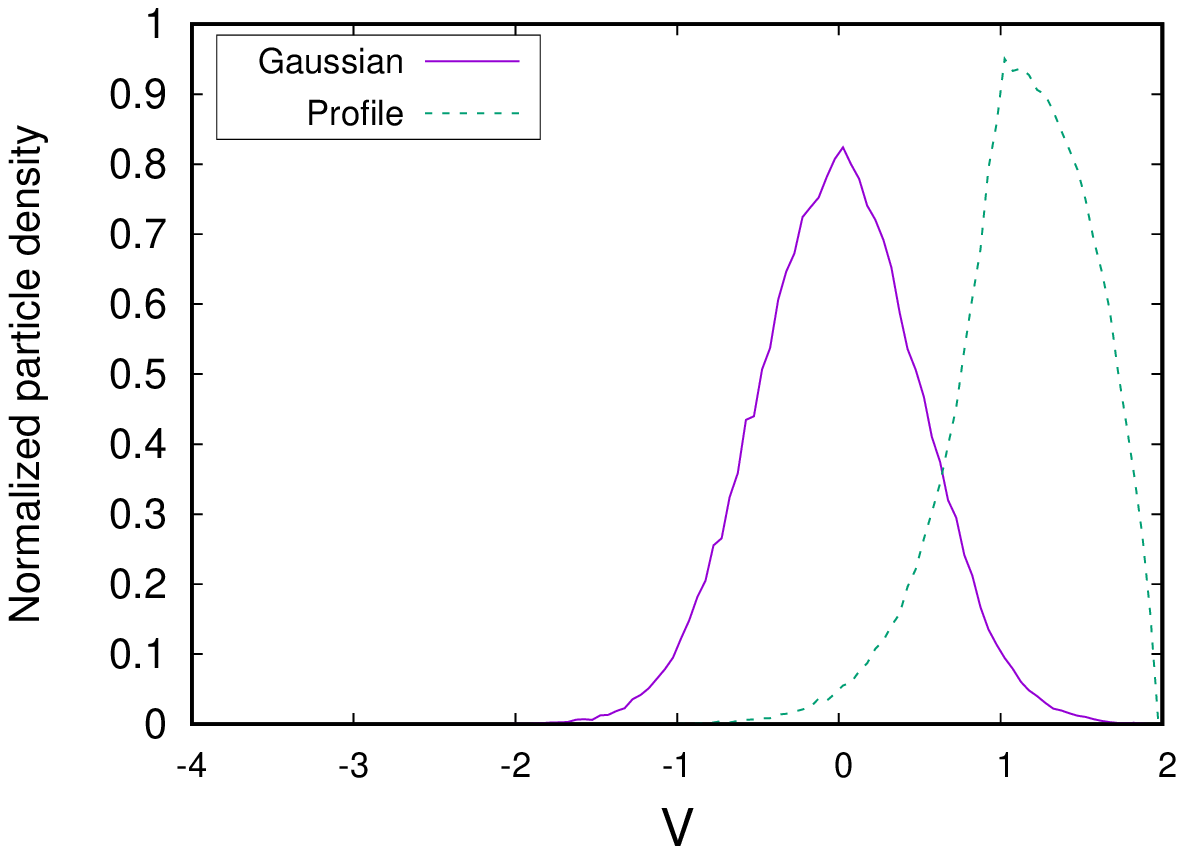}
		\caption{}
		\label{fig:initial-condition-b15}
	\end{subfigure}\hfil 
	\begin{subfigure}{0.38\textwidth}
		\includegraphics[width=\linewidth]{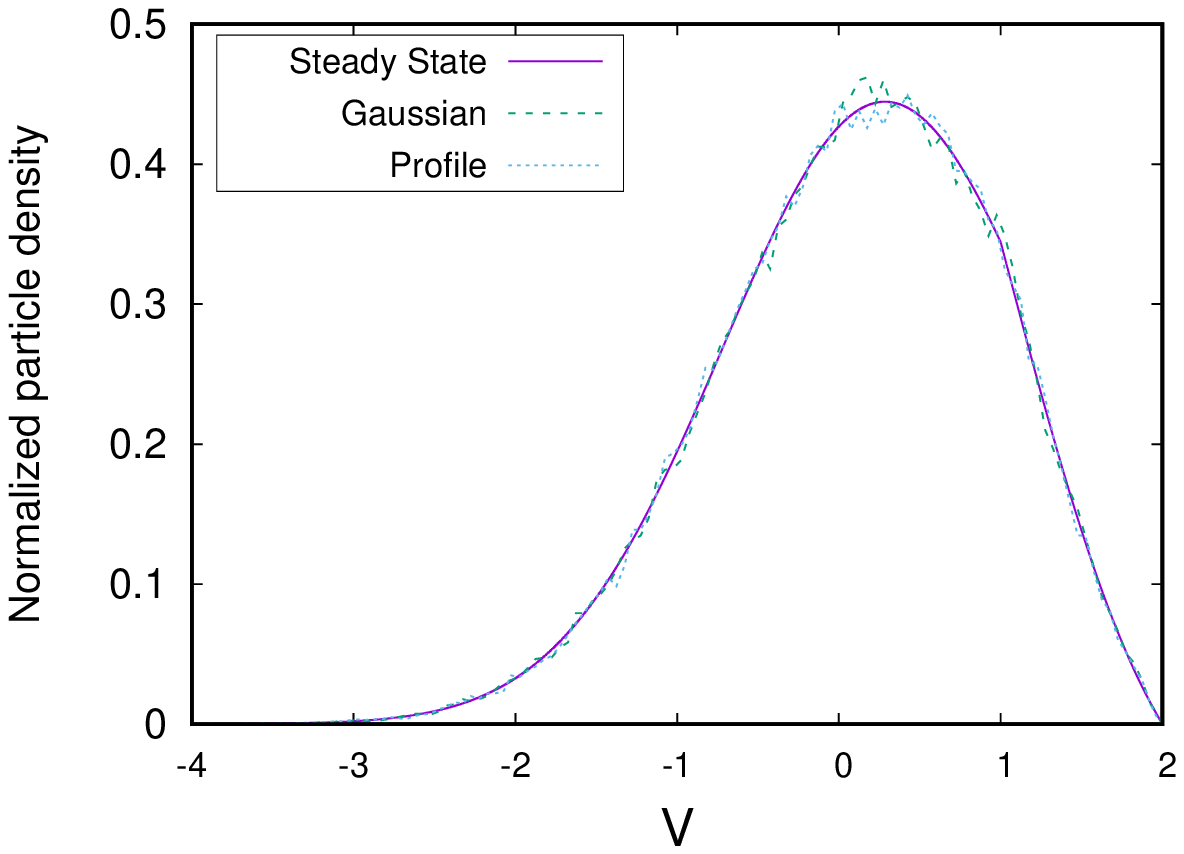}
		\caption{}
		\label{fig:comparison_edp_b15_stationary}
	\end{subfigure}\hfil 
	
	\medskip
	\begin{subfigure}{0.38\textwidth}
		\includegraphics[width=\linewidth]{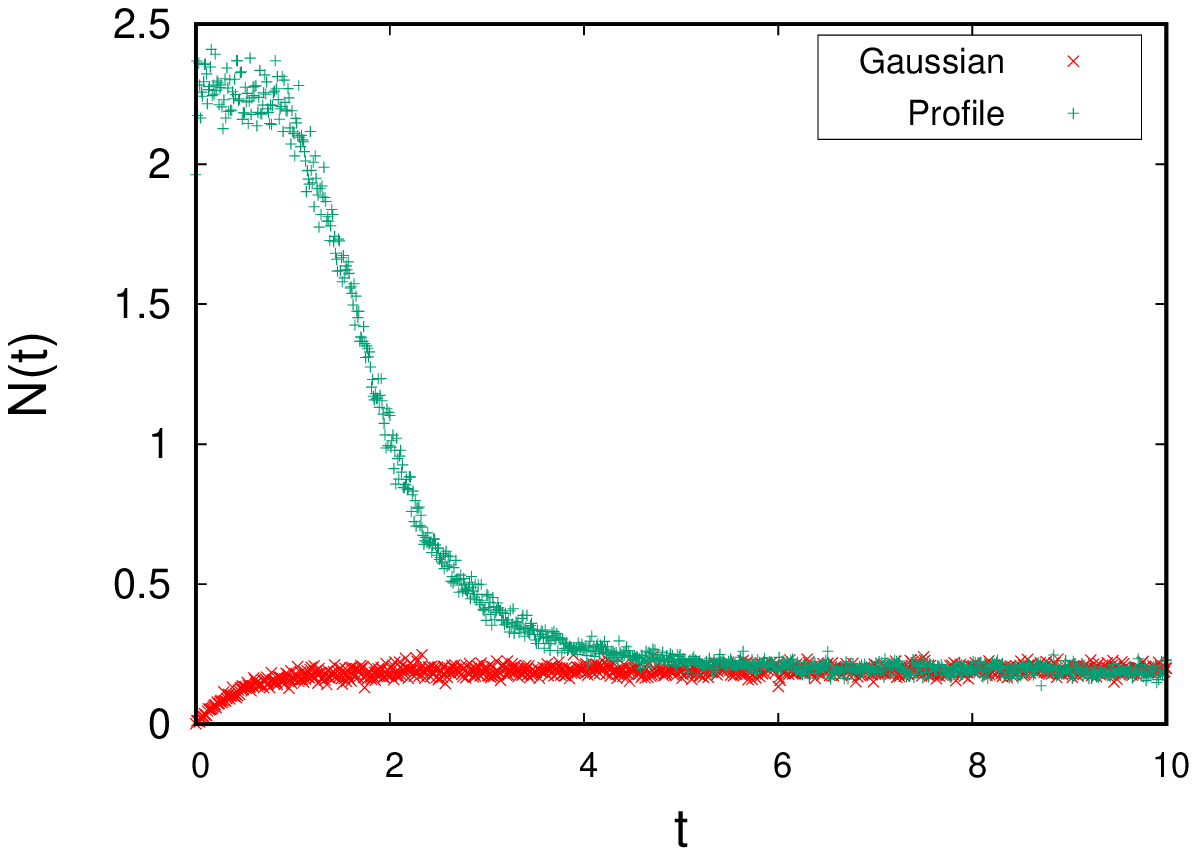}
		\caption{}
		\label{fig:firing-rate-b15-stationary}
	\end{subfigure}\hfil 
	\begin{subfigure}{0.38\textwidth}
		\includegraphics[width=\linewidth]{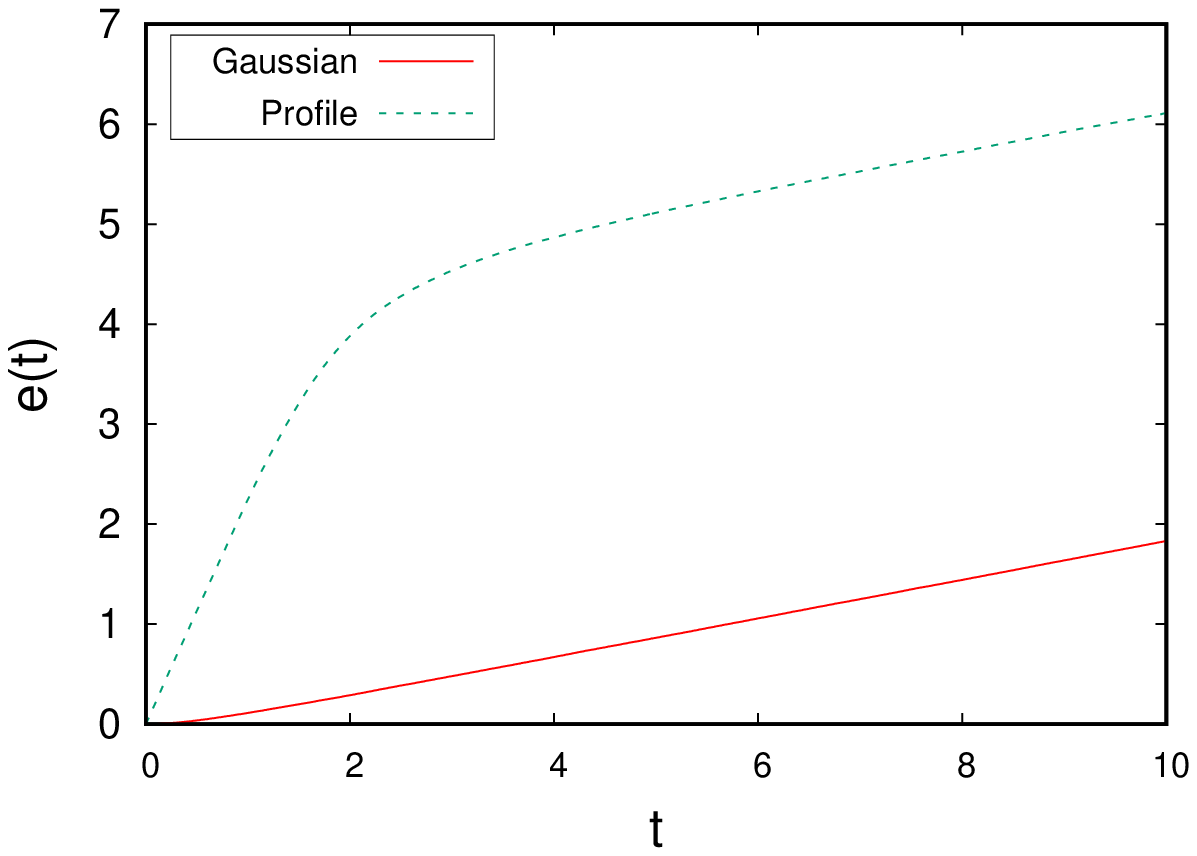}
		\caption{}
		\label{fig:expectation_b15_stationary}
	\end{subfigure}\hfil 
	
	\caption{          {\bf Convergence to the lowest steady state}.
		$\mathbf{b=1.5}$.
		The initial condition is a normal distribution with $ \nu_0 = 0 $ and $ \sigma^2 = 0.25 $ in the "Gaussian" case and a profile distribution given by Eq.\eqref{eq:profile} in the "Profile" one. 
		{\em (a)} Initial conditions used in the simulations.
		{\em (b)} Particle voltage distributions at time $5$ in 
		comparison with the lowest steady state of
		Fokker-Planck  equation.
		{\em (c)} Time evolution of the firing rate.
		{\em (d)} Time evolution of the expectation.} 
	\label{fig:steadystate-b1.5}
\end{figure}

\noindent{\em Convergence to the stationary solution with the lowest
  firing rate.-}
Let us consider  two different initial conditions for the
particle system voltages, which are far from the threshold $ V_F $,
(see Fig.\ref{fig:initial-condition-b15}): a normal distribution with
$ \nu_0 = 0 $ and $ \sigma^2 = 0.25 $, and the profile given by
\begin{equation}\label{eq:profile}
  p(v) = \frac{F_{rate}}{a}e^{-\frac{\left(v-bF_{rate}\right)^2}{2a}}
  \int_{max(v, V_R)}^{V_F}e^{\frac{\left(w-bF_{rate}\right)^2}{2a}}dw
\end{equation}
with $F_{rate} = 2.31901$, which approximates the highest stationary
firing rate.  The profile \eqref{eq:profile} is an approximation
of the steady state for the Fokker-Planck equation with
the highest firing rate (see \cite{caceres2011analysis} for details).

\

\newpage

In the first case, where the initial condition is a normal
distribution, the particle system tends to  
the 
steady state 
with lowest firing rate value, see Fig.\ref{fig:comparison_edp_b15_stationary}.
The evolution over time is described in terms of the firing rate
in Fig.\ref{fig:firing-rate-b15-stationary},
and also through the expectation in Fig.\ref{fig:expectation_b15_stationary}.
The firing rate  evolves towards the lowest stationary value,
and therefore, the expectation presents
a linear growth at the end of the simulation.

\

To describe the unstability of the steady state with the highest
firing rate value, we  consider the second initial datum,
profile \eqref{eq:profile}.
In Fig.\ref{fig:comparison_edp_b15_stationary}, we see that the distribution of voltages ends up in the same steady state as in the previous case.
Simulations starting with the profile
\eqref{eq:profile} and a little increase of the value 
$F_{rate}$ produce the explosion in finite time, as in
\cite{caceres2011analysis}.
Therefore, based on our results, and results in \cite{caceres2011analysis},
the steady state with the highest firing rate
appears to be a ``limit state''  between
situations of convergence towards the lowest steady state and those of
blow-up.

\

\begin{figure}[H]
  \centering 
        	\begin{subfigure}{0.38\textwidth}
		\includegraphics[width=\linewidth]{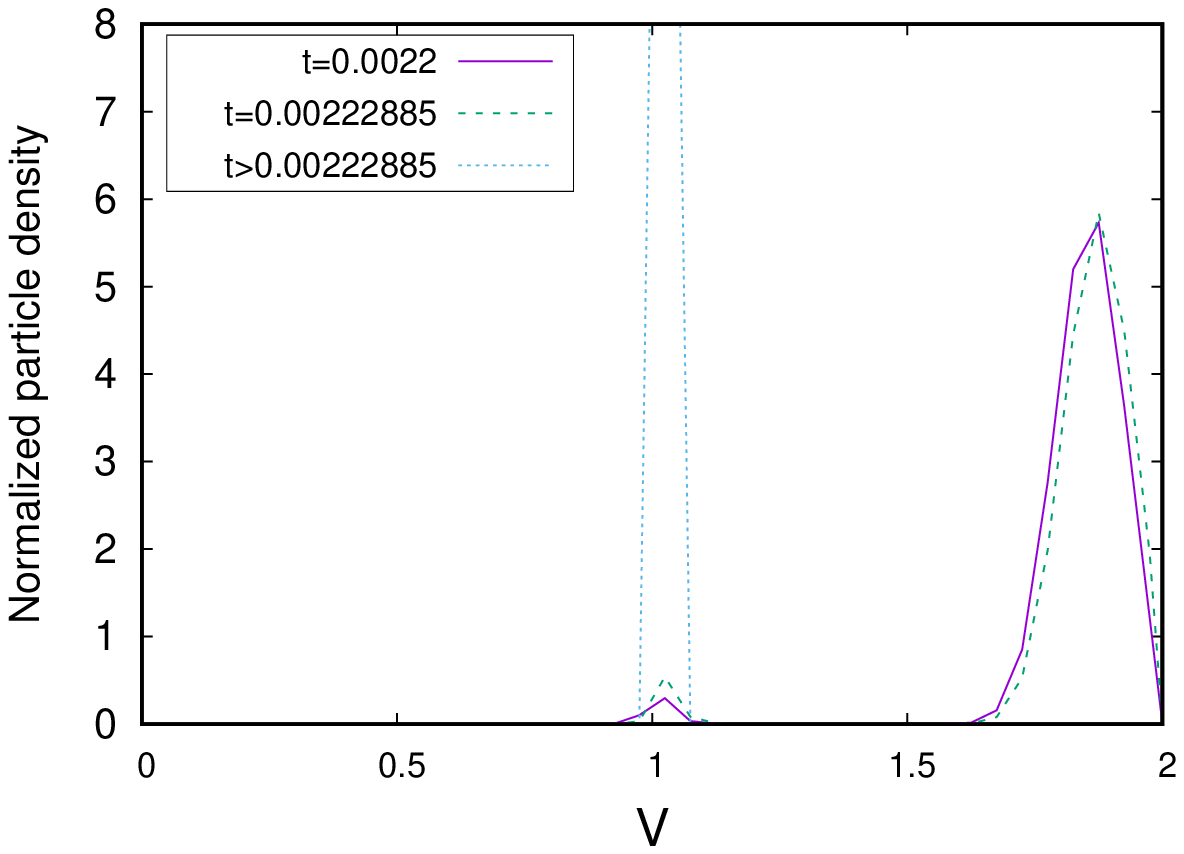}
		\caption{}
		\label{fig:b15_histograms}
	\end{subfigure}\hfil 
	\begin{subfigure}{0.38\textwidth}
		\includegraphics[width=\linewidth]{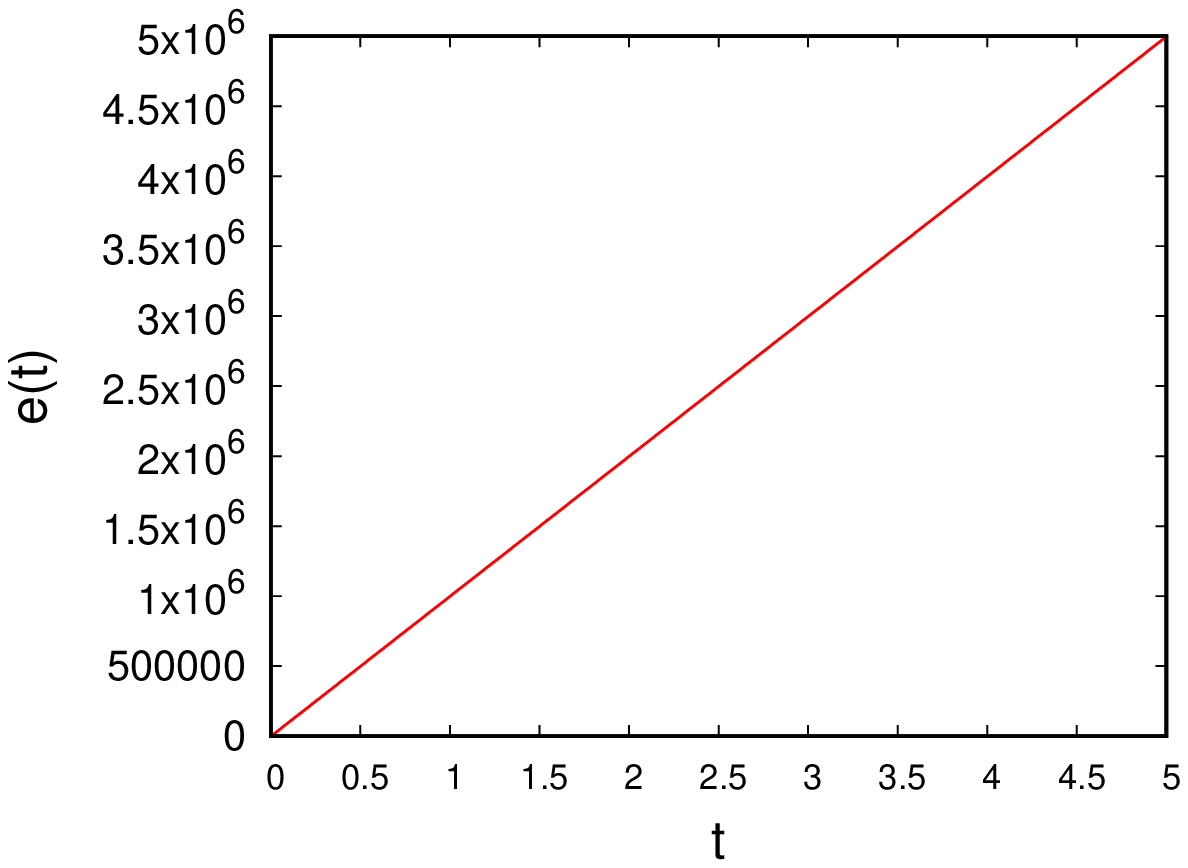}
		\caption{}
		\label{fig:expectation_b15_blow-up_longtime}
	\end{subfigure}\hfil 
	
	\medskip
	\begin{subfigure}{0.38\textwidth}
		\includegraphics[width=\linewidth]{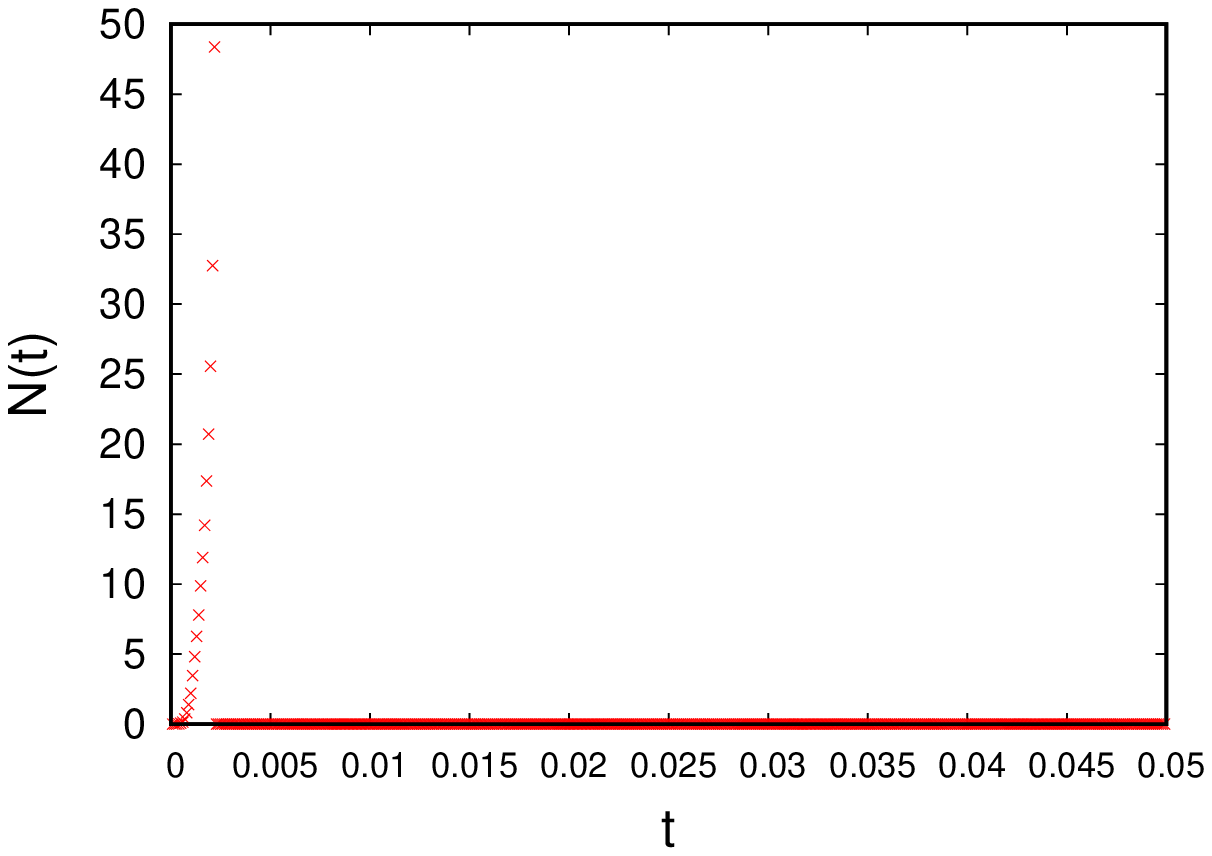}
		\caption{}
		\label{fig:firing_rate_b15_blow-up_zoom}
	\end{subfigure}\hfil 
		\caption{
		{\bf Blow-up phenomenon}.
		$\mathbf{b=1.5}$.
		The initial condition is a normal distribution with
		$ \nu_0 = 1.83 $ and $ \sigma = 0.003 $.
		{\em (a)} Time evolution  of the particle voltage distribution before and after the blow-up.
		{\em (b)} Time evolution  of the expectation.
		{\em (c)} Time evolution  of the firing rate.
	}
	\label{fig:b15-blowup}
\end{figure}

{\em Blow-up phenomenon.-}
We consider an initial condition given
by a normal distribution with $ \nu_0 = 1.83 $ and $ \sigma = 0.003 $
(Fig.\ref{fig:initial-conditions} Right), and we show 
the data obtained during the simulation  in Fig.\ref{fig:b15-blowup}.

\

After a short period of time, almost all neurons spike
simultaneously (blow-up phenomenon). Then neurons reset their
voltages to the value $ V_R $. However, the variation of the expectation
computed in the blow-up time is so large that the whole network can spike
again in the next step. This means that, if blow-up occurs at time $t_{j-1} $,
then $ b\frac{n^j}{\mathcal{N}}>V_F-V_R $, due to
the high value of the connectivity parameter $b$.
This phenomenon causes the system to be in a constant state of
blow-up and reset, which we call ``trivial solution''.
In Fig.\ref{fig:b15_histograms} we observe
the  voltages distributions  obtained before  and after the blow-up.
The final value of the voltage distribution shows
that  all neurons are locate in $ V_R $.
So that ``trivial solution''  appears to be a Dirac delta at $V_R$,
because neurons reset to $V_R$ at blow-up time. Consequently, in Fig.\ref{fig:expectation_b15_blow-up_longtime} we see that the final value of the expectation is near to $ T/\Delta t=5\times 10^6 $, which is the value that expectation would reaches if all neurons spike in each step from beginning to end of the simulation.
In Fig.\ref{fig:firing_rate_b15_blow-up_zoom} we see clearly the moment when the firing rate  diverges.

\

\begin{figure}[H]
	\centering 
	\begin{subfigure}{0.38\textwidth}
		\includegraphics[width=\linewidth]{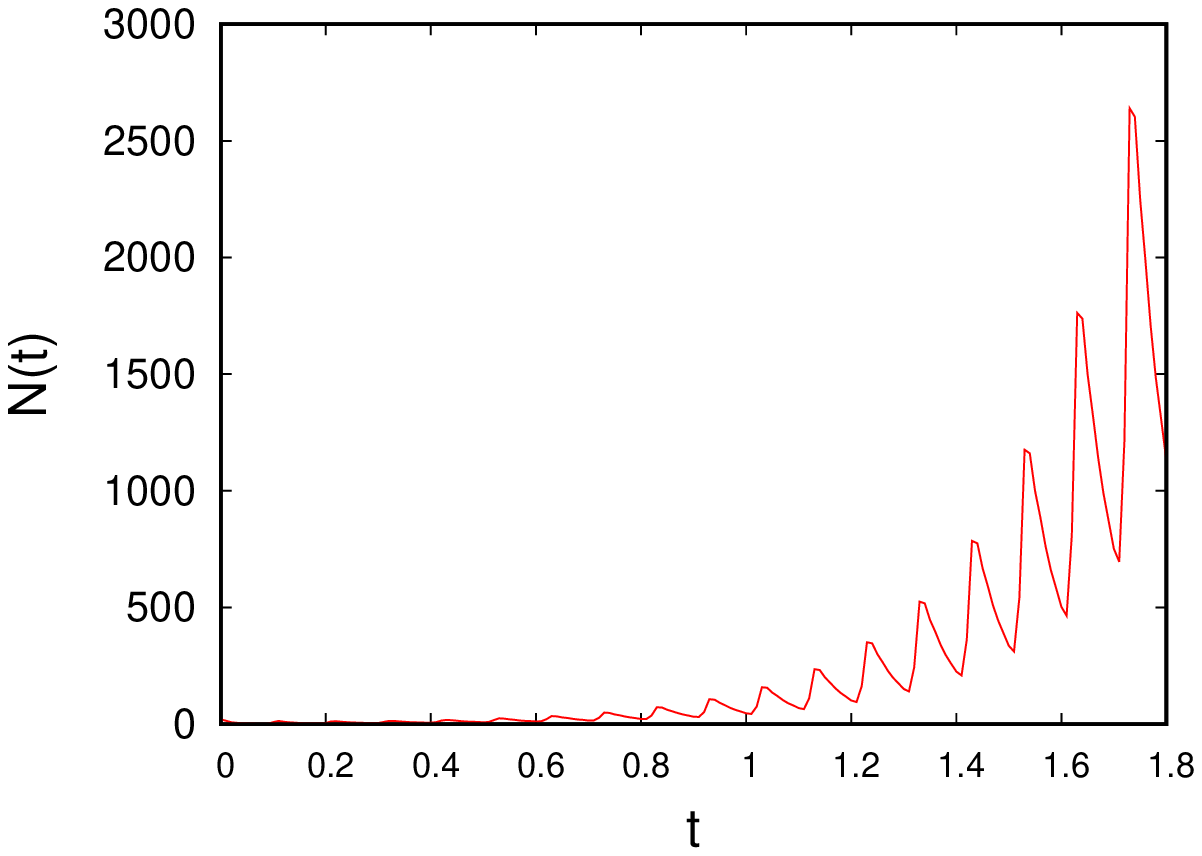}
		\caption{}
		\label{fig:firing_rate_b15_blow-up_longtime_delay}
	\end{subfigure}\hfil 
	\begin{subfigure}{0.38\textwidth}
		\includegraphics[width=\linewidth]{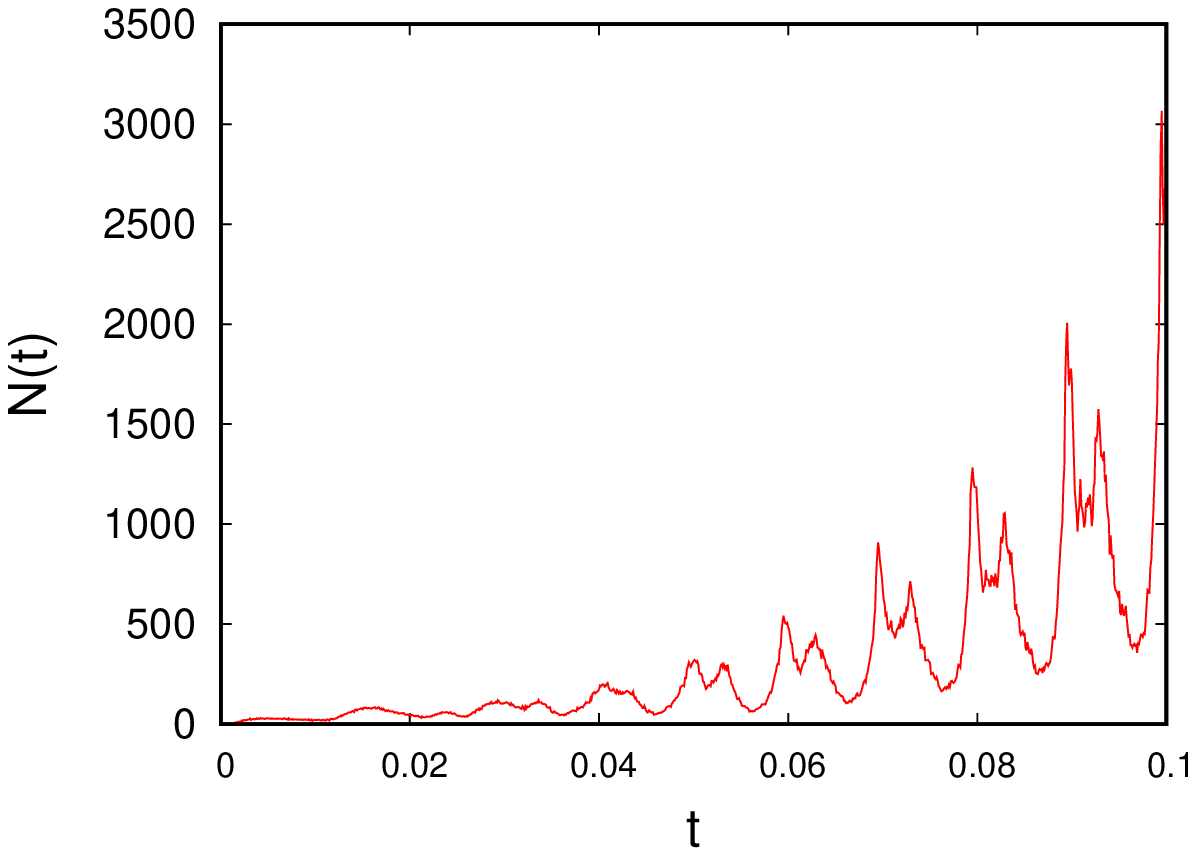}
		\caption{}
		\label{fig:firing_rate_b15_blow-up_zoom_delay}
	\end{subfigure}\hfil 
	
	\medskip
	\begin{subfigure}{0.38\textwidth}
		\includegraphics[width=\linewidth]{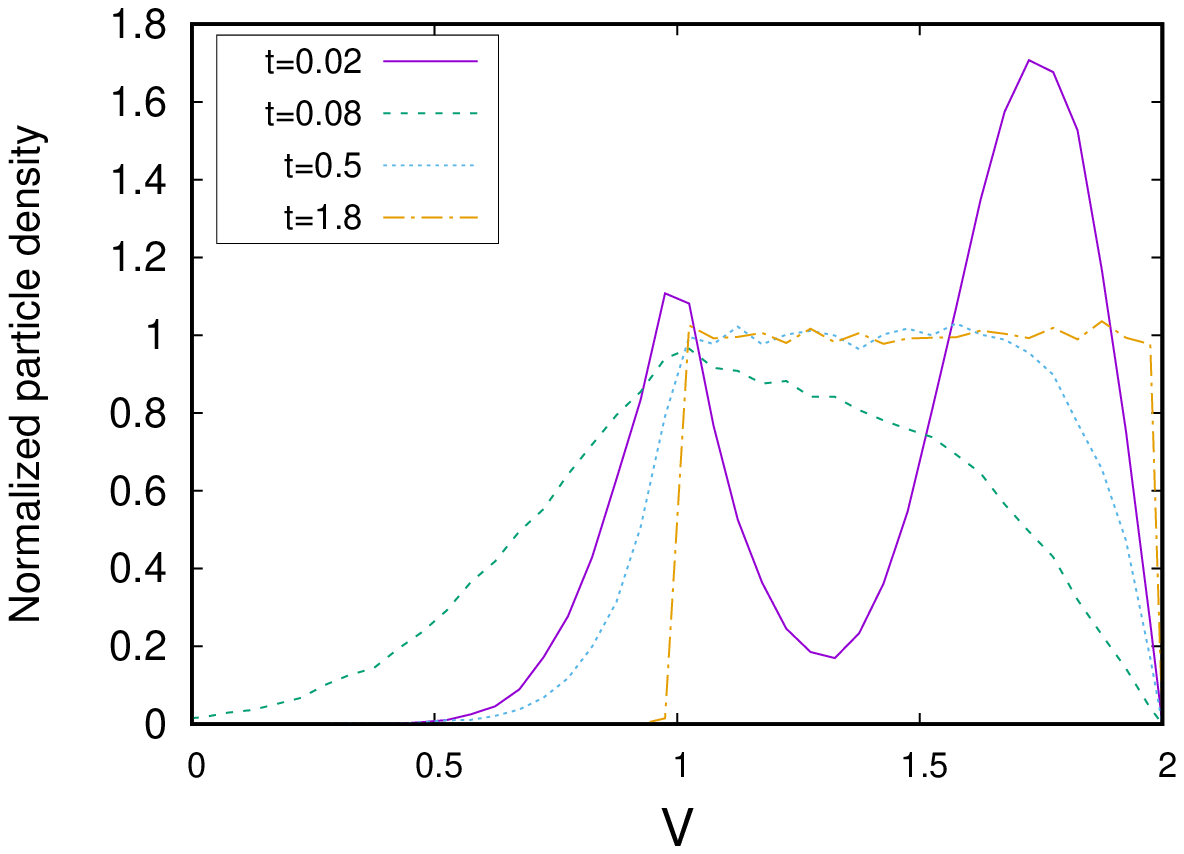}
		\caption{}
		\label{fig:b15_histograms-delay-01}
	\end{subfigure}\hfil 
	\begin{subfigure}{0.38\textwidth}
		\includegraphics[width=\linewidth]{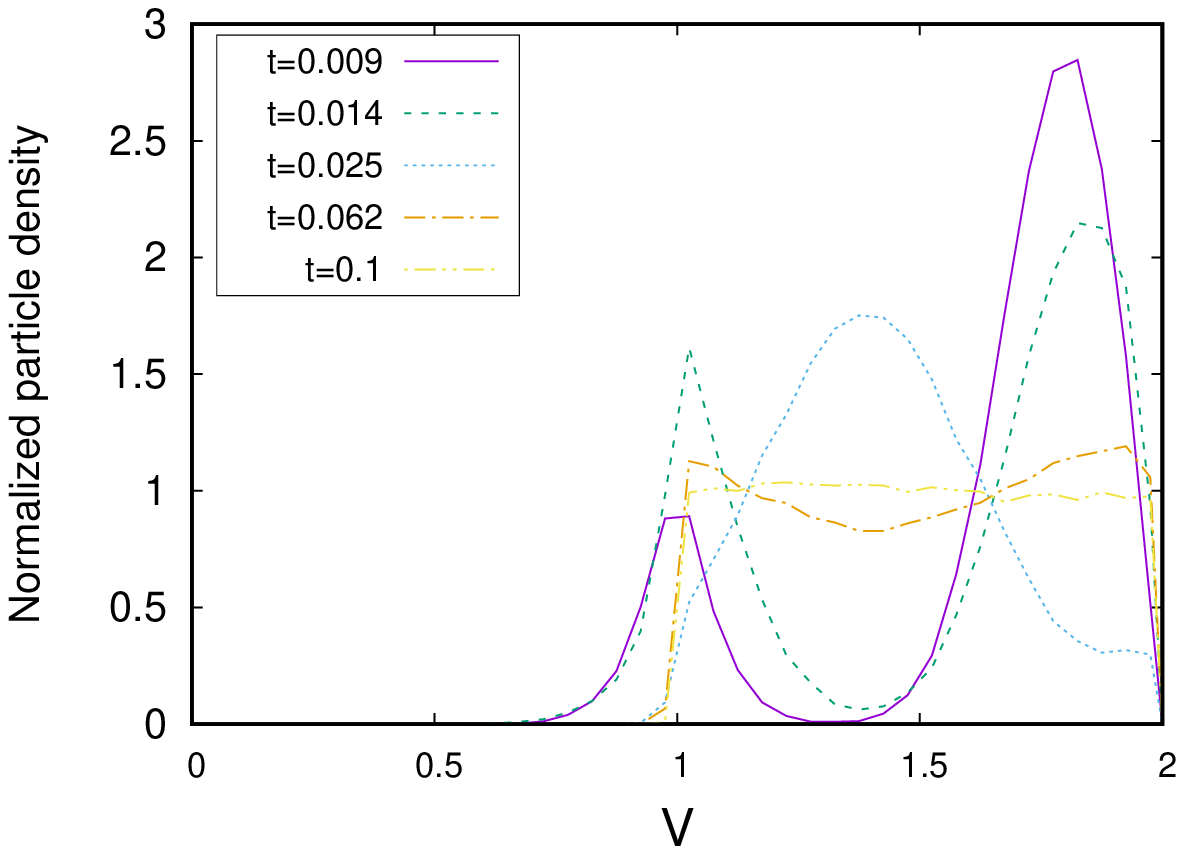}
		\caption{}
		\label{fig:b15_histograms-delay-001}
	\end{subfigure}\hfil 
	
	\caption{
		{\bf Blow-up situation with delay}.
		$\mathbf{b=1.5}$.
                	The initial condition is a normal distribution with
		$ \nu_0 = 1.83 $ and $ \sigma = 0.003 $.
                On the left  $ \delta=0.1 $. On the right $ \delta=0.01 $.
		{\em (a)} and {\em (b)} Time evolution  of the firing rate.  
		{\em (c)} and {\em (d)} Time evolution of the particle voltage distribution.
	}
	\label{fig:b15-blowup-delay}
\end{figure}

{\em Transmission delay.-}
We analyse here what happens in the blow-up situation, when transmission delays are taken into account and the system  avoids the explosion.
We consider the initial datum given in
Fig.\ref{fig:initial-conditions} Right and develop several simulations
with different transmission delay values: $ \delta = 0.1 $ and $ \delta=0.01 $.
In Fig.\ref{fig:b15-blowup-delay} we show both simulations. With a
transmission delay the voltage distribution avoids
blow-up and it converges to  a new state. We
have called to this state "plateau distribution", see Figs.\ref{fig:b15_histograms-delay-01} and  \ref{fig:b15_histograms-delay-001}.
 In this new distribution almost all neurons voltages are located between
$ V_R $ and $ V_F $. Moreover, the probability to find any voltage in $ v\in\left[1,2\right) $ is almost the same.

\
\\

\newpage

In both simulations, see Figs.\ref{fig:firing_rate_b15_blow-up_longtime_delay} and \ref{fig:firing_rate_b15_blow-up_zoom_delay},
the firing rates tend to increase. This behaviour is in full agreement with the formation of the ``plateau'' distributions observed in Figs.\ref{fig:b15_histograms-delay-01} and  \ref{fig:b15_histograms-delay-001}. On the other hand, the comparison between Figs.\ref{fig:firing_rate_b15_blow-up_longtime_delay} and \ref{fig:firing_rate_b15_blow-up_zoom_delay} shows
the influence of the delay size. 
As the delay gets smaller and smaller, the firing rate tends to diverge
in a finite time. We checked it for 
$\delta = 0.00001$, with which the system does not
avoid the blow-up. All neurons spike at once and the distribution converges
to the ``trivial solution''.  Finally, we remark  that the time needed
to get close to the ``plateau'' state is directly proportional to the value of the delay, see Figs.\ref{fig:b15_histograms-delay-01} and \ref{fig:b15_histograms-delay-001}.

\subsubsection{Case without steady states. Connectivity parameter
  $\mathbf{b=2.2}$}

\begin{figure}[H]
	\begin{center}
		\begin{minipage}[c]{0.38\linewidth}
			\begin{center}
				\includegraphics[width=\textwidth]{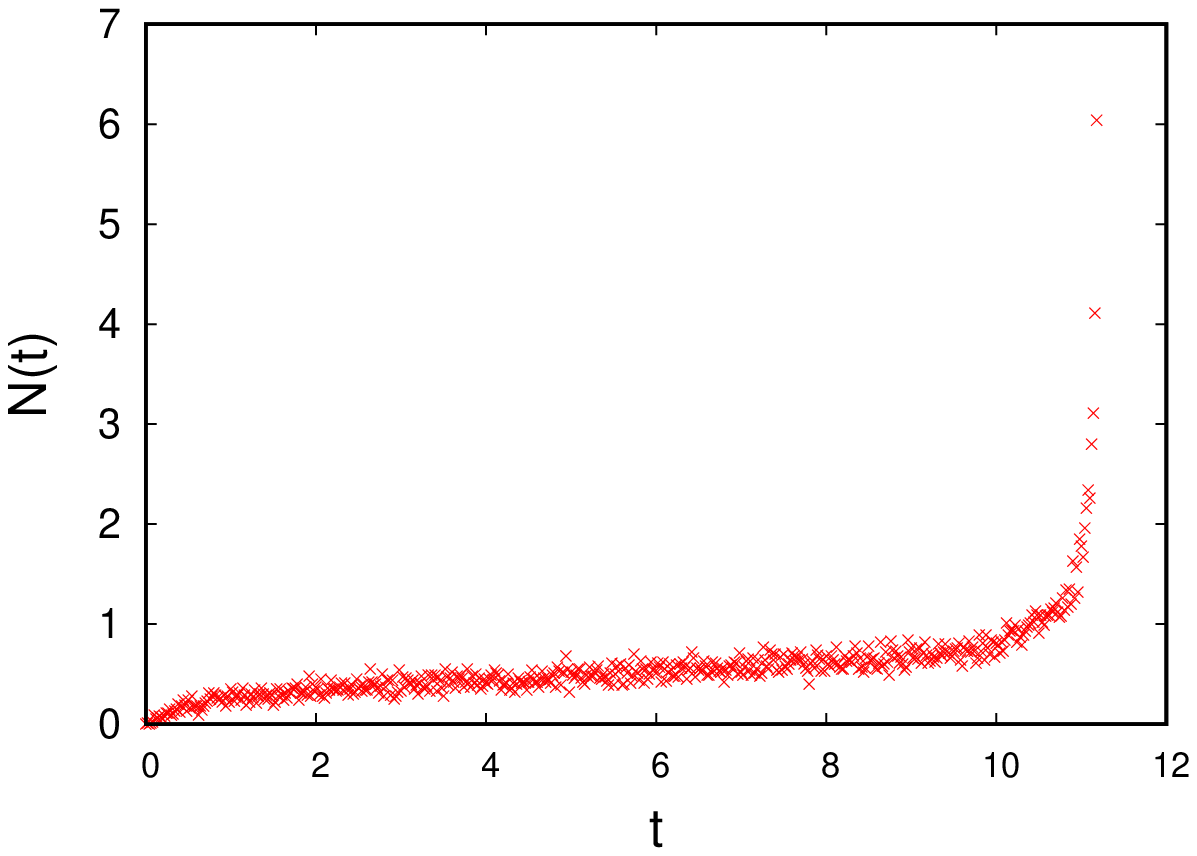}
			\end{center}
		\end{minipage}
		\begin{minipage}[c]{0.38\linewidth}
			\begin{center}
				\includegraphics[width=\textwidth]{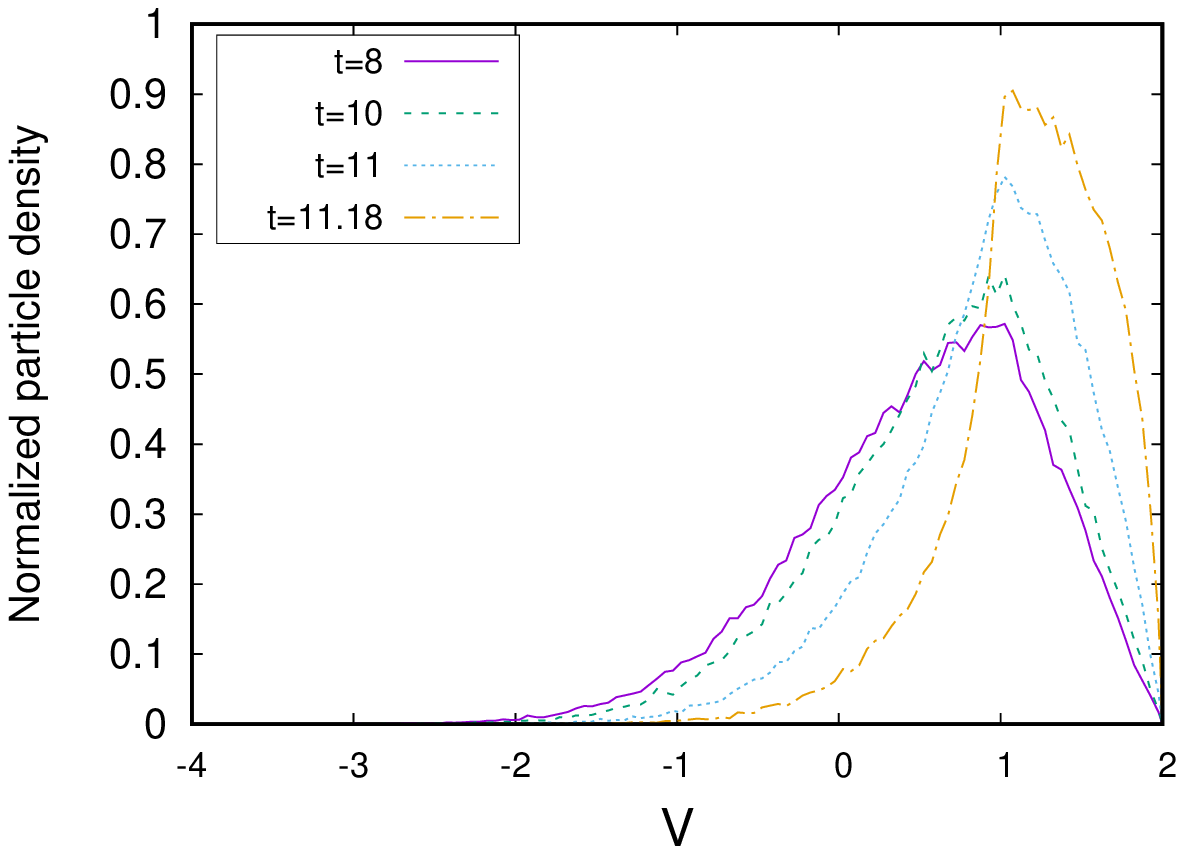}
			\end{center}
		\end{minipage}
	\end{center}
	\caption{{\bf Blow-up phenomenon without delay.}
          $\mathbf{b=2.2.}$
The initial condition is a  normal distribution with $ \nu_0 = 0 $ and $ \sigma^2 = 0.25$.
		{\em Left}: Time evolution of the firing rate.
		{\em Right}: Time evolution of the particle voltage
                distribution before the blow-up. 
	}
	\label{fig:b22_blowup}
\end{figure}

We also  study  the case with $ b=2.2 $, which represents
highly excited neural networks that do not exhibit stationary states
(see Fig.\ref{fig:steady-states}). In Fig.\ref{fig:b22_blowup} we can observe that the blow-up is guaranteed
regardless of the initial conditions,
without transmission delay (see \cite{roux2020towards}). In that simulation the initial
distribution is far from the firing threshold and the firing rate diverges in
finite time. On the right side of the figure we see the last output for the distribution, before all the neurons fire at once.

\

\begin{figure}[H]
	\centering 
	\begin{subfigure}{0.38\textwidth}
		\includegraphics[width=\linewidth]{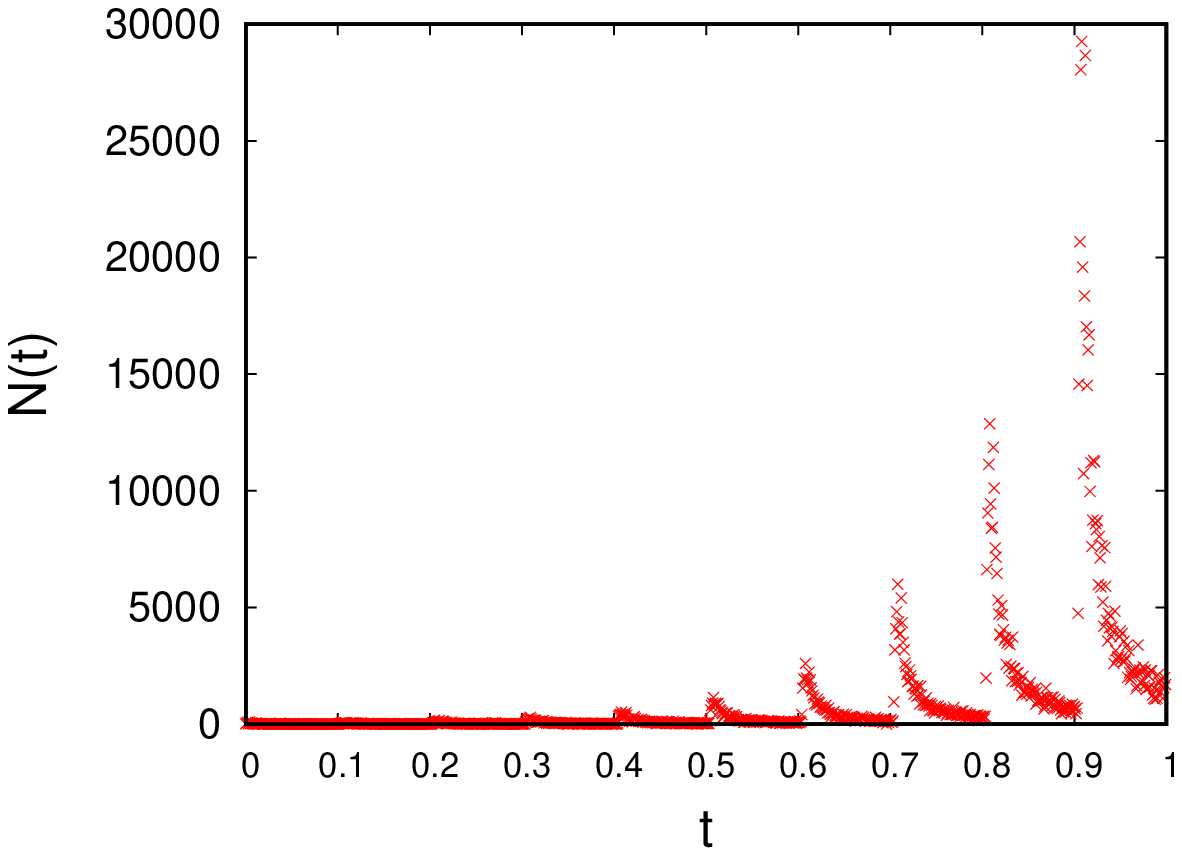}
		\caption{}
		\label{fig:firing_rate_b22_blow-up_zoom-delay}
	\end{subfigure}\hfil 
	\begin{subfigure}{0.38\textwidth}
		\includegraphics[width=\linewidth]{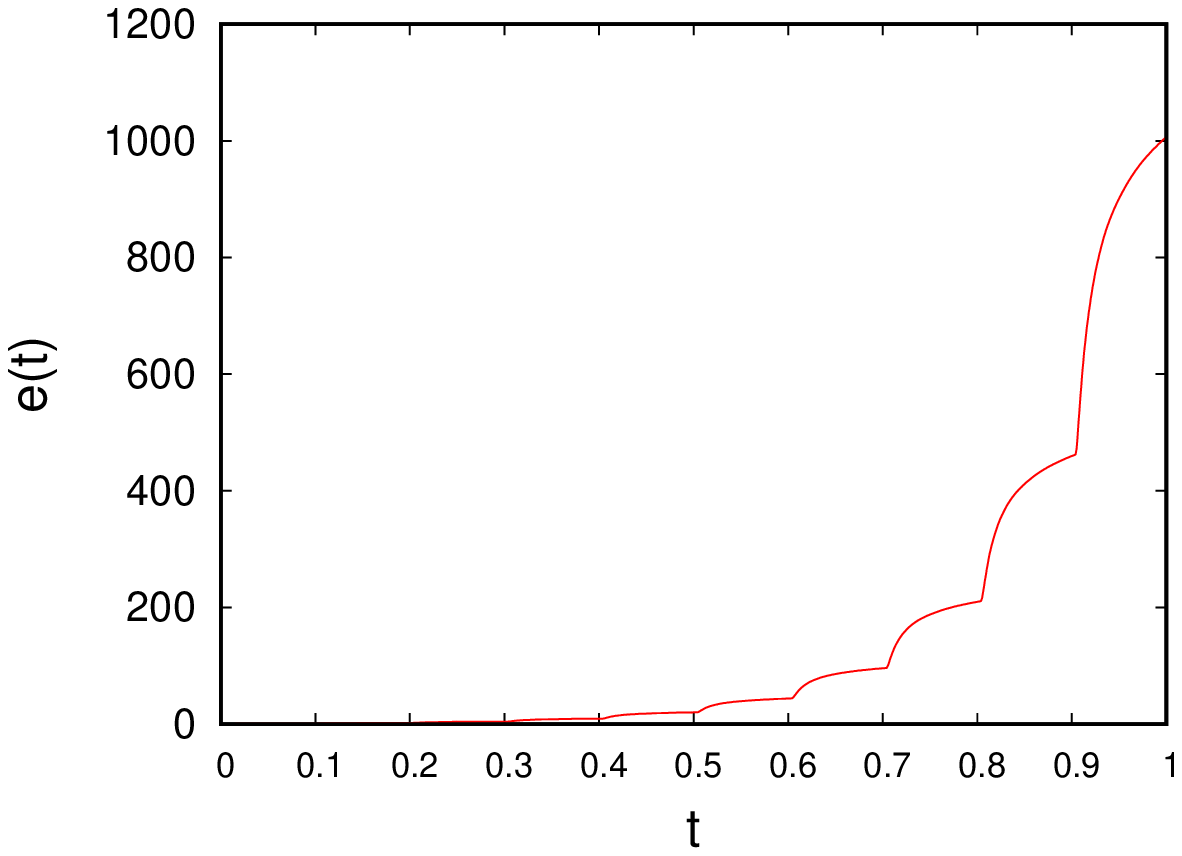}
		\caption{}
		\label{fig:expectation_b22_blow-up_zoom-delay}
	\end{subfigure}\hfil 
	
	\medskip
	\begin{subfigure}{0.38\textwidth}
		\includegraphics[width=\linewidth]{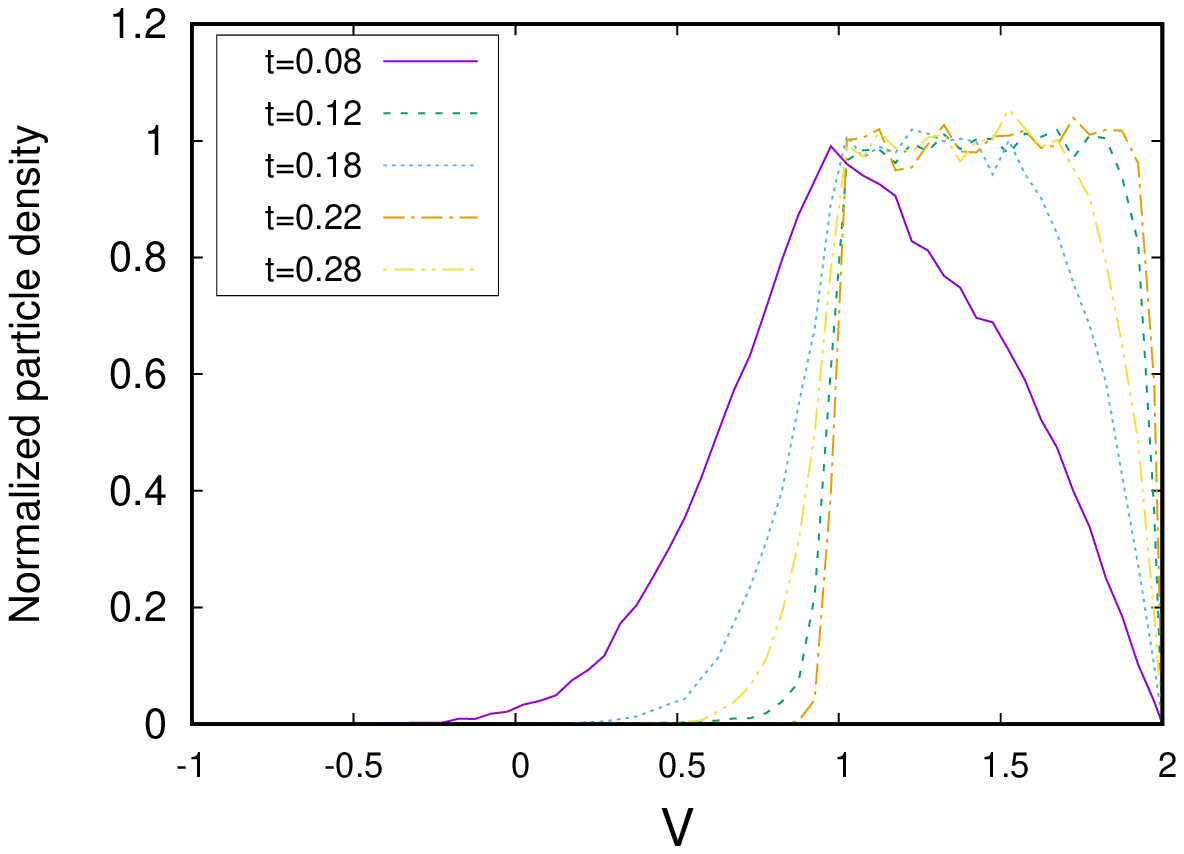}
		\caption{}
		\label{fig:b22_histograms-delay-01_1}
	\end{subfigure}\hfil 
	\begin{subfigure}{0.38\textwidth}
		\includegraphics[width=\linewidth]{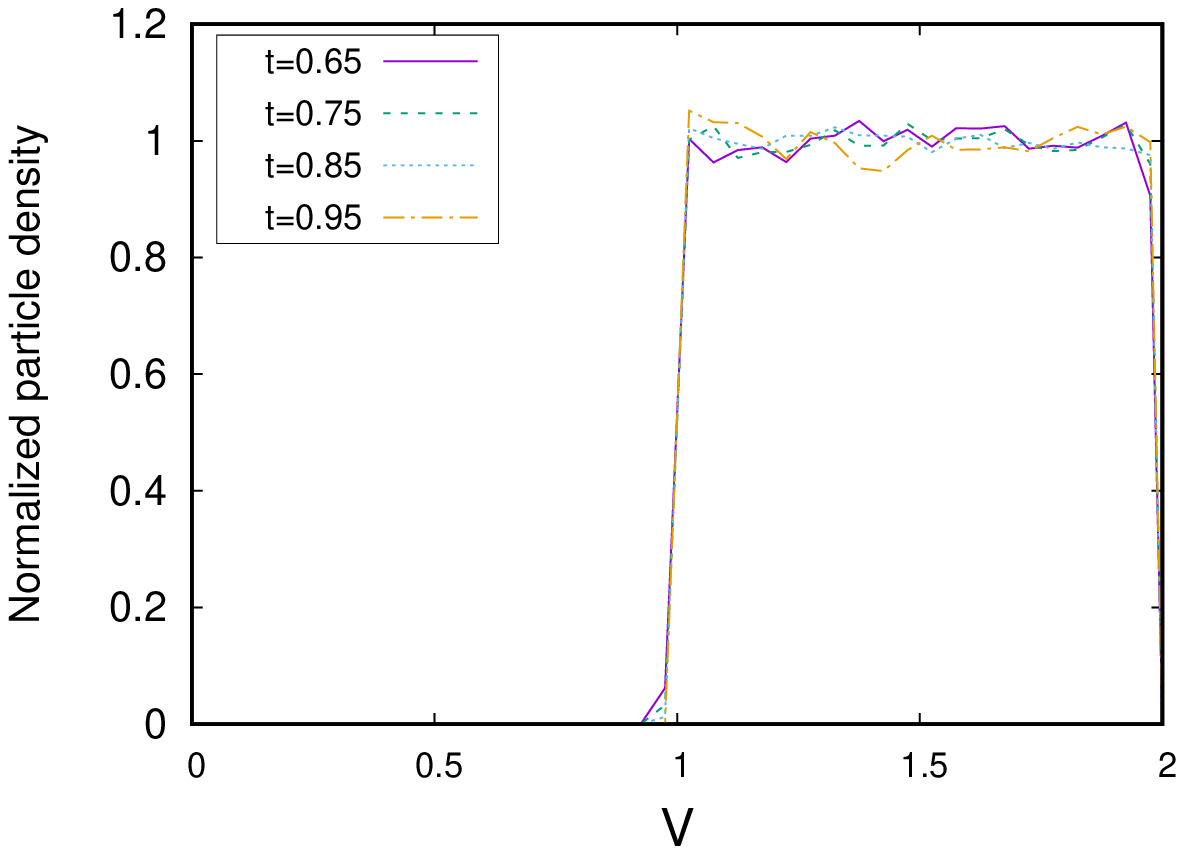}
		\caption{}
		\label{fig:b22_histograms-delay-01_2}
	\end{subfigure}\hfil 
	
	\caption{
		{\bf Blow-up situation with delay}.
		$\mathbf{b=2.2}$.
		The initial condition is a normal distribution with
		$ \nu_0 = 1.83 $ and $ \sigma = 0.003 $, and the delay value $ \delta=0.1 $.
		{\em (a)} Time evolution  of the firing rate.
		{\em (b)} Time evolution  of the expectation.
		{\em (c)} and {\em (d)} Time evolution  of the particle voltage distribution during the formation of the "plateau" state.
	}
	\label{fig:b22-blowup-delay}
\end{figure}

When a
transmission delay is considered, 
the system avoids blow-up and converges to a  "plateau" state.
Fig.\ref{fig:b22-blowup-delay} shows time evolution  of the particle
system, with an initial distribution close to $V_F$ and
with a transmission delay value $\delta=0.1$.
In this situation, the firing rate increases,
but does not diverge in finite time, as we can see in Fig\ref{fig:firing_rate_b22_blow-up_zoom-delay}.
The firing rate also represents the growth rate of expectation, as shown in Fig.\ref{fig:expectation_b22_blow-up_zoom-delay}.
As in the case $b=1.5$, that behaviour is reflected in the evolution of the potentials distribution. 
At the beginning of the simulation,  Fig.\ref{fig:b22_histograms-delay-01_1} ,
we see how the system oscillates
between a shape similar to that which occurs without delay and a ``plateau''
distribution.
Some time later, the system approaches the ``plateau'' distribution,
i.e., a uniform distribution in $(V_R, V_F)$, Fig.\ref{fig:b22_histograms-delay-01_2}.

\

This behaviour was observed for Fokker-Planck
equation solutions in \cite[(Fig.2)]{caceres2018analysis},  and although the evolution of the solutions was not shown
in said article, the construction of the "plateau"
distribution was already observed.

\

\subsection{Limiting case $b=V_F-V_R$}

In this section we analyse the limiting case for which the notion of
physical solution no longer makes sense (see Sect.\ref{sec:cascade}).
This case  is in between of two regimes: for values $b<1$, there is only
one steady state (see Fig.\ref{fig:steady-states})
and  simulations  show that all neurons in the system can fire at the same
time, and then converge towards the steady state (see  Sect.\ref{sec:results-fisical-solution}).
However, if  $b>1$, the system needs the transmission delay
to avoid the explosion, because the physical solutions do not
make sense
(see Sect.\ref{sec:results-non-physical-solutions}).
For $V_F=2$ and $V_R=1$,
the limiting value of the connectivity parameter is $b=1$.
For this value of $b$,  neurons 
can spike twice if their membrane potentials are close to  $V_F$.
Therefore, we focus on studying this case through the delayed
and non-delayed  algorithms,
without cascade mechanism.
As first test, we  checked the convergence to the unique
stationary state, starting with a distribution of neurons with membrane potentials far from the threshold value.
We skip these figures here, and concentrate on the results found for possible explosion situations. Throughout this section,
the initial condition will be always a normal distribution with mean
$ \nu_0 = 1.83 $ and variance $ \sigma = 0.003 $ (see Fig.\ref{fig:initial-conditions}). 

\

\begin{figure}[H]
	\centering 
	\begin{subfigure}{0.38\textwidth}
		\includegraphics[width=\linewidth]{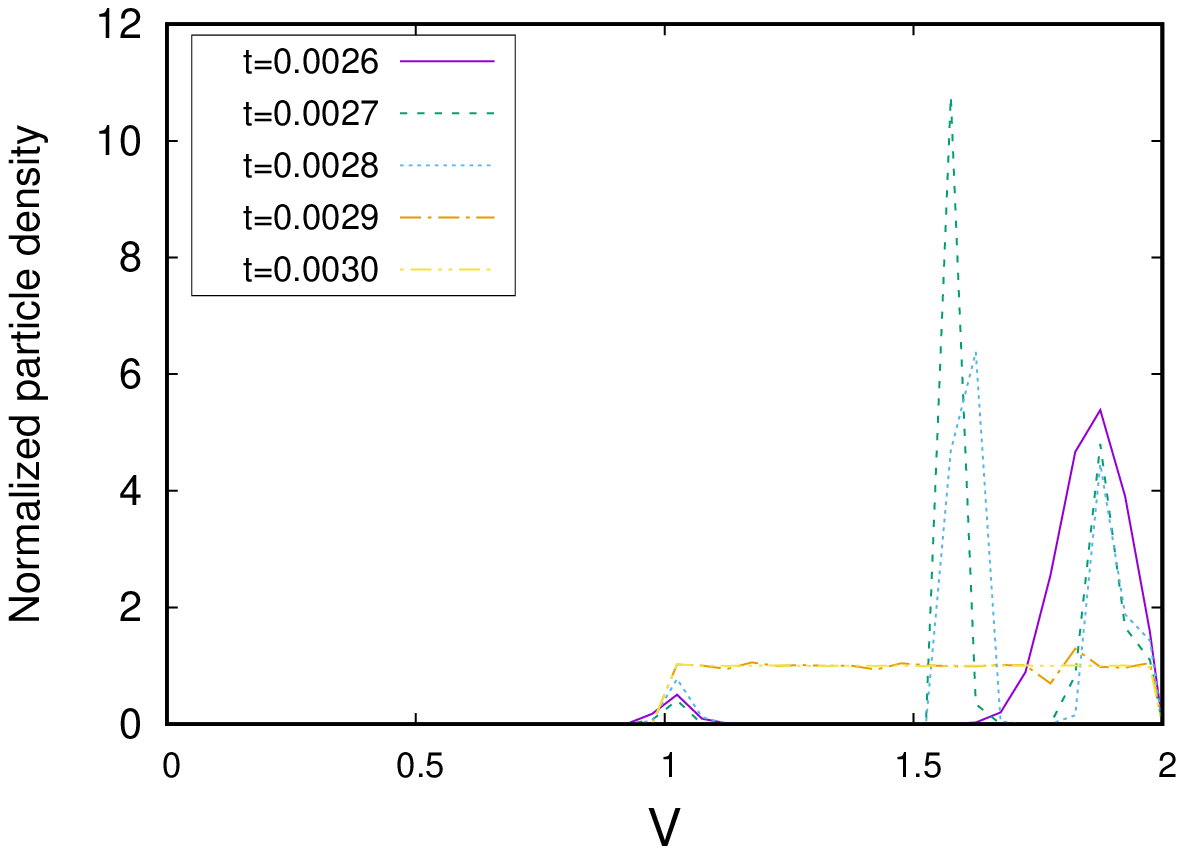}
		\caption{}
		\label{fig:b1-histograms-normal-1}
	\end{subfigure}\hfil 
	\begin{subfigure}{0.38\textwidth}
		\includegraphics[width=\linewidth]{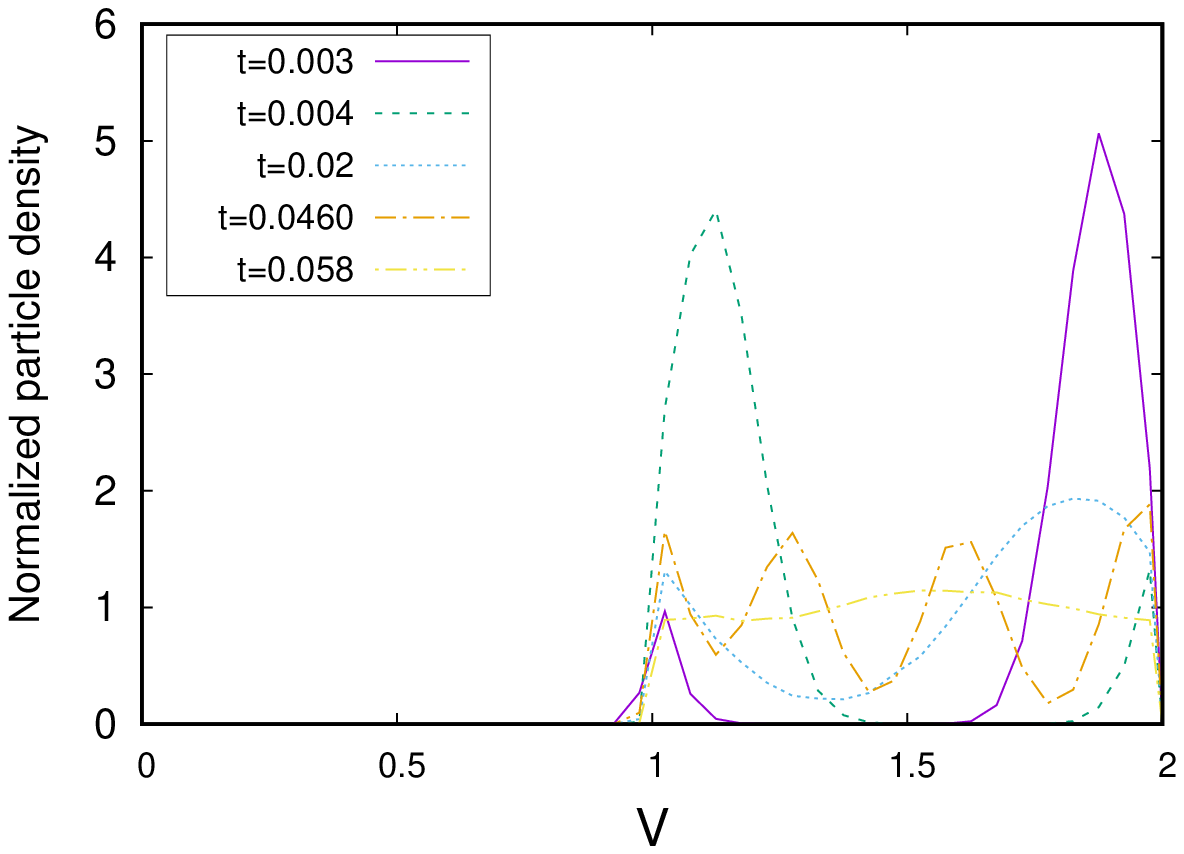}
		\caption{}
		\label{fig:b1-histograms-delay-1}
	\end{subfigure}\hfil 
	
	\medskip
	\begin{subfigure}{0.38\textwidth}
		\includegraphics[width=\linewidth]{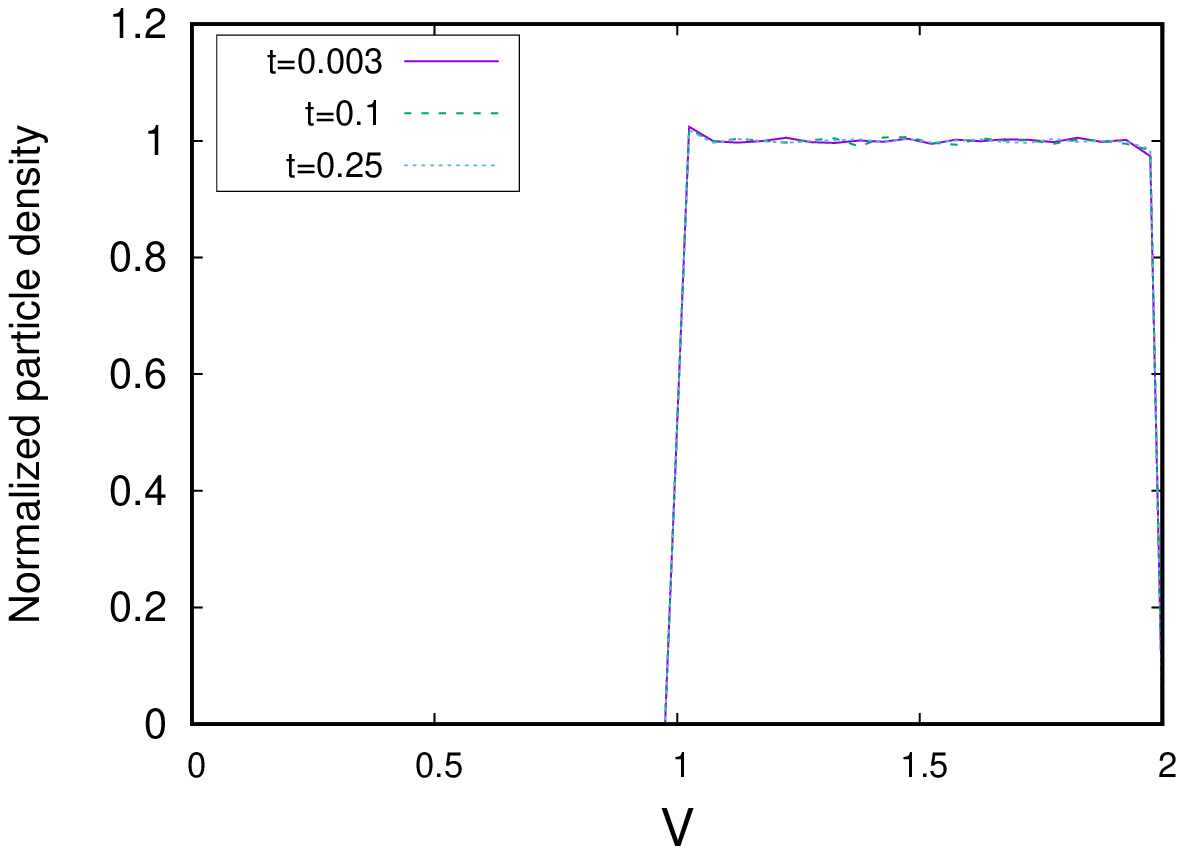}
		\caption{}
		\label{fig:b1-histograms-normal-2}
	\end{subfigure}\hfil 
	\begin{subfigure}{0.38\textwidth}
		\includegraphics[width=\linewidth]{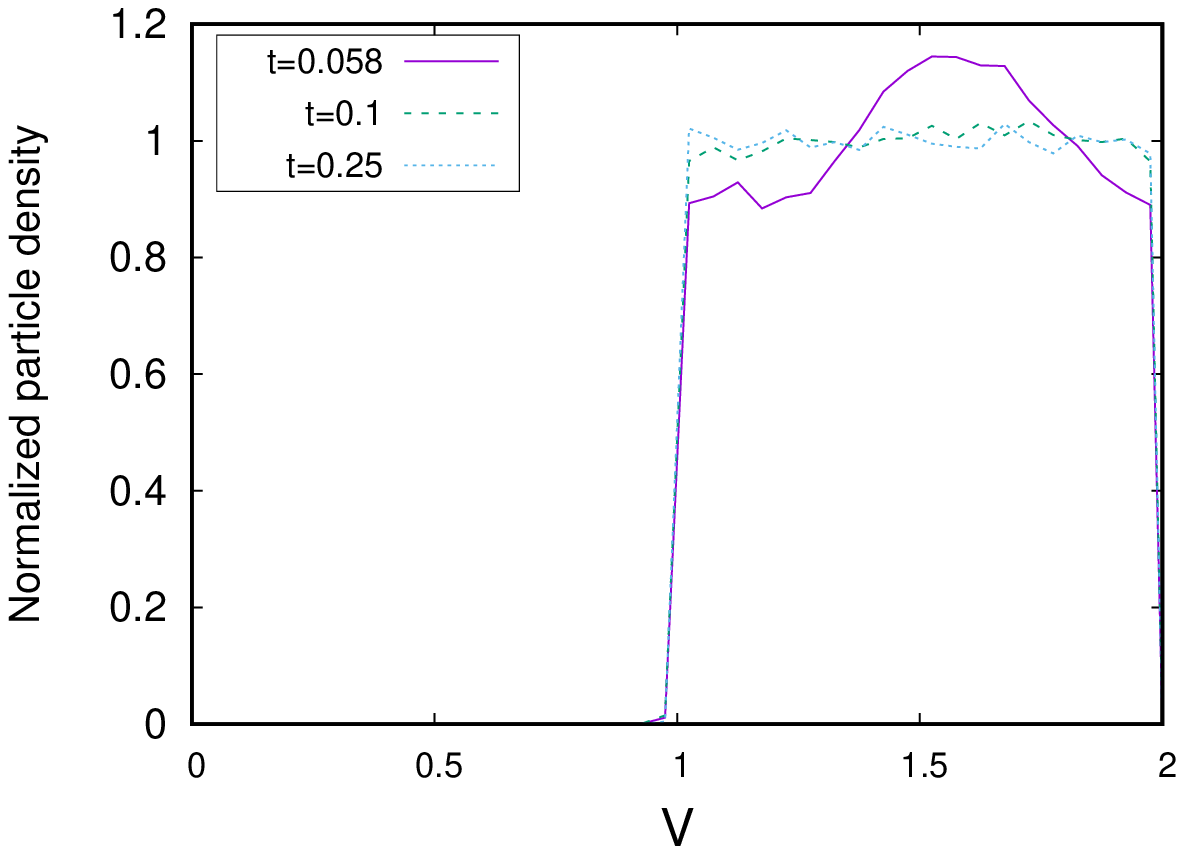}
		\caption{}
		\label{fig:b1-histograms-delay-2}
	\end{subfigure}\hfil 
	
	\medskip
	\begin{subfigure}{0.38\textwidth}
		\includegraphics[width=\linewidth]{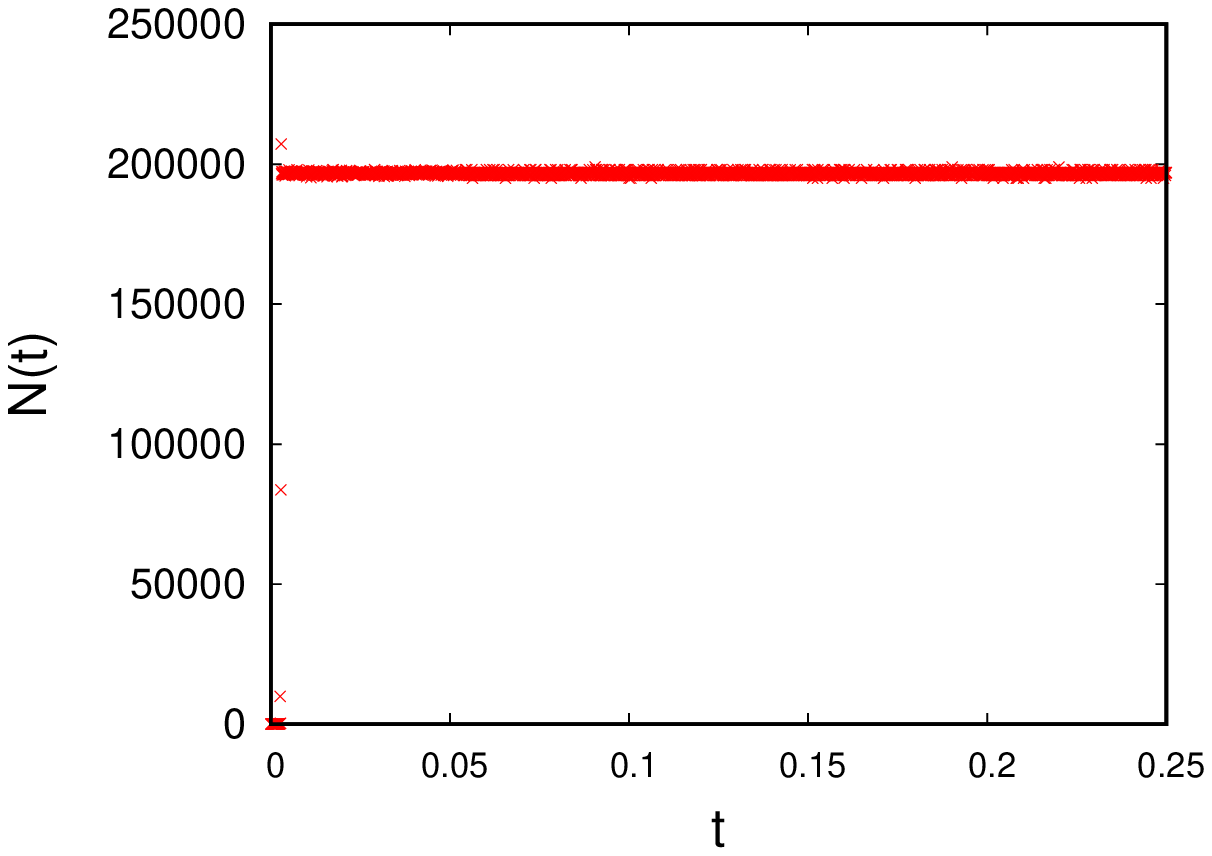}
		\caption{}
		\label{fig:b1-firing-rate-normal}
	\end{subfigure}\hfil 
	\begin{subfigure}{0.38\textwidth}
		\includegraphics[width=\linewidth]{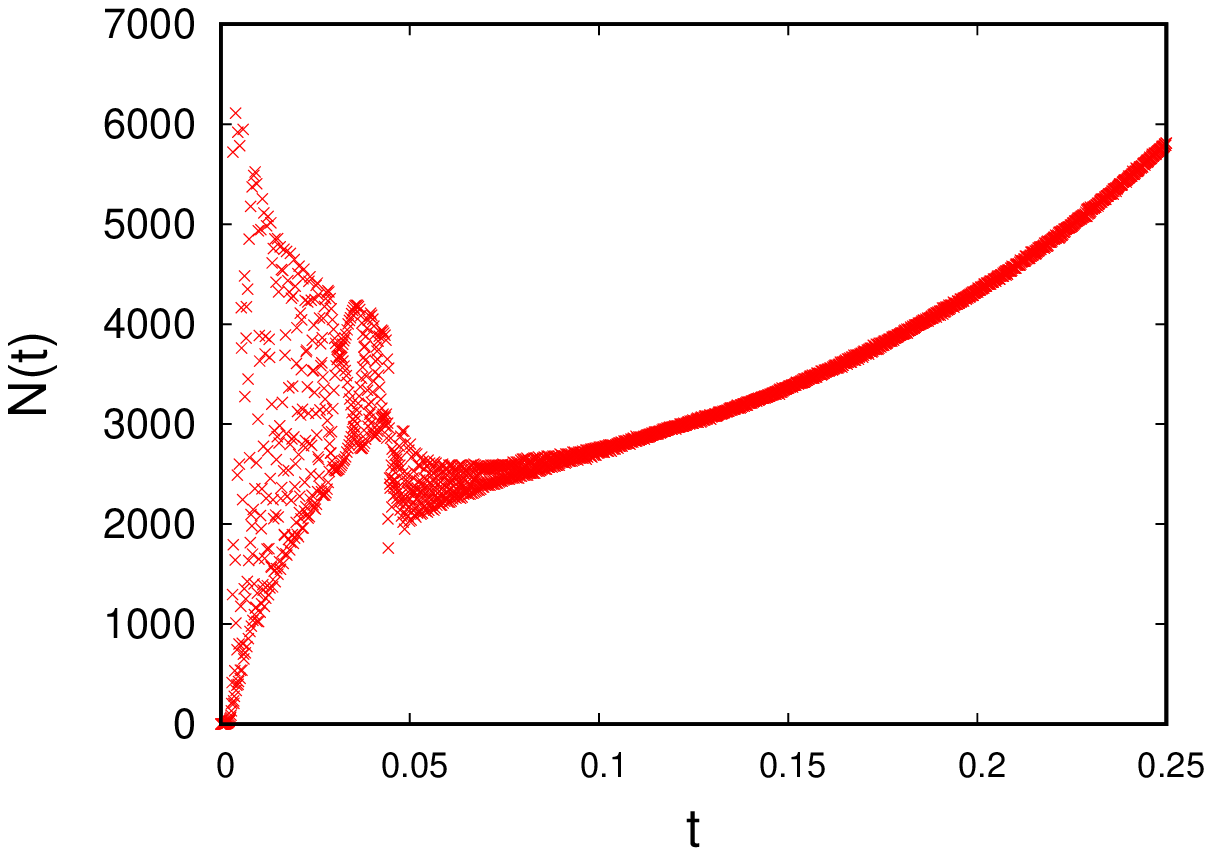}
		\caption{}
		\label{fig:b1-firing-rate-delay}
	\end{subfigure}\hfil 
	
	\caption{
		{\bf Time evolution of a particle system without transmission delay and with a very small delay}            
		$\mathbf{\delta=0.0001.}$
		$\mathbf{b=1.}$
		The initial condition is a normal distribution with
		$ \nu_0 = 1.83 $ and $ \sigma = 0.003 $ in both simulations.
		{\em (a)} Time evolution  of the particle voltage distribution without delay.		
		{\em (b)} Time evolution of the particle voltage distribution with delay.	
		{\em (c)} Time evolution  of the particle voltage distribution without delay.
		{\em (d)} Time evolution  of the particle voltage distribution with delay.						
		{\em (e)} Time evolution  of the firing rate without delay.
		{\em (f)} Time evolution  of the firing rate with delay.
	}
	\label{fig:b1-blowup-normal-delay}
\end{figure}

The delayed system approximates the  solution of the particle system
as the transmission delay $ \delta $ tends to zero. With this purpose,
we have made the comparison of the non-delayed system with the sufficiently
low delayed one ($ \delta=0.0001 $) in Fig.\ref{fig:b1-blowup-normal-delay}.

For the non-delayed case, we observe in Fig.\ref{fig:b1-histograms-normal-1}
how the voltages distribution a\-ppro\-aches the firing threshold and
the system achieves the "plateau" state, which appears to be a stable state,
as shown in Fig.\ref{fig:b1-histograms-normal-2}.
The transition between all neurons firing at once and the ``plateau''
distribution occurs extremely quickly.
However, the delayed system oscillates further until it approaches the
``plateau'' distribution
(see Fig.\ref{fig:b1-histograms-delay-1}).
This ``plateau'' state seems to be also stable
as shown in Fig.\ref{fig:b1-histograms-delay-2}.
The dynamics of the firing rates  
can be seen in the bottom of Fig.\ref{fig:b1-blowup-normal-delay}.
In the non-delayed case, the firing rate  remains constant from almost
the beginning in a very high value (see Fig.\ref{fig:b1-firing-rate-normal}).
This means that the system has converged towards the ``plateau'' distribution,
and that value is an approximation of an infinite firing rate.
In the case with delay the firing rate undergoes oscillations prior
to the formation of the ``plateau'' (see Fig.\ref{fig:b1-firing-rate-delay}).
Later, once the ``plateau'' is formed, its value tends to grow.
In the long term it should reach the value of the non-delayed case,
which would give a more defined form to the ``plateau''.

\

\begin{figure}[H]
	\centering 
	\begin{subfigure}{0.38\textwidth}
		\includegraphics[width=\linewidth]{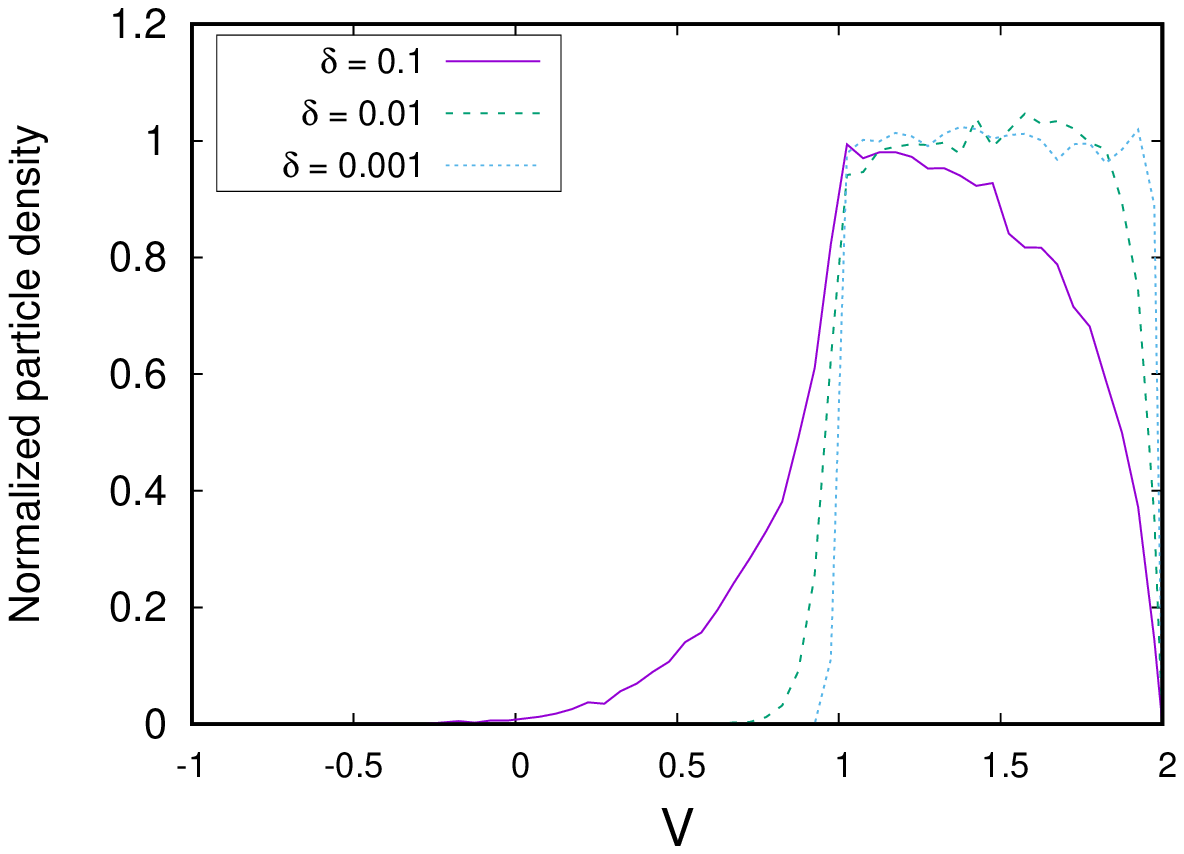}
		\caption{}
		\label{fig:b1-histograms-delay-sta-1}
	\end{subfigure}\hfil 
	\begin{subfigure}{0.38\textwidth}
		\includegraphics[width=\linewidth]{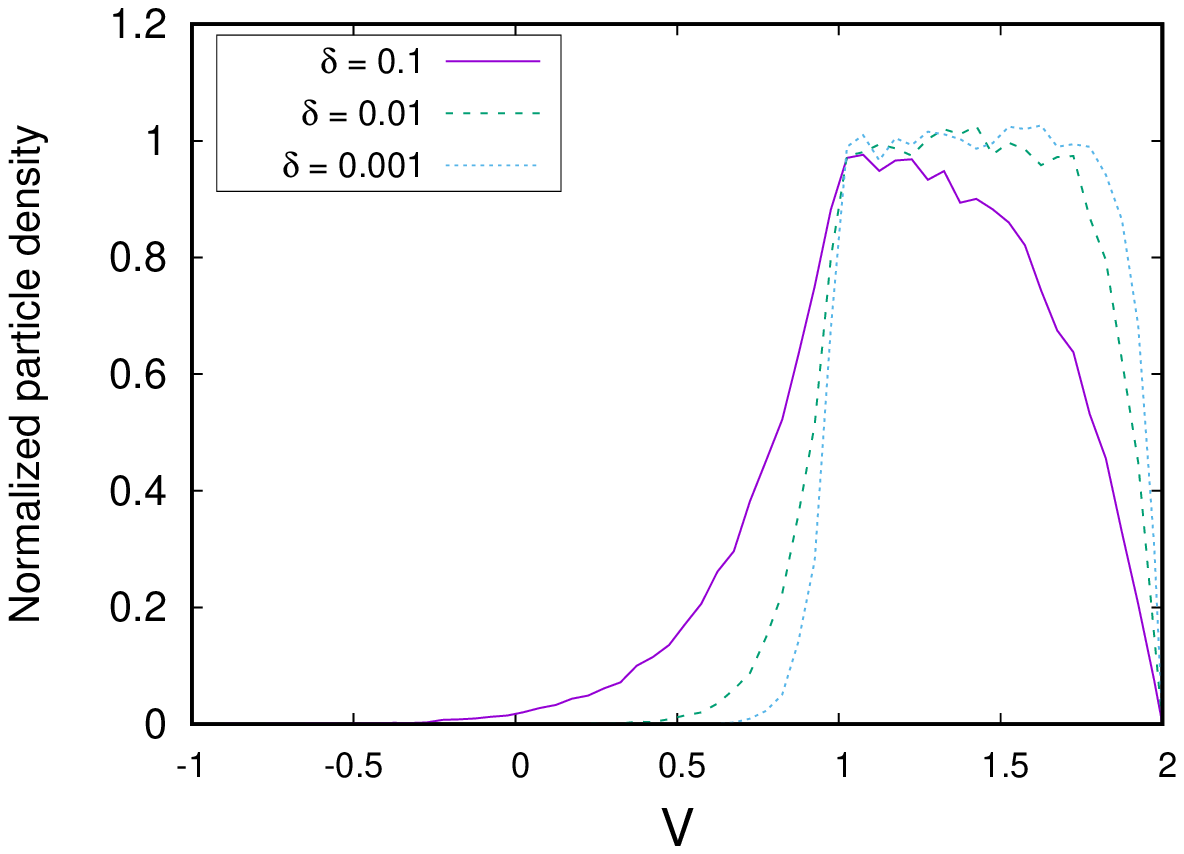}
		\caption{}
		\label{fig:b1-histograms-delay-sta-2}
	\end{subfigure}\hfil 
	
	\medskip
	\begin{subfigure}{0.38\textwidth}
		\includegraphics[width=\linewidth]{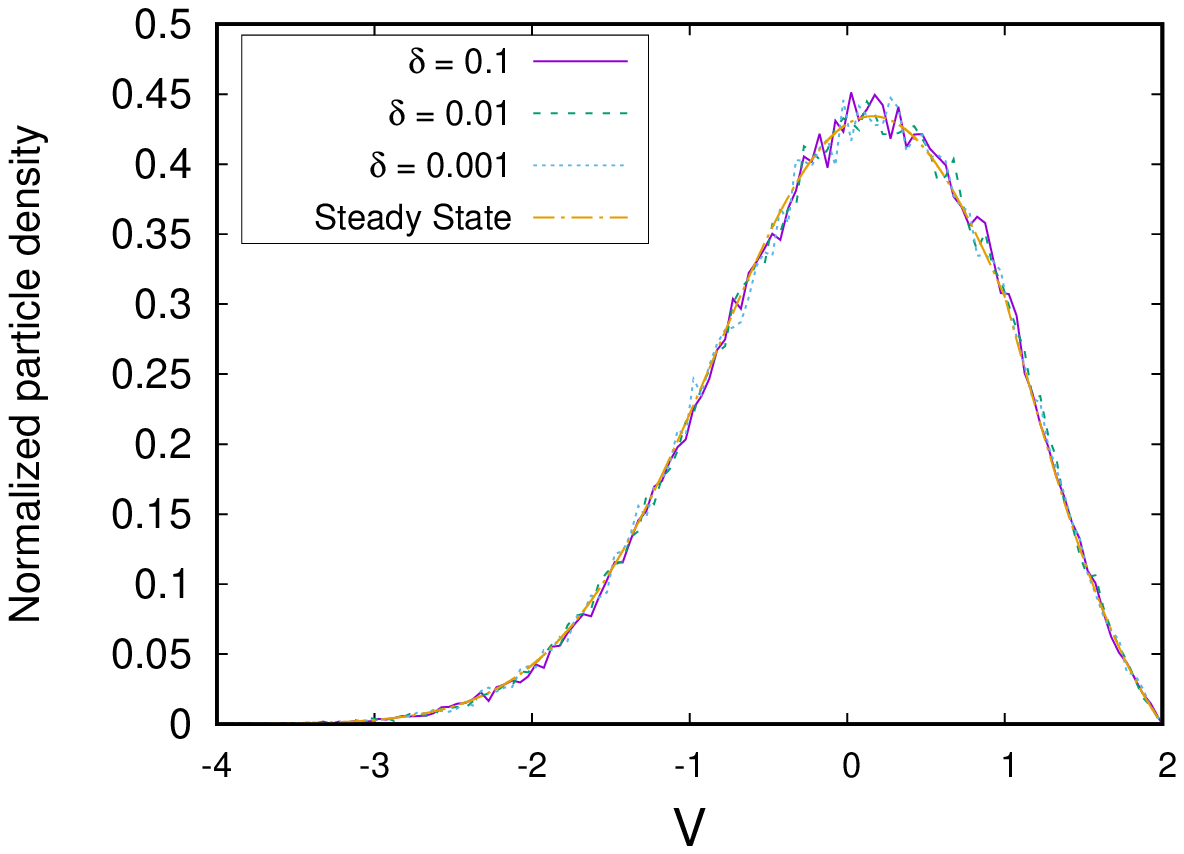}
		\caption{}
		\label{fig:b1-histograms-delay-sta-3}
	\end{subfigure}\hfil 
	
	\caption{
		{\bf Time evolution of a particle system with high transmission
			delays}.
		$\mathbf{b=1}$.
		The initial condition is a normal distribution with
		$ \nu_0 = 1.83 $ and $ \sigma = 0.003 $ 
		in all simulations for different delays.
		{\em (a)} Particle vol\-ta\-ge distribution at time $ t=0.0012 $ with different delay values.		
		{\em (b)} Particle vol\-ta\-ge distribution at time $ t=0.0025 $ with different delay values.	
		{\em (c)} Particle vol\-ta\-ge distribution at time $ t=5 $ with different delay values 
		compared to the steady state of the Fokker-Planck equation.
	}
	\label{fig:b1-delays}
\end{figure}

We have also studied the effect of higher values of delay for blow-up
situations.
As the delay value increases, the ``plateau'' state becomes unstable and
the system tends to the unique steady state, as shown in Fig.\ref{fig:b1-delays}. For the lower values $\delta=0.001$ and $\delta=0.01$ the system remains
close  to the "plateau" distribution for a while.
However, for $\delta=0.1$, the system does not even come close to
the "plateau" state,
see Figs.\ref{fig:b1-histograms-delay-sta-1} and
\ref{fig:b1-histograms-delay-sta-2}.
Finally, Fig.\ref{fig:b1-histograms-delay-sta-3}
shows how the system tends to the steady state for the three delay values
considered, even for the lowest ones,
for which the system was close to  the ``plateau'' state.

\

\subsection{Plateau formation}
\label{sec:result_justification-plateau}

In this section we try to understand the formation of
the observed ``plateau'' distributions.
Additionally,
we propose a possible explanation of that behaviour.

\begin{figure}[H]
	\begin{center}
		\begin{minipage}[c]{0.38\linewidth}
			\begin{center}
				\includegraphics[width=\textwidth]{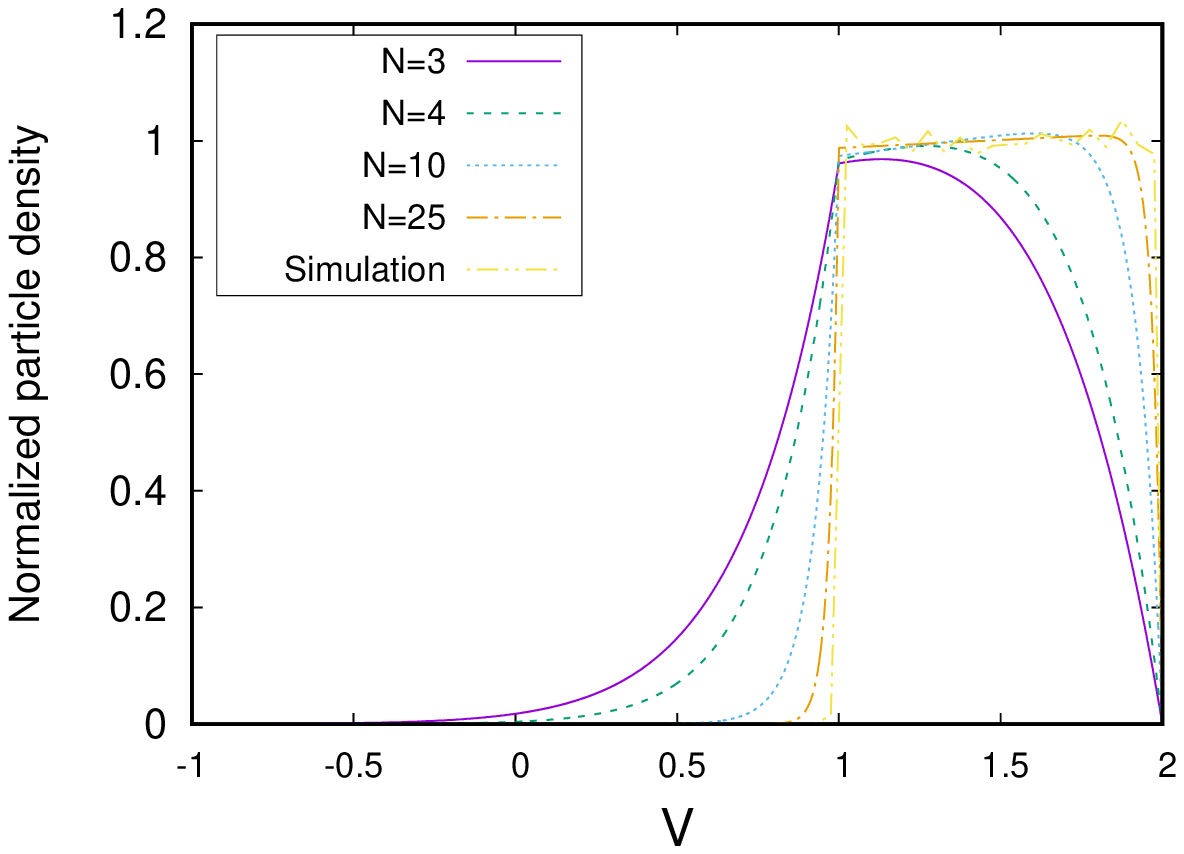}
			\end{center}
		\end{minipage}
		\begin{minipage}[c]{0.38\linewidth}
			\begin{center}
				\includegraphics[width=\textwidth]{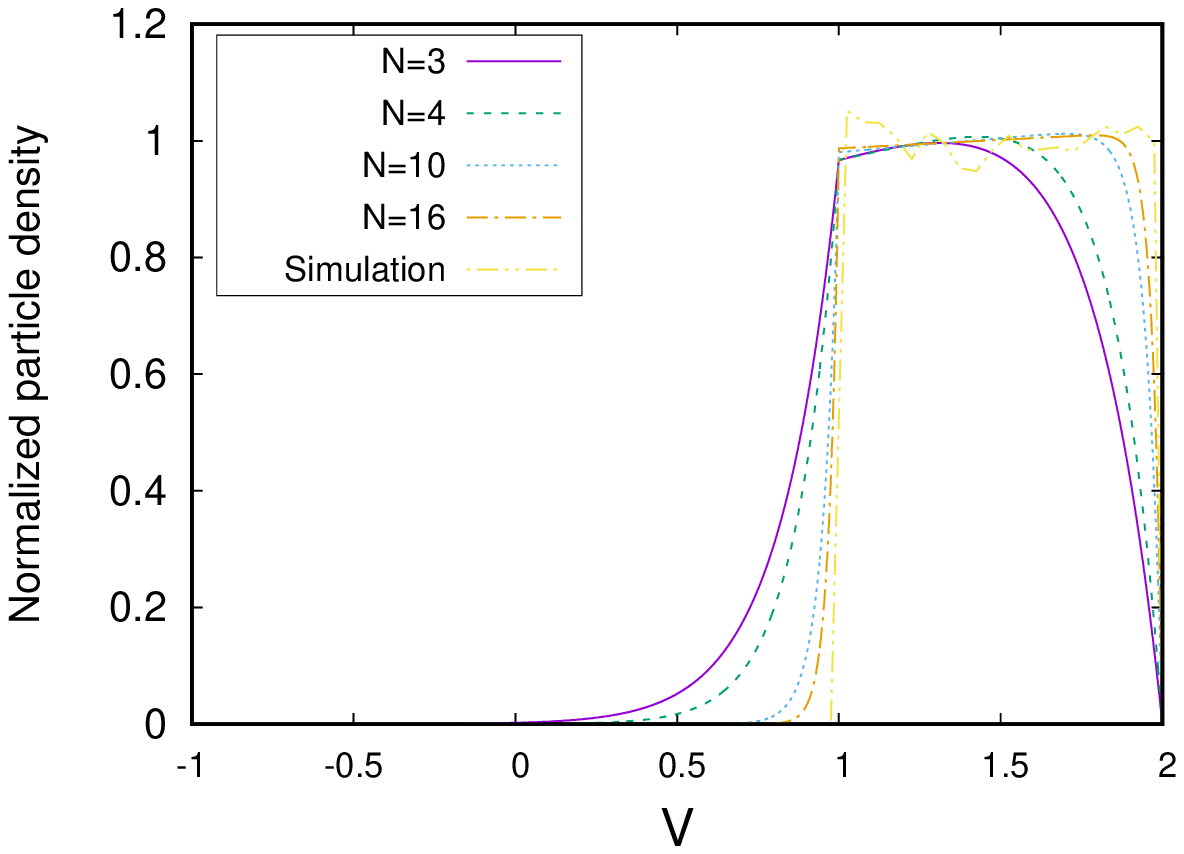}
			\end{center}
		\end{minipage}
	\end{center}
	\caption{{\bf Comparison  of ``plateau'' distributions with
			profiles \eqref{eq:profile-plateau}.}
		{\emph Left:} $b=1.5$.
		{\emph Right:}  $b=2.2$.}
	\label{fig:comparison_profiles-plateau}
\end{figure}

Fig.\ref{fig:comparison_profiles-plateau} shows the comparison
of  the ``plateau'' distributions found for $b=1.5$ and $b=2.2$,
and the following profiles:
\begin{equation}\label{eq:profile-plateau}
p(v) = \frac{\tilde{N}}{a}e^{-\frac{\left(v-bN\right)}{2a}}
\int_{max(v, V_R)}^{V_F}e^{\frac{\left(w-bN\right)^2}{2a}}dw,
\quad  N\in \R^+,
\end{equation}
where
$\tilde{N}=a\left(\int_{-\infty}^{V_F}
e^{-\frac{\left(v-bN\right)}{2a}}
\int_{max(v, V_R)}^{V_F}e^{\frac{\left(w-bN\right)^2}{2a}}dw dv
\right)^{-1}$, which guarantees the conservation law:
$\int_{-\infty}^{V_F} p(v) dv=1$.
There is a high coincidence
between the ``plateau'' distributions and the
profiles \eqref{eq:profile-plateau} as $N$ increases.
Furthermore, the dynamics of the particle system
appears to go through states with the form
\eqref{eq:profile-plateau}  with increasing $N$, during the formation of its
``plateau'' states (for instance, see Fig.\ref{fig:b22_histograms-delay-01_1}).

\

On the other hand,
we know that steady states of the Fokker-Planck equation are profiles
\eqref{eq:profile-plateau} as long as $\tilde{N}=N$, which is exactly the
Eq.(3.6) in \cite{caceres2011analysis}. That is, the
implicit equation  $NI(N)=1$, with $I(N)=a^{-1}\int_{-\infty}^{V_F}
e^{-\frac{\left(v-bN\right)}{2a}}
\int_{max(v, V_R)}^{V_F}e^{\frac{\left(w-bN\right)^2}{2a}}dw dv$.
Therefore, for any $N$ that is not a solution of that implicit equation,
the profile \eqref{eq:profile-plateau}   is not a stationary
distribution of the system. As a consequence, the profiles shown in Fig.\ref{fig:comparison_profiles-plateau} are not stationary states.
For $b=1.5$ (left plot) there are two steady states,
  but neither corresponds to the values shown in that figure,
  and for $b=2.2$ (right plot) there are no stationary
  solutions.

\

Therefore, the question is: does the system evolve towards profiles
\eqref{eq:profile-plateau} with $N$ tending to infinity?
Based on  our results, we think the answer could be
afirmative, because:
\begin{itemize}
	\item For the delayed systems,
	Figs.\ref{fig:b15-blowup-delay} and \ref{fig:b22-blowup-delay} show that the
	particle system appears to tend to different profiles
	\eqref{eq:profile-plateau},
	during the time intervals
	$J_m=\left(m\,\delta, (m+1)\,\delta\right)$, with $m\in\N$.
	In these intervals, the system seems to evolve towards "pseudo-stationary" states that verify the equation
	\begin{equation} 
	\frac{\partial}{\partial v}\left[ 
	(v-bN_\delta^m)p_m+a\frac{\partial}{\partial v}p_m(v) + \tilde{N}_\delta^m H(v-V_R) 
	\right]=0,  \nonumber
	\end{equation} 
	for certain values $N_\delta^m$ and
	$\tilde{N}_\delta^m$.
	Thus, taking into account  the boundary conditions,
	we  integrate the equation
	and find that 
	\begin{equation} 
	p_m(v)= \f{\tilde{N}_\delta^m}{a} e^{-\f{(v-bN_\delta^m)^2}{2a}}  \int_{\max(v, 
		V_R)}^{V_F} e^{\f{(w-bN_\delta^m)^2}{2a}} \,dw, 
	\label{stationarydelta}
	\end{equation}
	which corresponds to \eqref{eq:profile-plateau}.
	In certain sense, we can understand $N_\delta^m$ as an appro\-xi\-mation of
	a ``pseudo-stationary'' firing rate in the delay time.
	\item Moreover, 
	Fig.\ref{fig:b15-blowup-delay}(a) and Fig.\ref{fig:b22-blowup-delay}(a)
	show how  the firing rates increase over time. This also causes
	the sequences of $N_\delta^m$ and $\tilde{N}_\delta^m$ values to increase.
	As a result, profiles \eqref{stationarydelta}  look more and more
	like a ``plateau'' distribution, since
	$\lim_{N_\delta^m\to \infty} p_m(v)=\frac{1}{V_F-V_R}H(v-V_R)$, for $v<V_F$.
	
      \item In the case $ b=1 $ we also find a ``plateau'' distribution in
        the blow-up situations (see Fig.\ref{fig:b1-blowup-normal-delay}),
        even without the presence of transmission delay. In this case,
        there is only one steady state with finite firing rate.
        However, $ b=1 $ is the only connectivity value with which $\lim_{N\to \infty} NI(N)=1$ (see Fig.\ref{fig:steady-states} Left and
	\cite{caceres2011analysis}). The last statement leads us to think that profile \eqref{eq:profile-plateau} with
	$N\to \infty$, i.e. $\tilde{p}(v):=\frac{1}{V_F-V_R}H(v-V_R)$ for $v<V_F$
	and $\tilde{p}(V_F):=0$,
	is also a stationary solution. Although it would not be in the classical sense, since
	$\tilde{p}(V_R^-)\neq \tilde{p}(V_R^+)$.
\end{itemize}

\section{Conclusions}
\label{sec: Conclusions}

In this work we have made progress on the path towards a better
	understanding of the NNLIF models. We have numerically analysed
the particle system that originates these models.
First,
we have validated our numerical scheme by comparing our results with those
obtained for the Fokker-Planck equation in
\cite{caceres2011analysis,caceres2018analysis,caceres2019global}.
After which, we have discovered new properties of the solutions.
Specifically, we have shown what happens to the particle system when the
blow-up phenomenon occurs in a finite time. We have discovered that system behaviour after the blow-up depends on the connectivity parameter $ b $:
\begin{itemize}
	\item  For weakly connected systems, that is, $b <V_F-V_R$,
	  the system overcomes the explosion and tends to its
          unique steady state.
	This fact is in agreement with the global existence theory and
	with the notion of physical solution, instead of classical one
	\cite{delarue2015particle}. At the instant of the explosion,
	the firing rate diverges,
	but in such a way that the singularity of the expectation is controlled.
	Then the neurons tend to the equilibrium distribution of the system.
	\item For highly connected systems, understood as
	$b\ge V_F-V_R$, the concept of physical solution no longer makes sense,
	because neurons can fire twice at the same time.
	For this range of values,
	the system presents a wealth of properties, since
	it can have one, two or none steady state.
	Our numerical results show that:
	\begin{itemize}
		\item In situations where the blow-up would occur in a finite time, the system
		tends to a ``plateau'' distribution when synaptic delays are taken into
		account. That happens even with an
		initial distribution far from the threshold potential,
		in cases where there are no steady states.
		\item When the system exhibits two steady states, only the one
		with the lowest firing rate is stable, as shown in
		\cite{caceres2011analysis}.
		But our results also seem to indicate that ``plateau'' distributions
		are stable, when transmission delay is taken into account.
		Therefore, the system would exhibit bistability between
		the steady state with the lowest firing rate and the ``plateau''
		distribution.
		\item  The limiting case $b=1$ is especially
		interesting because the ``plateau'' profile \eqref{eq:profile-plateau}
		coincides with the stationary profile,
		with
		$N\to \infty$.
		Our results indicate that the system evolves to a "plateau" distribution
		under blow-up situations, either without synaptic delay or 
		with a very small delay value.
		Nevertheless, for a high enough delay the system avoids blow-up and tends to the stationary state, instead of the ``plateau''.
	\end{itemize}
\end{itemize}
From a neurophysiological point of view, the explosion phenomenon
can be understood as the synchronization of the neural network, where all
neurons spike at once.
This phenomenon occurs if synaptic delay is not taken into account. However, when some transmission delay is included in the model,
we show that the system moves into an asynchronous state
(stationary distribution away from the threshold value) or towards a
``plateau'' distribution, which means that
the membranes potential
tend to be uniformly distributed in the interval $ (V_R, V_F) $.
The  system seems to evolve towards these ``plateau'' distributions,
passing near the ``pseudo equilibria'' caused by the transmission delay. The limiting case $b=1$
has a ``pseudo equilibrium'' associated with $N \to \infty$,
that matches the profile of a ``limit steady-state'' (with $N\approx \infty$).
In this case, the ``plateau'' distribution is obtained without the need
to include any transmission delay.
The rigorous demonstration of the conjectures arising from our results is posed as future work.

\

\thanks{\em The authors acknowledge support from projects
MTM2017-85067-P
of Spanish Ministerio de Econom\'\i a, Industria y Competitividad and
the European Regional Development Fund (ERDF/FEDER).}

\bibliographystyle{acm}
\bibliography{neurons}
\end{document}